\documentclass [prc,twocolumn,superscriptaddress,floatfix,showpacs,amsmath,letterpaper] {revtex4}

\usepackage {bm}
\usepackage {graphicx}
\usepackage [bookmarks=true,bookmarksnumbered=true,colorlinks=false,pdfborder={0 0 0}] {hyperref}
\usepackage [final] {microtype}
\usepackage {dcolumn}

\newcommand {\MM} [1] {\ensuremath{#1}}
\newcommand {\tsp} [1] {\ensuremath{\mskip #1\thinmuskip}}
\newcommand {\IT} [1] {\ensuremath{#1}}
\newcommand {\RM}  [1] {\ensuremath{\textrm{#1}}}
\newcommand {\SQRT}[1] {\ensuremath{\sqrt{#1}}}

\newcommand {\collision} [2] {\MM{#1 + #2}}
\newcommand {\SUB} [2] {\MM{#1\ensuremath{_{#2}}}}
\newcommand {\SUP} [2] {\MM{#1\ensuremath{^{#2}}}}
\newcommand {\ABS} [1] {\MM{\ensuremath{|}#1\ensuremath{|}}}
\newcommand {\FRAC}[2] {\ensuremath{\frac{#1}{#2}}}
\newcommand {\DIV} [2] {\MM{#1/#2}}

\newcommand {\DIVe}[2] {\MM{#1\tsp{-1}/\tsp{-0.3}#2}}

\newcommand {\capsword} [1] {\textsc{#1}}

\newcommand {\NLO} {\capsword{NLO}}
\newcommand {\QCD} {\capsword{QCD}}
\newcommand {\pQCD} {p\capsword{QCD}}

\newcommand {\BNL} {\capsword{BNL}}
\newcommand {\RHIC} {\capsword{RHIC}}
\newcommand {\RCF} {\capsword{RCF}}
\newcommand {\NERSC} {\capsword{NERSC}}
\newcommand {\LBNL} {\capsword{LBNL}}

\newcommand {\PHENIX} {\capsword{PHENIX}}
\newcommand {\STAR} {\capsword{STAR}}

\newcommand {\CTEQ} {\capsword{CTEQ}}

\newcommand {\BEMC} {\capsword{BEMC}}

\newcommand {\SMD} {\capsword{SMD}}
\newcommand {\BBC} {\capsword{BBC}}
\newcommand {\BBCs} {\BBC{}s}
\newcommand {\TPC} {\capsword{TPC}}
\newcommand {\FTPC} {\capsword{FTPC}}

\newcommand {\MB} {\capsword{MB}}
\newcommand {\HT} {\capsword{HT}}
\newcommand {\NSD} {\capsword{NSD}}

\newcommand {\ZDC} {\capsword{ZDC}}
\newcommand {\ZDCs} {\ZDC{}s}
\newcommand {\ADC} {\capsword{ADC}}
\newcommand {\MIP} {\capsword{MIP}}
\newcommand {\CPV} {\capsword{CPV}}
\newcommand {\SVT} {\capsword{SVT}}
\newcommand {\SSD} {\capsword{SSD}}
\newcommand {\IFC} {\capsword{IFC}}
\newcommand {\DAQ} {\capsword{DAQ}}
\newcommand {\KKP} {\capsword{KKP}}
\newcommand {\GRV} {\capsword{GRV}}

\newcommand {\etameson} {\ensuremath{\eta}}
\newcommand {\etamesonprime} {\ensuremath{\eta\,'}}
\newcommand {\omegameson} {\ensuremath{\omega}}
\newcommand {\rhomeson} {\ensuremath{\rho}}
\newcommand {\lambdabaryon} {\ensuremath{\Lambda}}
\newcommand {\antilambdabaryon} {\ensuremath{\bar{\Lambda}}}
\newcommand {\mass} {\IT{m}}
\newcommand {\xvar} {\IT{x}}
\newcommand {\momentum} {\IT{p}}
\newcommand {\energy} {\IT{E}}

\newcommand {\Number} {\IT{N}}
\newcommand {\nucleon} {\IT{N}}

\newcommand {\transverse} {\IT{T}}

\newcommand {\xT} {\SUB{\xvar}{\transverse}}
\newcommand {\pT} {\SUB{\momentum}{\transverse}}
\newcommand {\kT} {\SUB{\IT{k}}{\transverse}}
\newcommand {\mT} {\SUB{\mass}{\transverse}}

\newcommand {\eT} {\SUB{\energy}{\transverse}}

\newcommand {\cspeed} {\IT{c}}

\newcommand {\pico} {\RM{p}}

\newcommand {\micro} {\ensuremath{\mu}}
\newcommand {\milli} {\RM{m}}
\newcommand {\centi} {\RM{c}}

\newcommand {\Mega} {\RM{M}}
\newcommand {\Giga} {\RM{G}}

\newcommand {\second} {\RM{s}}

\newcommand {\degree} {\SUP{}{\circ}}

\newcommand {\meter} {\RM{m}}

\newcommand {\mum}{\micro\meter}
\newcommand {\mm} {\milli\meter}
\newcommand {\cm} {\centi\meter}

\newcommand {\cmsquare} {\SUP{\cm}{2}}
\newcommand {\Volt} {\RM{V}}

\newcommand {\barn} {\RM{b}}
\newcommand {\mb} {\milli\barn}

\newcommand {\pb} {\pico\barn}

\newcommand {\pbinv} {\SUP{\pb}{-\tsp{-0.4}1}}

\newcommand {\eV} {\tsp{0.1}\RM{e}\tsp{-0.3}\Volt}
\newcommand {\GeV} {\Giga\eV}
\newcommand {\GeVc} {\DIVe{\GeV}{\cspeed}}

\newcommand {\GeVcc} {\DIVe{\GeV}{\SUP{\cspeed}{2}}}

\newcommand {\MeV} {\Mega\eV}

\newcommand {\MeVcc} {\DIVe{\MeV}{\SUP{\cspeed}{2}}}

\newcommand {\Tesla} {\RM{T}}

\newcommand {\rad} {\RM{rad}}
\newcommand {\mrad} {\milli\rad}

\newcommand {\unitns} [1] {\MM{#1}}
\newcommand {\unit} [1] {\MM{\tsp{1.5} #1}}

\newcommand {\xcoord} {\IT{x}}

\newcommand {\zcoord} {\IT{z}}
\newcommand {\der} {\IT{d}}
\newcommand {\dEdx} {\DIV{\der\energy\tsp{-0.2}}{\der\xcoord}}
\newcommand {\zvertex} {\SUB{\zcoord}{\RM{vert}}}
\newcommand {\BR} {\RM{\BR}}
\newcommand {\DCA} {\RM{\DCA}}

\newcommand {\Ncoll} {\SUB{\Number}{\RM{coll}}}
\newcommand {\Ncollmean} {\ensuremath{\langle}\Ncoll\ensuremath{\rangle}}

\newcommand {\rapidity} {\IT{y}}
\newcommand {\pseudorapidity} {\ensuremath{\eta}}

\newcommand {\phiangle} {\ensuremath{\varphi}}

\newcommand {\etacoord} {\ensuremath{\eta}}
\newcommand {\etacoordabs} {\ABS{\etacoord}}
\newcommand {\zvertexabs} {\ABS{\zvertex}}
\newcommand {\gama} {\ensuremath{\gamma}}
\newcommand {\EPS} {\ensuremath{\varepsilon}}
\newcommand {\Mgg} {\SUB{\mass}{\gama\gama}}
\newcommand {\APPROX} {\ensuremath{\approx}}
\newcommand {\LESS} {\ensuremath{<}}
\newcommand {\LESSEQ} {\ensuremath{\leq}}
\newcommand {\LESSAPPROX} {\ensuremath{\lesssim}}
\newcommand {\GREATER} {\ensuremath{>}}
\newcommand {\GREATEREQ} {\ensuremath{\geq}}
\newcommand {\GREATERAPPROX} {\ensuremath{\gtrsim}}
\newcommand {\PLMN} {\ensuremath{\pm}}
\newcommand {\DELTA} {\ensuremath{\Delta}}
\newcommand {\KAPPA} {\ensuremath{\kappa}}
\newcommand {\TIMES} {\ensuremath{\times}}
\newcommand {\PI} {\ensuremath{\pi}}
\newcommand {\TO} {\ensuremath{\to}}
\newcommand {\SIGMA} {\ensuremath{\sigma}}
\newcommand {\piplus} {\SUP{\PI}{+}}
\newcommand {\piminus} {\SUP{\PI}{-}}
\newcommand {\piplusminus} {\SUP{\PI}{\PLMN}}
\newcommand {\pizero} {\SUP{\PI}{0}}
\newcommand {\svar} {\IT{s}}

\newcommand {\sqrts} {\SQRT{\tsp{-0.5}\svar\tsp{0.5}}}
\newcommand {\sNN} {\SQRT{\tsp{-0.3}\SUB{\svar}{\tsp{-0.5}\nucleon\tsp{-1}\nucleon}\tsp{0.5}}}
\newcommand {\momentumthree}{\textbf{p}}
\newcommand {\muscale} {\ensuremath{\mu}}

\newcommand {\electron} {\IT{e}}
\newcommand {\proton} {\IT{p}}
\newcommand {\neutron} {\IT{n}}
\newcommand {\antiproton} {\ensuremath{\bar\proton}}
\newcommand {\antineutron} {\ensuremath{\bar\neutron}}
\newcommand {\deuteron} {\IT{d}}

\newcommand {\gold} {\RM{Au}}

\newcommand {\nucleus} {\IT{A}}

\newcommand {\protonproton} {\collision{\proton}{\proton}}

\newcommand {\deuterongold} {\collision{\deuteron}{\gold}}
\newcommand {\goldgold} {\collision{\gold}{\gold}}

\newcommand {\nucleonnucleon} {\collision{\nucleon}{\nucleon}}
\newcommand {\antineutronproton} {\collision{\antineutron}{\proton}}

\newcommand {\protonnucleus} {\collision{\proton}{\nucleus}}

\newcommand {\yr} [1] {#1}

\newcommand {\SMDe} {\SMD-\etacoord}
\newcommand {\SMDp} {\SMD-\phiangle}

\newcommand {\LUT} {\RM{\LUT}}
\newcommand {\PED} {\RM{\PED}}

\newcommand {\rBeamBg} {\IT{r}}

\newcommand {\FTPCRefMult} {\SUB{\Number}{\RM{\FTPC}}}

\newcommand {\Rcp} {\SUB{\IT{R}}{\IT{\tsp{0.4}C\tsp{-0.5}P}}}
\newcommand {\RAB} {\SUB{\IT{R}}{\IT{AB}}}

\newcommand {\RdA} {\SUB{\IT{R}}{\IT{\tsp{0.2}d\tsp{-0.1}A}}}
\newcommand {\TAB} {\SUB{\IT{T}}{\IT{AB}}}
\newcommand {\TdA} {\SUB{\IT{T}}{\IT{dA}}}
\newcommand {\TABmean} {\ensuremath{\langle}\TAB\tsp{-0.5}\ensuremath{\rangle}}
\newcommand {\TdAmean} {\ensuremath{\langle}\TdA\tsp{-0.4}\ensuremath{\rangle}}

\newcommand {\ee} [1] {\SUP{10}{#1}}
\newcommand {\e} [1] {\MM{\TIMES\ee{#1}}}

\newcommand {\Eseed}{\SUB{\energy}{\RM{seed}}}
\newcommand {\Eadd} {\SUB{\energy}{\RM{add}}}
\newcommand {\Emin} {\SUB{\energy}{\RM{min}}}
\newcommand {\Nmax} {\SUB{\Number}{\RM{max}}}

\newcommand {\etatopi} {\DIV{\etameson}{\pizero}}
\newcommand {\etatopisub} {\tsp{0.3}\DIV{\etameson\tsp{-0.5}}{\tsp{-0.6}\PI}}

\newcommand {\omegatopisub} {\tsp{0.3}\DIV{\omegameson\tsp{-0.6}}{\tsp{-0.6}\PI}}
\newcommand {\epluseminus} {\MM{\SUP{\electron}{+}\SUP{\electron}{-}}}

\newcommand {\HighTowerOne} {HighTower-\tsp{-0.4}1}
\newcommand {\HighTowerTwo} {HighTower-\tsp{-0.3}2}

\newcommand {\Rgamma} {\SUB{\IT{R}}{\tsp{0.3}\gama}}

\newcommand {\program} [1] {\textsc{#1}}

\newcommand {\PYTHIA} {\program{pythia}}
\newcommand {\GEANT} {\program{geant}}
\newcommand {\INCNLO} {\program{incnlo}}
\newcommand {\FLUKA} {\program{fluka}}

\newcommand {\opdash} {-}

\hypersetup {
pdfauthor = {STAR Collaboration},
pdftitle = {Inclusive neutral pion, eta meson, and direct photon production in p+p and d+Au collisions at sqrt(s_NN) = 200 GeV},
pdfsubject = {Neutral pion, eta meson, and direct photon production in high energy proton-proton and deuteron-gold collisions measured by the STAR experiment at the Relativistic Heavy Ion Collider (RHIC)},
pdfkeywords = {relativistic heavy ion collisions,high transverse momentum,neutral pion production,eta meson production,direct photon production,nuclear modification factor}
}

\begin {document}

\title {Inclusive $\protect\bm\pizero$, $\protect\bm\etameson$, and direct photon production at high transverse momentum \\
in $\protect\bm{\protonproton}$ and $\protect\bm{\deuterongold}$ collisions at $\protect\bm{\sNN{} = 200\unit{\GeV}}$
\enlargethispage{2\baselineskip}
}

\affiliation{Argonne National Laboratory, Argonne, Illinois 60439, USA}
\affiliation{University of Birmingham, Birmingham, United Kingdom}
\affiliation{Brookhaven National Laboratory, Upton, New York 11973, USA}
\affiliation{University of California, Berkeley, California 94720, USA}
\affiliation{University of California, Davis, California 95616, USA}
\affiliation{University of California, Los Angeles, California 90095, USA}
\affiliation{Universidade Estadual de Campinas, Sao Paulo, Brazil}
\affiliation{University of Illinois at Chicago, Chicago, Illinois 60607, USA}
\affiliation{Creighton University, Omaha, Nebraska 68178, USA}
\affiliation{Czech Technical University in Prague, FNSPE, Prague, 115 19, Czech Republic}
\affiliation{Nuclear Physics Institute AS CR, 250 68 \v{R}e\v{z}/Prague, Czech Republic}
\affiliation{University of Frankfurt, Frankfurt, Germany}
\affiliation{Institute of Physics, Bhubaneswar 751005, India}
\affiliation{Indian Institute of Technology, Mumbai, India}
\affiliation{Indiana University, Bloomington, Indiana 47408, USA}
\affiliation{University of Jammu, Jammu 180001, India}
\affiliation{Joint Institute for Nuclear Research, Dubna, 141 980, Russia}
\affiliation{Kent State University, Kent, Ohio 44242, USA}
\affiliation{University of Kentucky, Lexington, Kentucky, 40506-0055, USA}
\affiliation{Institute of Modern Physics, Lanzhou, China}
\affiliation{Lawrence Berkeley National Laboratory, Berkeley, California 94720, USA}
\affiliation{Massachusetts Institute of Technology, Cambridge, MA 02139-4307, USA}
\affiliation{Max-Planck-Institut f\"ur Physik, Munich, Germany}
\affiliation{Michigan State University, East Lansing, Michigan 48824, USA}
\affiliation{Moscow Engineering Physics Institute, Moscow Russia}
\affiliation{City College of New York, New York City, New York 10031, USA}
\affiliation{NIKHEF and Utrecht University, Amsterdam, The Netherlands}
\affiliation{Ohio State University, Columbus, Ohio 43210, USA}
\affiliation{Old Dominion University, Norfolk, VA, 23529, USA}
\affiliation{Panjab University, Chandigarh 160014, India}
\affiliation{Pennsylvania State University, University Park, Pennsylvania 16802, USA}
\affiliation{Institute of High Energy Physics, Protvino, Russia}
\affiliation{Purdue University, West Lafayette, Indiana 47907, USA}
\affiliation{Pusan National University, Pusan, Republic of Korea}
\affiliation{University of Rajasthan, Jaipur 302004, India}
\affiliation{Rice University, Houston, Texas 77251, USA}
\affiliation{Universidade de Sao Paulo, Sao Paulo, Brazil}
\affiliation{University of Science \& Technology of China, Hefei 230026, China}
\affiliation{Shandong University, Jinan, Shandong 250100, China}
\affiliation{Shanghai Institute of Applied Physics, Shanghai 201800, China}
\affiliation{SUBATECH, Nantes, France}
\affiliation{Texas A\&M University, College Station, Texas 77843, USA}
\affiliation{University of Texas, Austin, Texas 78712, USA}
\affiliation{Tsinghua University, Beijing 100084, China}
\affiliation{United States Naval Academy, Annapolis, MD 21402, USA}
\affiliation{Valparaiso University, Valparaiso, Indiana 46383, USA}
\affiliation{Variable Energy Cyclotron Centre, Kolkata 700064, India}
\affiliation{Warsaw University of Technology, Warsaw, Poland}
\affiliation{University of Washington, Seattle, Washington 98195, USA}
\affiliation{Wayne State University, Detroit, Michigan 48201, USA}
\affiliation{Institute of Particle Physics, CCNU (HZNU), Wuhan 430079, China}
\affiliation{Yale University, New Haven, Connecticut 06520, USA}
\affiliation{University of Zagreb, Zagreb, HR-10002, Croatia}

\author{B.~I.~Abelev}\affiliation{University of Illinois at Chicago, Chicago, Illinois 60607, USA}
\author{M.~M.~Aggarwal}\affiliation{Panjab University, Chandigarh 160014, India}
\author{Z.~Ahammed}\affiliation{Variable Energy Cyclotron Centre, Kolkata 700064, India}
\author{A.~V.~Alakhverdyants}\affiliation{Joint Institute for Nuclear Research, Dubna, 141 980, Russia}
\author{B.~D.~Anderson}\affiliation{Kent State University, Kent, Ohio 44242, USA}
\author{D.~Arkhipkin}\affiliation{Brookhaven National Laboratory, Upton, New York 11973, USA}
\author{G.~S.~Averichev}\affiliation{Joint Institute for Nuclear Research, Dubna, 141 980, Russia}
\author{J.~Balewski}\affiliation{Massachusetts Institute of Technology, Cambridge, MA 02139-4307, USA}
\author{O.~Barannikova}\affiliation{University of Illinois at Chicago, Chicago, Illinois 60607, USA}
\author{L.~S.~Barnby}\affiliation{University of Birmingham, Birmingham, United Kingdom}
% ...

\collaboration{\STAR\ Collaboration}\noaffiliation

\begin {abstract}
We report a measurement of high-\pT\ inclusive \pizero\!, \etameson, and direct photon production in \protonproton\ and \deuterongold\ collisions 
at \MM{\sNN{} = 200\unit{\GeV}} at midrapidity (\MM{0 \LESS{} \pseudorapidity{} \LESS{} 1})\@.
Photons from the decay \MM{\pizero{} \TO{} \gama\gama} were detected in the Barrel Electromagnetic Calorimeter of the \STAR\ experiment 
at the Relativistic Heavy Ion Collider.
The \MM{\etameson{} \TO{} \gama\gama} decay was also observed and constituted the first \etameson\ measurement by \STAR\@.
The first direct photon cross section measurement by \STAR\ is also presented, 
the signal was extracted statistically by 
subtracting the \pizero\!, \etameson, and \omegameson(782) decay background from the inclusive photon distribution observed
in the calorimeter.
The analysis is described in detail, and the results are found to be in good
agreement with earlier measurements and with next-to-leading order perturbative \QCD\ calculations.
\looseness=-1
\enlargethispage{2\baselineskip}
\end {abstract}


\maketitle

\tableofcontents

\newpage

\section {Introduction \label {se:introduction}}

The high center-of-mass energy (\MM{\sNN{} = 200\unit{\GeV}}) of the Relativistic Heavy Ion Collider (\RHIC)
opens up the hard scattering regime, which is accessed by measuring particle
production at high transverse momentum \pT\@. The high-\pT\ particles (\MM{\pT{} \GREATERAPPROX{} 3\unit{\GeVc}})
originate from the fragmentation of partons that have scattered in the early stage of the collisions.
Hence, in heavy-ion collisions the high-\pT\ particles can be used to
probe the produced medium of strongly interacting matter.
A significant suppression of high-\pT\ hadron production
relative to a simple binary collision scaling from \protonproton\ has been observed at \RHIC\ in 
central \goldgold\ collisions~\cite {ref_star_AuAu_suppression}\@.
Furthermore, it was found that jet-like correlations
opposite to trigger jets are suppressed, and that
the azimuthal anisotropy in hadron emission persists out
to very high \pT~\cite {ref_star_disap_corr_AuAu,ref_star_azim_aniz,ref_star_azim_aniz_pp_AuAu}\@. 
In contrast, no suppression effects were seen in \deuterongold\ collisions~\cite {ref_star_dAu_evidence,ref_phenix_pi0_dAu,ref_phobos_dAu,ref_brahms_dAu}, 
which has led to the conclusion that the observations made in \goldgold\ are due to the high-density medium
produced in such collisions and not to initial state effects.
The most probable explanation to date is that the suppression is due to parton
energy loss from induced gluon radiation (jet quenching)
in the extremely hot and dense medium~\cite{2004qgpconf123G}\@.
To quantitatively understand this behavior and, in particular, 
to separate hot from cold nuclear matter effects, such as Cronin effect~\cite{ref_cronin} and 
parton shadowing and antishadowing~\cite{ref_HIJING_shadowing,ref_EKS_shadowing,ref_nDS_shadowing}, precise measurements of identified
hadrons at high \pT\ in \protonproton\ and \deuterongold\ collisions are required~\cite{ref_dAu_1}\@.

Prompt photons have long been proposed as a powerful tool for studying the jet quenching via photon-jet correlations~\cite{ref_gammajet}\@.
In the dominant hard photon production processes (quark-gluon Compton scattering and quark-antiquark annihilation),
the outgoing photon balances the momentum of its partner parton
and has large enough mean free path to escape the collision system,
providing a calibrated probe for studying the energy loss and mean free path of the parton in the medium.
In addition, prompt photons constitute a background for measuring the medium-induced production of photons
in response to the energy deposited by that parton~\cite{PhysRevLett.90.132301}\@.

The thermal photon spectrum is directly related to the temperature of the hot and dense medium created in the heavy-ion collision,
provided that it is in thermal equilibrium~\cite{Peitzmann2002175}\@.
The measurement of such a spectrum requires a knowledge of the prompt photon background,
which can be measured in \protonproton\ and \deuterongold\ systems that share multiple sources of photons with heavy-ion collisions
but are not expected to produce an extended thermal system.

The measurements of \pizero's and direct photons in \protonproton\ collisions are also of specific interest 
for studies of the proton spin structure (see, e.g., Ref.~\cite {ref_rhic_spin_prospects}), which are underway at \RHIC\@.
A main objective of the \RHIC\ spin program is to constrain 
the polarization of the gluons inside the proton [\MM{\DELTA\IT{G}(\IT{x})}]\@.
The unpolarized cross sections provide a test of the next-to-leading order perturbative \QCD\ (\NLO\ \pQCD) framework, 
which is used to interpret the measured spin-dependent observables.

In this paper, we present the first results for the high-\pT\ \pizero\!, \etameson, and direct photon
production in \protonproton\ and \deuterongold\ collisions at \MM{\sNN{} = 200\unit{\GeV}}
in the pseudorapidity range \MM{0 \LESS{} \pseudorapidity{} \LESS{} 1}, measured by the \STAR\ experiment at \RHIC\ 
(except the cross section for \pizero\ production in \protonproton\ collisions, first presented in Ref.~\cite {ref_star_pi0ALL})\@.
The \STAR\ Barrel Electromagnetic Calorimeter was used to detect high-\pT\ \pizero\ and \etameson\ mesons via their \MM{\gama\gama} decays. 
The direct photon signal was extracted statistically by subtracting the \pizero\tsp{-1.5}, \etameson, and \omegameson(782) decay background
from the inclusive photon distribution observed in the calorimeter. 
The presented data constitute a necessary baseline for the measurements of \pizero\tsp{-1.5}, \etameson, 
and direct photon production in heavy-ion collisions at \RHIC\@.
Inclusive \pizero\ production was previously measured in \STAR\ for low \pT\ at midrapidity in \goldgold\ collisions 
at \MM{\sNN = 130\unit{\GeV}}~\cite {ref_star_auau130_pi0_tpc} and \MM{200\unit{\GeV}}~\cite {ref_star_auau200_pi0}, 
and at the forward rapidities in \protonproton\ and \deuterongold\ collisions at \MM{\sNN = 200\unit{\GeV}}~\cite {ref_star_fpd_pi0_pp_dAu}\@.
\STAR\ has also measured the production of other identified particles, such as \piplusminus, \SUP{\IT{K}}{\PLMN}, \proton/\antiproton, and 
hadronic resonances~\cite {ref_star_idhadrons1,ref_star_idhadrons,ref_dAu_1,ref_star_resonances_dAu}\@.
The \PHENIX\ experiment at \RHIC\ has also measured the \pizero, \etameson, and direct photon production at \MM{\sNN{} = 200\unit{\GeV}} 
in a variety of collision systems, including \protonproton\ and 
\deuterongold~\cite{Adare:2007dg,ref_phenix_pi0_dAu,ref_phenix_eta,ref_phenix_eta_pi0_dAu,ref_phenix_photons_pp}\@.

The paper is organized as follows.
In section~\ref {section_experiment}, we describe the detectors that were used in this analysis. 
In section~\ref {section_data_reconstruction}, we describe the data processing chain used to reconstruct photon candidates in the raw data.
Sections~\ref {section_pizero_analysis} and~\ref {sec_direct_photons} show how these photon candidates were used 
to calculate the yields of \pizero\ and \etameson, and direct photons, respectively.
Finally, in section~\ref {sec_results}, we present the results and compare our data to the theoretical calculations 
and to the measurements by other experiments.

\section {Experimental setup}
\label {section_experiment}

The \STAR\ detector (Solenoidal Tracker At \RHIC)~\cite {ref_star_overview_nim} was designed primarily for measurements of hadron production
in heavy-ion and proton\opdash{}proton collisions over a large solid angle.
For this purpose, tracking detectors with large acceptance and high granularity were placed inside a large-volume solenoidal magnetic field (0.5\unit{\Tesla})\@.
The detector subsystems relevant for the present analysis are briefly described in the following sections.

\subsection {Time Projection Chamber}

The Time Projection Chamber (\TPC)~\cite {ref_star_tpc} is the central tracking device in \STAR\@.
It allows one to track charged particles, measure their momenta, and identify the
particle species by measuring the ionization energy loss \dEdx\@.

The \TPC\ barrel measures \MM{4.2\unit{\meter}} in length, and has an inner radius of \MM{0.5\unit{\meter}}
and an outer radius of \MM{2\unit{\meter}}\@.
The \TPC\ acceptance covers \MM{\PLMN1.8} units in pseudorapidity and full azimuth.
Particle momentum is measured in the range \MM{0.1}--\MM{30\unit{\GeVc}}\@.
In this analysis, \TPC\ tracks were used to reconstruct the interaction vertex 
and to identify the energy deposits of charged particles in the calorimeter.

\subsection {Forward \TPC\ modules}

Two Forward Time Projection Chambers (\FTPC)~\cite {ref_star_ftpc} extend the \STAR\ tracking
capability to the pseudorapidity range \MM{2.5 \LESS{} \etacoordabs{} \LESS{} 4}\@.
Each \FTPC\ is a cylindrical volume with a 
diameter of \MM{75\unit{\cm}} and a length of \MM{120\unit{\cm}}, with radial drift field and
pad readout chambers mounted on the outer cylindrical surface.
Two such detectors were installed partially inside the main \TPC, on both sides of the interaction point.
In this analysis, the forward charged-track multiplicity recorded in the \FTPC\ in the gold beam direction 
served as a measure of the centrality in \deuterongold\ collisions.

\subsection {Barrel Electromagnetic Calorimeter}

A Barrel Electromagnetic Calorimeter (\BEMC)~\cite {ref_emc_nim} was incrementally added to the \STAR\ setup 
in \yr{2001}--\yr{2005} to measure the energy deposited by high-\pT\ photons and electrons and to provide a trigger signal.
The calorimeter is located inside the magnet coil and surrounds the \TPC,
covering a pseudorapidity range \MM{\etacoordabs{} \LESS{} 1} and full azimuth.

The full calorimeter consists of two contiguous half\tsp{0.4}-barrels, located east and west of the nominal interaction point, 
each of which is azimuthally segmented into \MM{60} modules.
Each module is approximately \MM{26\unit{\cm}} wide and covers \MM{6\unitns{\degree}} (\MM{105\unit{\mrad}})
in azimuth and one unit in pseudorapidity.
The active depth is \MM{23.5\unit{\cm}}, to which \MM{6.6\unit{\cm}} of structural elements are added at the outer radius.
Results presented in this paper used only the west calorimeter half\tsp{0.4}-barrel (\MM{0 \LESS{} \etacoord{} \LESS{} 1}), 
which was fully installed and calibrated in \yr{2003}--\yr{2005}\@.

The modules are segmented into \MM{40} projective towers of lead\opdash{}scintillator stacks, \MM{2} in \phiangle\ and \MM{20} in \etacoord\ direction\@.
A tower covers \MM{0.05\unit{\RM{rad}}} in \MM{\DELTA\phiangle} and \MM{0.05} units in \MM{\DELTA\etacoord}\@.
Each calorimeter half\tsp{0.4}-barrel is thus segmented into a total of \MM{2400} towers.
Each tower consists of a stack of \MM{20} layers of lead and \MM{21} layers of scintillator.
All these layers are \MM{5\unit{\mm}} thick, except the first two scintillator layers, which are \MM{6\unit{\mm}} thick.
A separate readout of these two layers provides the calorimeter preshower signal, which was not used in this analysis.
A Shower Maximum Detector (see below) is positioned behind the fifth scintillator layer.
The whole stack is held together by mechanical compression and friction between layers.
From layer-by-layer tests of the \BEMC\ optical system,
together with an analysis of cosmic ray and beam test data,
the nominal energy resolution of the calorimeter is estimated to be
\MM{\DIV{\ensuremath{\delta}\tsp{-0.3}\energy\tsp{-0.5}}{\tsp{-0.5}\energy} = \DIV{14\unitns{\%}}{\SQRT{\energy\,(\unitns{\GeV})}} \ensuremath{\oplus} 1.5\unitns{\%}}~\cite {ref_emc_nim}\@.

\subsection {Shower Maximum Detector}

The Shower Maximum Detector (\SMD) is a multi\tsp{0.2}-\tsp{-0.2}wire proportional counter with strip readout.
It is located at a depth of approximately \MM{5.6} radiation lengths at \MM{\etacoord{} = 0},
increasing to \MM{7.9} radiation lengths at \MM{\etacoord{} = 1},
including all material immediately in front of the calorimeter.
The purpose of the \SMD\ is to improve the spatial resolution of the calorimeter 
and to measure the shower profile.
This is necessary because the transverse dimension of each tower (\MM{\APPROX 10 \TIMES{} 10\unit{\cmsquare}})
is much larger than the lateral spread of an electromagnetic shower.
The improved resolution is essential to separate the two photon showers originating from the decay of high-momentum \pizero\ and \etameson\ mesons.

Independent cathode planes with strips along \etacoord\ and \phiangle\ directions allow the reconstruction of 
two projections of a shower.
The coverage is \MM{\DELTA\etacoord{} \TIMES{} \DELTA\phiangle = 0.0064 \TIMES{} 0.1\unit{\RM{rad}}}
for the \etacoord-strips and \MM{0.1 \TIMES{} 0.0064\unit{\RM{rad}}} for the \phiangle-strips, 
each group of \MM{2 \TIMES 2} towers covers \MM{15} strips in each \SMD\ plane behind it.
In total, \SMD\ contains \MM{36000} strips.

\subsection {Trigger detectors}

In addition to the \STAR\ barrel detectors, sampling hadronic calorimeters were placed at a distance of \MM{18\unit{\meter}} from the interaction point, 
on both sides of the experimental hall.
In heavy-ion collisions, these Zero Degree Calorimeters (\ZDC)~\cite {ref_rhic_beam_instrumentation,ref_rhic_zdc} 
measure the total energy of the unbound neutrons emitted from the nuclear fragments after a collision.
The charged fragments of the collision are bent away by the \RHIC\ dipole magnets upstream of the \ZDCs\@.
For the \deuterongold\ data used in this analysis, the \ZDC\ provided a collision trigger
by requiring the detection of at least one neutron in the gold beam~direction.

To provide a collision trigger in \protonproton\ collisions, Beam\opdash{}Beam Counters (\BBC)~\cite {ref_star_rel_lum_measur_bbc,ref_star_local_polar_bbc}
were mounted around the beam pipe beyond both poletips of the \STAR\ magnet at a distance of \MM{3.7\unit{\meter}} from
the interaction point.
The detector consists of two sets of small and large hexagonal scintillator tiles arranged into a ring 
that covers pseudorapidities between \MM{2.1} and \MM{5.0}\@.
The minimum bias trigger required a coincidence of signals in at least one of the \MM{18} small \BBC\ tiles on each side of the interaction region.

The two \BBC\ counters record timing signals that can be used to determine the time of flight for the forward fragments. 
The difference between these two flight times provides a measurement of the \zcoord\ position of the interaction vertex (\zvertex) 
to an accuracy of about \MM{40\unit{\cm}}~\cite{ref_grebenyuk_thesis}\@.
Events with large values of the time-of-flight difference, which indicate the passage of beam background, were rejected at the trigger level.
The \BBCs\ also served to measure the beam luminosity in \protonproton\ runs.

\section {Data reconstruction}
\label {section_data_reconstruction}

\subsection {Datasets and statistics}
\label {subsec_datasets}

The data used in this analysis were taken in the \deuterongold\ run of \yr{2003}\ and 
in the \protonproton\ run of \yr{2005}, both at \MM{\sNN = 200\unit{\GeV}}\@.
The integrated luminosity was \MM{0.66\unit{\pbinv}} for the \protonproton\ data 
and the equivalent nucleon\opdash{}nucleon luminosity was \MM{0.22\unit{\pbinv}} for the \deuterongold\ data.
The following trigger conditions were used:

\textit {Minimum bias (MinBias) trigger in \textit{d}\,+\,Au collisions}$\quad$\\
This condition required the presence of at least one neutron signal in the \ZDC\ in the gold beam direction.
As determined from detailed simulations of the \ZDC\ acceptance~\cite {ref_star_dAu_evidence}, this trigger captured \MM{(95 \PLMN{} 3)\unitns{\%}}
of the total \deuterongold\ hadronic cross section of \MM{\SUP{\SUB{\SIGMA}{\RM{hadr}}}{\deuterongold} = 2.21 \PLMN{} 0.09\unit{\barn}}\@.

\textit {MinBias trigger in \textit{p}\,+\,\textit{p} collisions}$\quad$\\
This condition required the coincidence of signals from two \BBC\ tiles on opposing sides of the interaction point.
Due to the dual-arm configuration, this trigger was sensitive to the non--singly diffractive (\NSD) cross section,
which is a sum of the non-diffractive and doubly diffractive cross sections.
The total inelastic cross section is a sum of the \NSD\ and singly diffractive cross sections.
A MinBias cross section of
\MM{\SUB{\SIGMA}{\RM{\BBC}} = 26.1 \PLMN{} 0.2\,(\RM{stat.}) \PLMN{} 1.8\,(\RM{syst.})\unit{\mb}}
was independently measured via van der Meer scans in dedicated accelerator runs~\cite {ref_bbc_vernierscan}\@.
This trigger captured \MM{(87 \PLMN{} 8)\unitns{\%}} of the \protonproton\ \NSD\ cross section, 
as was determined from a detailed simulation of the \BBC\ acceptance~\cite {ref_star_AuAu_suppression}\@.
Correcting the \BBC\ cross section for the acceptance, we obtained the \NSD\ cross section \MM{\SUP{\SUB{\SIGMA}{\RM{\NSD}}}{\protonproton} = 30.0 \PLMN{} 3.5\unit{\mb}}\@.

\textit {HighTower trigger}$\quad$\\
This condition required a transverse energy deposit \eT\ above a predefined threshold in at least one calorimeter tower,
in addition to satisfying the MinBias condition.
This trigger enriched the recorded dataset with events that had a large \eT\@.
Two different thresholds were applied, defining the \HighTowerOne\ and \HighTowerTwo\ datasets.
The nominal values of these thresholds were set to \MM{2.6} and \MM{3.5\unit{\GeV}} in \protonproton, and to \MM{2.5} and \MM{4.5\unit{\GeV}} in \deuterongold\ runs.
Prior to each run, all towers were equalized to give the uniform transverse energy response, 
by adjusting the high-voltage settings of the individual photomultipliers.

\textit {HighTower software filter}\\
The HighTower-triggered data were additionally filtered using a software implementation of the HighTower trigger.
In this filter, the highest tower \ADC\ value found in the event was required to exceed
the same \HighTowerOne\ (\HighTowerTwo) threshold as the one that was used during the run.
This filter was needed to remove events that were falsely triggered due to the presence of noisy channels (hot towers)\@.
Such channels were identified offline in a separate analysis and recorded in a database.
In addition, the highest calibrated transverse energy of a tower in the event 
was required to exceed slightly higher thresholds (\MM{\eT{} + 0.5\unit{\GeV}}) than those used during the run,
to account for possible inaccuracy of the online calibration of the towers.
This software filter also served to make the trigger efficiency for the Monte Carlo detector simulation and for the real data as close as possible.

\subsection {Beam background rejection}

During the data taking in \yr{2003}--\yr{2005}, interactions of beam ions with material
approximately \MM{40\unit{\meter}} upstream from the interaction region
gave rise to particles that traversed the detector almost parallel to the beam direction.
This source of background was eliminated by installing additional shielding in the \RHIC\ tunnel for the subsequent runs.

\begin {figure} [tb]
\centerline {\hbox {
\includegraphics {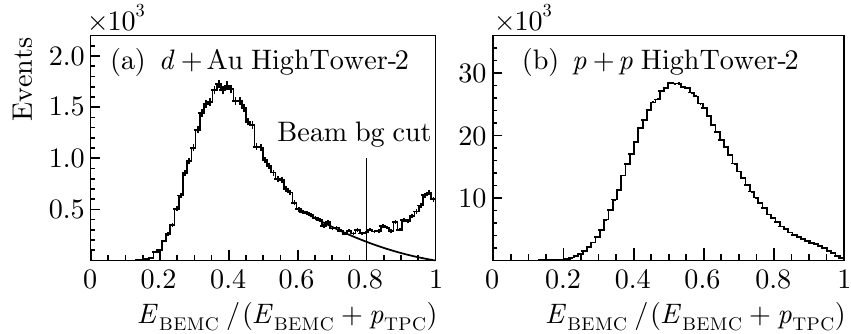}
}}
\caption {(a) Distribution of \MM{\rBeamBg{} = \DIV{\SUB{\energy}{\RM{\BEMC}}}{(\SUB{\energy}{\RM{\BEMC}} + \SUB{\momentum}{\RM{\TPC}})}} in \deuterongold\ events,
which shows beam background at \MM{\rBeamBg{} \GREATER{} 0.8}\@.
The curve corresponds to a second order polynomial fit,
constrained to pass through zero at \MM{\rBeamBg{} = 1},
used to estimate the false rejection rate.
(b) Distribution of \rBeamBg\ in \protonproton\ events\@.
\label {fig_cuts_beambg_1}
}
\end {figure}

To identify events containing such background tracks, the ratio
\begin {equation} \label {eq:rBeamBg_def}
\rBeamBg{} = \FRAC {\SUB{\energy}{\RM{\BEMC}}} {\SUB{\energy}{\RM{\BEMC}} + \SUB{\momentum}{\RM{\TPC}}}
\end {equation}
was calculated, where \SUB{\energy}{\RM{\BEMC}} is the total transverse energy recorded in the \BEMC\ and 
\SUB{\momentum}{\RM{\TPC}} is the transverse momentum sum of all charged tracks reconstructed in the \TPC\@.
In events containing background, \rBeamBg\ was large (close to 1)
because photons from these background events deposited a large amount of energy in the calorimeter,
while the accompanying charged tracks were not reconstructed in the \TPC, because they did not point to the vertex.
Figure~\ref {fig_cuts_beambg_1}
shows the distributions of \rBeamBg\ for the \deuterongold\ and \protonproton\ data.
The peak near unity in panel (a) indicates the presence of beam background in \deuterongold\ collisions.
Events with \MM{\rBeamBg{} \GREATER{} 0.8} were removed from the \deuterongold\ analysis.
This cut rejected \MM{3.4\unitns{\%}} of MinBias and \MM{13\unitns{\%}} of \HighTowerTwo\ events.
From a polynomial fit to the \deuterongold\ distribution in the region \MM{\rBeamBg{} = 0.6}--\MM{0.8}
[curve in Fig.~\ref {fig_cuts_beambg_1}(a)],
the false rejection rate was estimated to be \MM{3.6\unitns{\%}} in the \deuterongold\ \HighTowerTwo\ data
and less than \MM{1\unitns{\%}} in the other datasets.
By studying this rejection rate as a function of \SUB{\energy}{\RM{\BEMC}}, 
we estimated the potential distortion of the \pizero, \etameson, and photon spectra 
due to the removal of these events to be below \MM{1\unitns{\%}} in all datasets.

Figure~\ref {fig_cuts_beambg_1}(b) shows the distribution of \rBeamBg\ for the HighTower \protonproton\ data. 
The background was negligible because of the \BBC\ coincidence requirement in 
the trigger and the timing cut on the \BBC\ vertex position.
Therefore, no cut on \rBeamBg\ was applied to the \protonproton\ data.

The residual beam background contamination in the \deuterongold\ MinBias trigger was estimated from an
analysis of the empty \RHIC\ bunches to be \MM{(5 \PLMN{} 1)\unitns{\%}}~\cite {ref_dAu_1}\@.
To estimate the residual background in our data, we analyzed a sample of \MM{3\e{5}} MinBias triggers from unpaired \RHIC\ bunches.
These events were passed through the same reconstruction procedure as other data.
We observed that \APPROX\MM{10\unitns{\%}} of the fake triggers passed all cuts, and that none of these contained a reconstructed \pizero\@.
The residual beam background contamination in the \pizero\ yield was thus estimated 
to be \MM{0.1 \TIMES{} 5\unitns{\%} = 0.5\unitns{\%}} and~considered~to~be~negligible.

\subsection {Determination of centralities}
\label {subsec_centralities}

To measure the centrality in \deuterongold\ collisions, we used the correlation between
the impact parameter of the collision and the charged-track multiplicity in the forward direction.
This correlation was established from a Monte Carlo Glauber simulation~\cite {ref_glauber_annrev,ref_star_AuAu130,ref_star_idpart_pp_dAu_AuAu} using,
as an input, the Woods\opdash{}Saxon nuclear matter density for the gold ion~\cite {ref_woods_saxon}
and the Hulth\'{e}n wave function of the deuteron~\cite {ref_hulthen_2}\@.
In this simulation, the inelastic cross section for a nucleon\opdash{}nucleon collision
was taken to be \MM{\SUP{\SUB{\SIGMA}{\RM{inel}}}{\nucleon\tsp{-1}\nucleon} = 42\unit{\mb}}\@.
The produced particles were propagated through a full \GEANT~\cite {ref_geant} simulation of the \STAR\ detector. 
Both the charged track multiplicity and the number of nucleon\opdash{}nucleon collisions simulated by the event generator were recorded.

\begin {figure} [tb]
\centerline {\hbox {
\includegraphics {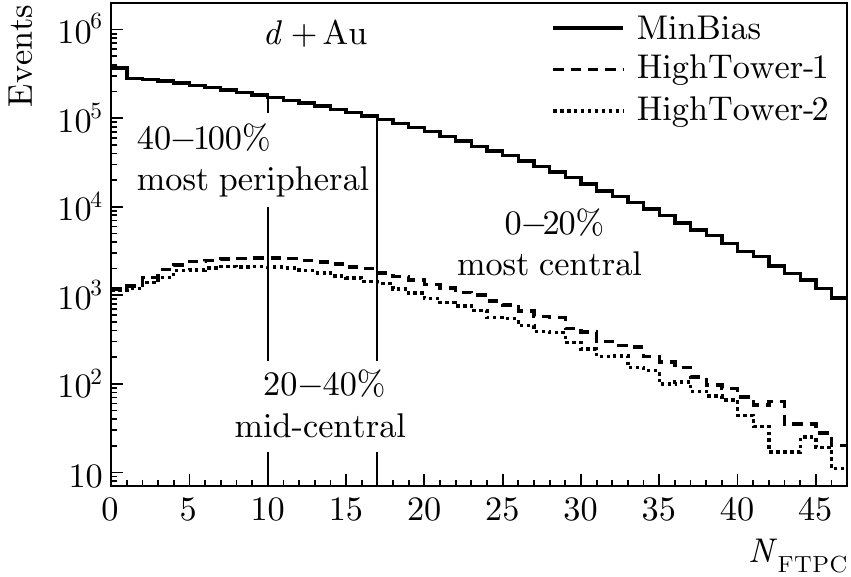}
}}
\caption {Centrality selection in the \deuterongold\ data, based on the \FTPC\ multiplicity \FTPCRefMult\@.
Three centrality classes were defined, containing \MM{0}--\tsp{-0.3}\MM{20\unitns{\%}} most central,
\MM{20}--\MM{40\unitns{\%}} mid-central, and \protect
\MM{40}--\tsp{-0.4}\MM{100\unitns{\%}} most peripheral events, respectively.
\label {fig_ftpc_centrality}
}
\end {figure}

For the event-by-event centrality determination, we measured the multiplicity of tracks
reconstructed in the \FTPC\ module in the gold beam direction (\FTPCRefMult)\@.
Centrality bins were defined following the scheme used in other \STAR\ publications~\cite {ref_star_dAu_evidence}\@.
The following quality cuts were applied to the reconstructed 
tracks: (i) at least \MM{6} hits were required on the track;
(ii) \MM{\pT{} \LESS{} 3\unit{\GeVc}}, which guaranteed 
that the track was fully contained in the \FTPC\ acceptance; and
(iii) distance of closest approach to the vertex had to be less than \MM{3\unit{\cm}}\@.
The multiplicity distributions obtained from the \deuterongold\ data are shown in Fig.~\ref {fig_ftpc_centrality}
for the MinBias, \HighTowerOne, and \HighTowerTwo\ triggers.

Based on \FTPCRefMult, the events were separated into three centrality classes:
\MM{0}--\tsp{-0.3}\MM{20\unitns{\%}} most central, \MM{20}--\MM{40\unitns{\%}} mid-central, and \MM{40}--\tsp{-0.4}\MM{100\unitns{\%}} most peripheral,
as indicated by the vertical lines in Fig.~\ref {fig_ftpc_centrality}\@.
Table~\ref {tab_centrality_classes}
\begin {table} [tbp]
\begin {center}
\caption {Centrality classes defined for the \deuterongold\ data and the corresponding \Ncollmean\ values~\protect\cite {ref_star_dAu_evidence}\@.
The errors given for \Ncollmean\ indicate the systematic uncertainty.}
\label {tab_centrality_classes}
\begin {ruledtabular} 
\begin {tabular} {@{\extracolsep{\fill}}D{.}{\ }{8.15}cD{,}{\,\PLMN\,}{4.3}}
\noalign{\smallskip}
\multicolumn {1} {c} {Centrality class} & \FTPCRefMult\ range & \multicolumn {1} {c} {\Ncollmean} \\
\noalign{\smallskip}\hline\noalign{\smallskip}
\multicolumn {1} {c} {\deuterongold\ MinBias}              & --               &  7.5 , 0.4 \\
\text{\MM{0}--\tsp{-0.3}\MM{20\unitns{\%}}}.\text{most central}      & $\geq 17$        & 15.0 , 1.1 \\
\text{\MM{20}--\MM{40\unitns{\%}}}.\text{mid-central}                & 10--16           & 10.2 , 1.0 \\
\text{\MM{40}--\tsp{-0.4}\MM{100\unitns{\%}}}.\text{most peripheral} & $< 10$           &  4.0 , 0.3 \\
\noalign{\smallskip}\hline\noalign{\smallskip}
\multicolumn {1} {c} {\protonproton\ \ \ \ }                   & --               & \multicolumn {1} {c} {\ \,1} \\
\end {tabular}
\end {ruledtabular}
\end {center}
\end {table}
lists the \FTPCRefMult\ ranges and the corresponding
mean numbers of binary collisions (\Ncollmean) obtained from the Glauber model, for each centrality class.
\label {syst_glauber_model}
The systematic uncertainties on \Ncollmean\ were estimated by varying the Glauber model parameters.

\subsection {Vertex finding efficiency}

In \protonproton\ data, a vertex was reconstructed based on the tracking information for \MM{65\unitns{\%}} of the MinBias events. 
For the remaining events, the vertex position in \zcoord\ was determined using the time information from the \BBCs\@.

In the \deuterongold\ HighTower data, the charged track multiplicities were large enough to always have a reconstructed vertex.
However, a vertex was missing in about \MM{7\unitns{\%}} of the MinBias events, and cannot be recovered from \BBC\ information
because the \BBC\ was not included in the \deuterongold\ MinBias trigger.
Events without a vertex have low charged track multiplicity, and the contribution from these
events to the \pizero\ yield above \MM{1\unit{\GeV}} was assumed to be negligible~\cite {ref_vertex_efficiency}\@.
\label {syst_vertex_eff}
Therefore, a correction for vertex inefficiency was applied as a constant normalization factor to the yield
and its uncertainty contributed to the total normalization uncertainty of the measured cross sections.

The vertex reconstruction efficiency in triggered \deuterongold\ MinBias events 
was \MM{\SUB{\EPS}{\RM{vert}} = 0.93 \PLMN{} 0.01}~\cite {ref_star_dAu_evidence}\@.
However, this efficiency depends on the collision centrality, and we assumed that it was \MM{100\unitns{\%}} for central events\@.
Scaling the efficiency above by the ratio of peripheral to total number of \deuterongold\ events, 
we obtained an efficiency correction factor of \MM{0.88 \PLMN{} 0.02} for the sample of peripheral events.

Events with \MM{\zvertexabs{} \GREATER{} 60\unit{\cm}} were rejected in the analysis, 
because the amount of material traversed by a particle increases dramatically at large values of \zvertexabs\@.
As a consequence, the \TPC\ tracking efficiency is reduced for vertices located far from the center of the detector.

\subsection {Energy calibration of the calorimeter}
\label {energy_calibration}

In the first step of the calorimeter calibration, the gains of the individual 
towers were matched to achieve an overall uniform response of the detector. 
For this purpose, minimum ionizing particles (\MIP) were used, by selecting the \TPC\ tracks of 
sufficiently large momentum (greater than \APPROX\MM{1\unit{\GeVc}})\@. 
These tracks were extrapolated to the \BEMC\ and the response spectra were accumulated, 
provided that the track extrapolation was contained within one tower and that the track was isolated.
For \deuterongold\ data, the isolation criterion meant that no other tracks were found in a \MM{3 \TIMES{} 3} patch around the tower;
for \protonproton\ data, these neighboring towers were required to have no signal above noise. 
The peak positions of such \MIP\ signals were used to calculate 
the tower-by-tower gain corrections needed to equalize the detector response~\cite {ref_star_bemc_mip_calibration}\@.

\begin {figure} [tb]
\centerline {\hbox {
\includegraphics {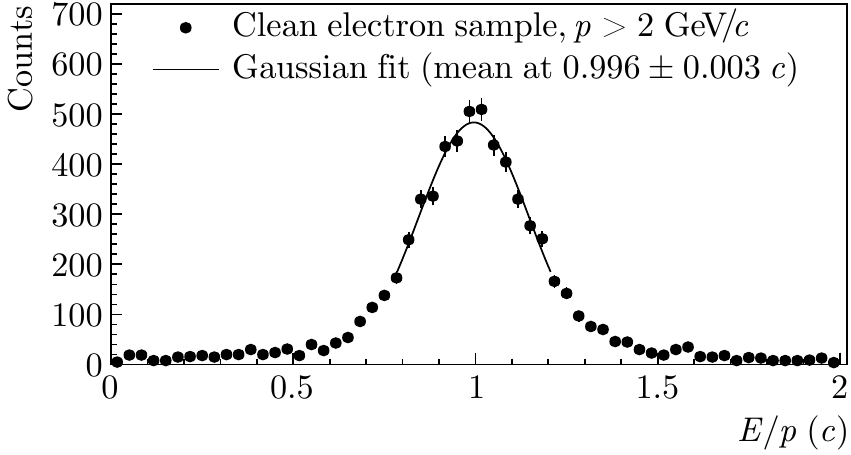}
}}
\caption {Electron energy measured in the \BEMC\ after calibration, 
divided by the momentum measured in the \TPC, in the \protonproton\ data\@.
Solid line is a Gaussian fit, which shows that the peak is centered at unity.
\label {fig_electron}
}
\end {figure}

In the second step, the energy scale was determined by comparing the momenta \momentum\ of 
identified electrons in the \TPC\ with the energies \energy\ recorded in the \BEMC\ using 
the relation \MM{\energy{} = \momentum} for ultra-relativistic electrons. 
Figure~\ref {fig_electron}
shows the distribution of \DIV{\energy\tsp{-0.2}}{\momentum} for a selected sample of 
at least \MM{90\unitns{\%}} pure electrons in the \protonproton\ data at \MM{\momentum \GREATER 2\unit{\GeVc}},
after the calibration has been performed. 
The Gaussian fit to the central part of the electron peak demonstrates that the mean has been placed at unity.
From a variation of the peak position with \momentum,
the systematic uncertainty of the electron calibration was conservatively estimated to be \MM{5\unitns{\%}}\@.
Within the present statistics, that calibration covers the momentum range only up to \MM{\momentum{} = 6\unit{\GeVc}}\@.
Because the peak position is close to unity at \MM{\momentum{} \GREATER{} 3.5\unit{\GeVc}},
we assume that the assigned systematic uncertainty covers possible non-linearities at higher
photon momenta \MM{\momentum{} \LESSAPPROX{} 15\unit{\GeVc}} probed in the present measurements.

This calibration method takes advantage of the well understood \TPC\ detector for the precise measurement 
of the electron track momentum in a wide range. 
A disadvantage is that it takes large statistics to calibrate the high-energy part of the spectrum. 
For this reason, only one global calibration constant was obtained.
It was found that the current calibration is less reliable at the edges of the half\tsp{0.4}-barrel. 
Therefore, the signals from the two \etacoord\tsp{0.5}-\tsp{0.2}rings at each side were removed from the analysis.

The absolute energy calibration of the \SMD\ was determined using the beam test data to an accuracy of about \MM{20\unitns{\%}}\@.
This analysis is not very sensitive to the absolute energy scale of the \SMD, because the main energy mesurement was done with the towers. 

\subsection {Particle reconstruction in the \BEMC}

The first step in the photon reconstruction was to find clusters of energy deposits in the calorimeter
by grouping adjacent hits that were likely to have originated from a single incident particle.
The cluster finding algorithm was applied to the signals from \BEMC\ towers and from each of the two \SMD\ layers.

The clustering started from the most energetic hit (seed) in a module and added neighboring hits of decreasing energy to the cluster, 
until either a predefined maximal cluster size or a lower hit energy threshold were reached. 
The algorithm then proceeded to process the next seed. 
The threshold values are listed in Table~\ref {tab:default_thresholds}\@.
\begin {table} [tbp]
\begin {center}
\caption {\label {tab:default_thresholds}Cluster finder threshold values used in the analysis.
\Eseed, \Eadd, and \Emin\ are the minimal energies for the seed hits, regular hits, and entire clusters, respectively.
\Nmax\ is the maximal cluster size.}
\begin {ruledtabular} 
\begin {tabular} {@{\extracolsep{\fill}}lD{.}{.}{1.2}D{.}{.}{1.4}D{.}{.}{1.2}c}
\noalign{\smallskip}
\multicolumn {1} {c} {Detector}  & \multicolumn {1} {c} {\Eseed\ (\unitns{\GeV})} & \multicolumn {1} {c} {\Eadd\ (\unitns{\GeV})} & \multicolumn {1} {c} {\Emin\ (\unitns{\GeV})} & \multicolumn {1} {c} {\Nmax} \\
\noalign{\smallskip}
\hline
\noalign{\smallskip}
Towers    & 0.35 & 0.035  & 0.02 & 4 \\
\SMD      & 0.2  & 0.0005 & 0.1  & 5 \\
\end {tabular}
\end {ruledtabular} 
\end {center}
\end {table}
By construction, the clusters were confined within a module and could not be shared by adjacent modules. 
However, the likelihood of shower sharing between modules is considered to be low, 
since the modules are physically separated by \APPROX\MM{12\unit{\mm}} wide air gaps. 
The \etacoord{}-\phiangle\ position of each cluster was calculated as the energy-weighted mean 
of the individual hit positions within the cluster.

After the tower and \SMD\ clusters were found, they were combined into \BEMC\ points,
which closely corresponded to the impact points and energy deposits of particles that traversed the calorimeter.
The procedure for forming the \BEMC\ points is described in detail in Ref.~\cite {ref_grebenyuk_thesis}\@.
The \SMD\ information was essential because the minimal opening angle of the decay photons 
decreases with increasing energy of the parent \pizero\@. 
The spatial resolution of the \BEMC\ towers alone is not sufficient to efficiently resolve the decay photons of 
\pizero's with \MM{\momentum{} \GREATER{} 5\unit{\GeVc}}\@.
For this reason, only the \BEMC\ points that contained tower, \SMDe, and \SMDp\ clusters were kept 
for the further analysis of the HighTower data.
In the analysis of MinBias data, used to obtain the \pizero\ signal at \MM{\pT \LESS 4\unit{\GeVc}}, 
all reconstructed \BEMC\ points were used, even when they did not contain \SMD\ clusters.

The \SMD\ efficiency decreases rapidly and its energy resolution becomes poor with decreasing energy of the traversing particle, 
leading to significant fluctuations in the strip readout for \MM{\energy{} \ensuremath{\lesssim} 2\unit{\GeV}}\@.
Therefore, in the \HighTowerOne\ data the \SMD\ clusters were accepted only when they contained signals from at least two strips.
This cut rejected a large fraction of the distorted and falsely split \SMD\ clusters,
and reduced a possible effect of poor \SMD\ response simulation at low energies.

\subsection {Charged particle veto using \TPC}
\label {subsec:cpv}

A charged particle veto (\CPV) cut was applied to reject the charged hadrons that were detected in the calorimeter.
A charged hadron was recognized as a \BEMC\ cluster with a \TPC\ track pointing to it.
The cluster was rejected if the distance \IT{D} between the \BEMC\ point and the closest \TPC\ track in the \etacoord\opdash\phiangle\ coordinates was
\begin {equation} \label {eq:Ddef}
\IT{D} = \SQRT{\SUP{(\DELTA\etacoord)}{2} + \SUP{(\DELTA\phiangle)}{2}} \LESS{} 0.04
.
\end {equation}
When a track was projected to the calorimeter surface, at a radius \MM{\IT{R} = 220\unit{\cm}}, 
this cut corresponded to a linear separation \MM{\IT{R}\IT{D} \APPROX{} 10\unit{\cm}} in the pseudorapidity range of this measurement\@.
The efficiency of this cut was \MM{35\unitns{\%}} in the MinBias data and \MM{71\unitns{\%}} in the \HighTowerOne\ and \HighTowerTwo\ data, 
in all \protonproton\ and \deuterongold\ datasets.
The \BEMC\ points remaining after this cut were considered to be photon candidates,
and were combined into pairs, defining the set of \pizero\ candidates.

\begin {figure} [tb]
\centerline {\hbox {
\includegraphics {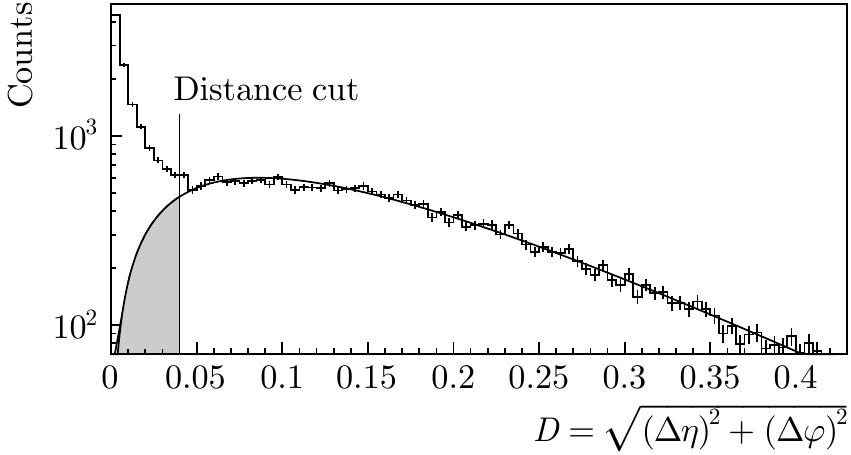}
}}
\caption {Distribution of the distances \IT{D} between \BEMC\ points and their closest tracks,
obtained from \protonproton\ \HighTowerOne\ data in the bin \MM{4 \LESS{} \pT{} \LESS{} 5\unit{\GeVc}}\@.
The curve shows a fit to Eq.~(\ref {eq:cpv_distr}), the vertical line indicates the \CPV\ cut.
\label {fig_cpv_cut}
}
\end {figure}

This veto introduced a false rejection of photon clusters
if an unrelated charged particle happened to hit the calorimeter close to the cluster.
Figure~\ref {fig_cpv_cut} shows 
the distribution of \IT{D} observed in the \protonproton\ data.
In this plot, one distinguishes the peak of real charged particles at small distances,
superimposed on a random component seen as a shoulder at larger distances.
Assuming a uniform distribution of track projections in \etacoord\ and \phiangle\ around the \BEMC\ point,
the radial distribution is given by
\begin {equation} \label {eq:cpv_distr}
\IT{f}(\tsp{-0.5}\IT{D}) =
\IT{D}
\tsp{1.0}
\SUP{\IT{e}}{-\IT{D} \ensuremath{\rho}}
.
\end {equation}
Here \ensuremath{\rho} is the charged track density in the vicinity of the photon.
This parameter was obtained from a simultaneous fit to the data in all bins of the event multiplicity \IT{M} measured in the \TPC,
assuming a linear dependence on \IT{M}, \MM{\ensuremath{\rho} = \IT{a} + \IT{b}\IT{M}}\@.
The parametrization given by Eq.~(\ref {eq:cpv_distr}) describes well the random component, as shown by the curve in Fig.~\ref {fig_cpv_cut}\@.
The relative number of random coincidences that were falsely rejected was obtained by integrating the fitted curve up to the distance cut
and weighting with the multiplicity distribution observed in each \pT\ bin.
The resulting correction factor was \MM{\SUP{\SUB{\EPS}{\RM{cpv}}}{\PI} = 0.94 \PLMN{} 0.02} for the \protonproton\ data
and \MM{0.89 \PLMN{} 0.02} for the \deuterongold\ data.
\label {syst_cpv}

In the direct photon analysis, the purity of the photon candidate sample was more important than in the \pizero\ analysis, 
therefore, a stronger cut \MM{\IT{R}\IT{D} \LESS{} 15\unit{\cm}} was used.
The correction factors  were calculated to be \MM{\SUP{\SUB{\EPS}{\RM{cpv}}}{\gama} = 0.95 \PLMN{} 0.02} for \protonproton\ and 
\MM{0.93 \PLMN{} 0.02} for \deuterongold\ data.
The residual contamination by charged particles (\SUB{\IT{C}}{\PLMN}) was estimated from the integrated excess of the
\IT{D} distribution over the fit to the random associations in the interval \MM{15 \LESS{} \IT{R}\IT{D} \LESS{} 25\unit{\cm}},
and was less than \MM{5\unitns{\%}} for all \pT\ bins.

The uncertainties of these corrections contributed to a \pT-independent
systematic uncertainty of the \pizero\!, \etameson, and direct photon yields.

\subsection {Photon conversions} 

A separate study was done to determine the degree to which the \GEANT\ geometry described the distribution of material in the real \STAR\ detector, 
and the corresponding correction factors \SUB{\IT{c}}{\RM{loss}} were extracted to account for any differences.

The photon conversion probability \SUB{\IT{P}}{\RM{conv}} as a function of the depth \IT{d} traversed in a material is given by
\begin {equation}\label {eq:conversion_gamma} 
\SUB{\IT{P}}{\RM{conv}} = 1 - \ensuremath{\exp{(-\IT{d}/\SUB{\IT{d}}{0})}},
\end {equation}
where \SUB{\IT{d}}{0} is the mean free path of the photon in that material. 
The probablity that a \pizero\ was not detected because at least one of its decay 
photons has converted is
\begin {equation}\label {eq:pions:23} 
\SUP{\SUB{\IT{P}}{\RM{loss}}}{\PI} = 2 \SUB{\IT{P}}{\RM{conv}} (1 - \SUB{\IT{P}}{\RM{conv}}) + \SUP{\SUB{\IT{P}}{\mathrm{conv}}}{2}
.
\end {equation} 
The \pizero\ losses due to conversions were in principle taken into account in the simulations mentioned in the sections below, 
because the material traversed by the photons was included in the \GEANT\ model of the detector. 
However, it was observed that the simulation failed to reproduce the number of photon conversions 
in the inner tracking system (\SVT, \SSD), and in the \TPC\ Inner Field Cage (\IFC), all of which have a very complicated geometry of silicon sensors, 
readout electronics, and support structures~\cite{ref_star_auau130_pi0_tpc,ref_star_auau200_pi0,ref_wetzler_thesis}. 
The number of conversions in the simulated \SVT, \SSD, and \IFC\ were underestimated 
by factors of \MM{\KAPPA{} = 2}, \MM{2}, and \MM{1.2}, respectively, compared to that in the real data~\cite {ref_russcher_thesis}\@.
In simulations, the photon conversion probability in these detectors was in the range \MM{\SUB{\IT{P}}{\RM{conv}} = 0.3\text{--}3.3\unitns{\%}}\@.
To account for the missing material in the \GEANT\ model,
the photon spectra were corrected by factors 
\MM{\SUP{\SUB{\IT{c}}{\RM{loss}}}{\gama} = \DIV{(1 - \ensuremath{\sum}\SUB{\IT{P}}{\RM{conv}})}{(1 - \ensuremath{\sum}\KAPPA\SUB{\IT{P}}{\RM{conv}})}},
with the values of \MM{1.06 \PLMN{} 0.02} and \MM{1.03 \PLMN{} 0.02} for the \protonproton\ and \deuterongold\ data, respectively.
Using Eq.~(\ref {eq:pions:23}), this corresponds to correction factors of \MM{\SUP{\SUB{\IT{c}}{\RM{loss}}}{\PI} = 1.12 \PLMN{} 0.03} (\protonproton) 
and \MM{1.07 \PLMN 0.03} (\deuterongold) for the \pizero\ spectra. 
Because the photon attenuation length in most absorbers rapidly approaches a constant for energies larger than \APPROX\MM{100\unit{\MeV}},
the correction factors were assumed to be independent of the photon \pT\@. 

\section {Neutral pion and eta meson analysis}
\label {section_pizero_analysis}

The \pizero\ and \etameson\ were identified by their decays
$$
\pizero{} \TO{} \gama\gama{} \quad \mbox{and}\quad \etameson{} \TO{} \gama\gama
.
$$
These decay modes have branching ratios of \MM{0.988} and \MM{0.392}, respectively~\cite {ref_pdg}\@.
The \pizero\ lifetime is \MM{\ensuremath{\tau} = 8.4\e{-\tsp{-0.4}17}\unit{\second}}, 
corresponding to a decay length \MM{\cspeed\ensuremath{\tau} = 0.025\unit{\mum}}\@.
The \etameson\ lifetime is even shorter (\MM{7\e{-\tsp{-0.4}19}\unit{\second}})\@.
Therefore, we assumed that the decay photons originated from the primary vertex. 
For each event, the invariant mass
\begin {equation} \label {eq:inv_mass}
\Mgg{} = \SQRT{\tsp{0.2} 2 \tsp{0.7} \SUB{\energy}{1} \SUB{\energy}{2} (1 - \ensuremath{\cos\psi})}
\end {equation}
was calculated for all pairs of photons detected in the \BEMC\@. 
Here \SUB{\energy}{1} and \SUB{\energy}{2} are the energies of the decay photons and \ensuremath{\psi} is the opening angle in the laboratory system.
The reconstructed masses were accumulated in invariant mass spectra, where the \pizero\ and the \etameson\ showed up as peaks around their 
nominal masses (\MM{\SUB{\mass}{\pizero} = 0.135\unit{\GeVcc}} and \MM{\SUB{\mass}{\etameson} = 0.547\unit{\GeVcc}}). 
These peaks were superimposed on a broad distribution of combinatorial background,
which originated from photon pairs that were not produced by the decay of a single parent particle.

\subsection {Asymmetry of photon pairs}

\begin {figure} [tb]
\centerline {\hbox {
\includegraphics {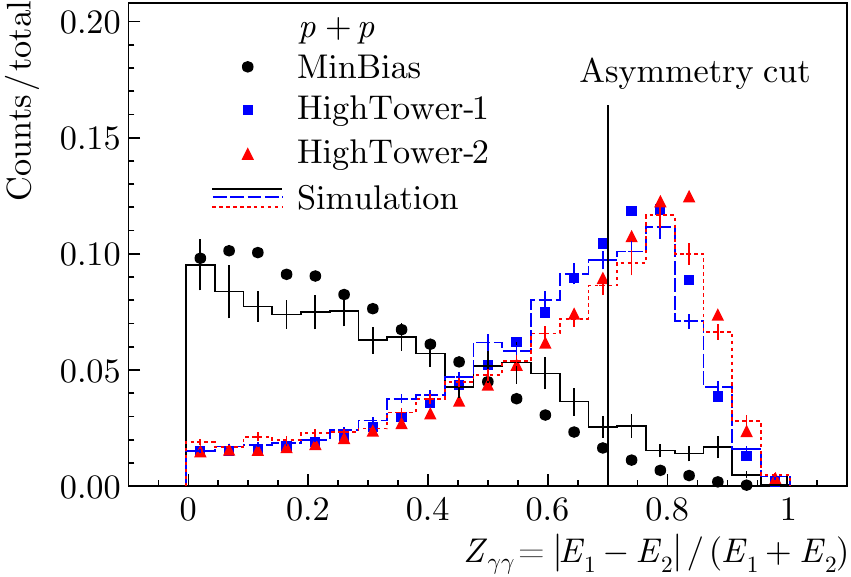}
}}
\caption {(color online) The energy asymmetry \SUB{\IT{Z}}{\gama\gama} of photon pairs reconstructed in \protonproton\ data (symbols) 
and in Monte Carlo simulation (histograms) for various triggers, 
normalized to unity for each trigger.
\label {fig_asymmetry}
}
\end {figure}

The energy asymmetry of the two\tsp{0.3}-body decay of neutral mesons is defined as
\begin {equation}
\SUB{\IT{Z}}{\gama\gama} \ensuremath{\equiv} \FRAC {\left| \SUP{\SUB{\energy}{1}}{\phantom{1}} - \SUB{\energy}{2} \right|} {\SUB{\energy}{1} + \SUB{\energy}{2}}
.
\end {equation}
From the decay kinematics it follows that the probability for a given \SUB{\IT{Z}}{\gama\gama} is independent on \SUB{\IT{Z}}{\gama\gama}\@.
Figure~\ref {fig_asymmetry}
shows the distribution of the asymmetry of photon pairs reconstructed in \protonproton\ data, including both \pizero\ and \etameson\ signals and background.
In the MinBias data the distribution is not flat because of the acceptance effects\tsp{0.5}---\tsp{0.5}photons from an asymmetric
decay have a large opening angle and one of them is likely to escape the \BEMC\@.
It is also seen that the HighTower energy threshold biases the asymmetry towards larger values,
because it is easier for an asymmetric decay to pass the trigger.
The corresponding asymmetry distributions obtained from the Monte Carlo simulation, which represented the pure \pizero\ signal, are also shown. 
The details of the simulation are given in section~\ref {subsect_eff_corr} below.
Asymmetries observed in the simulation are in general agreement with those in the real data, 
considering the presence of background in the data.

In this analysis, the \pizero\ and \etameson\ candidates were only accepted if the asymmetry was less than \MM{0.7}\@.
This cut rejected very asymmetric decays, where one of the \BEMC\ points had low energy.
It also rejected a significant part of the low-mass background (this background will be described in section~\ref {sec:lowmassbg})\@.
The asymmetry cut improved the \pizero\ signal-to-background ratio by a factor of \APPROX\MM{1.5}\@.

\subsection {Kinematic distributions}

\begin {figure} [tb]
\centerline {\hbox {
\includegraphics {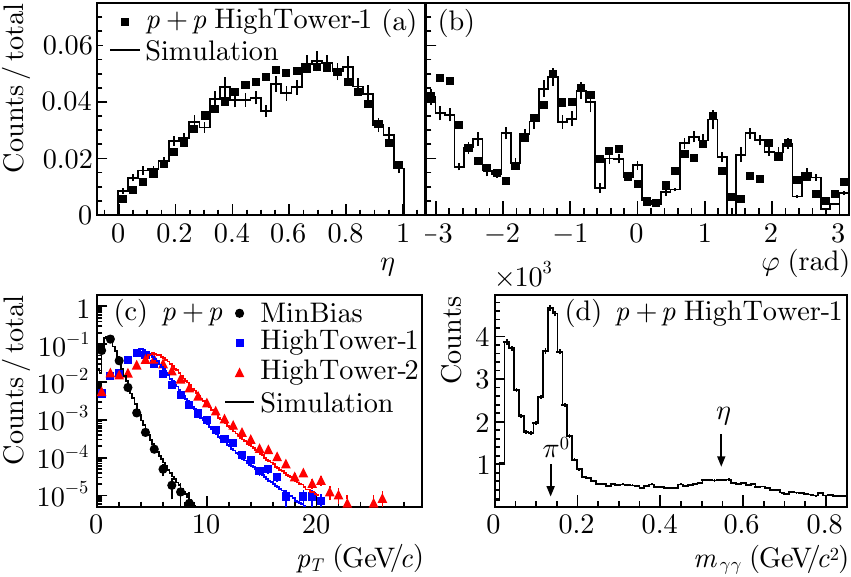}
}}
\caption {(color online) Distribution of \pizero\ candidates obtained from the \protonproton\ data, as a function of (a) \etacoord\ and (b) \phiangle,
and (c) \pT\ and (d) \Mgg\@.
\label {fig_data_candidates}
}
\end {figure}

For each \pizero\ candidate, the pseudorapidity \etacoord, the azimuth \phiangle,
the transverse momentum \pT, and the invariant mass \Mgg\ [Eq.~(\ref {eq:inv_mass})] were calculated.
Figure~\ref {fig_data_candidates}
shows the \etacoord, \phiangle, \pT, and \Mgg\ distributions of the \pizero\ candidates in the \protonproton\ data.
For the \deuterongold\ data these distributions are similar to those shown for \protonproton\@.
The corresponding \etacoord, \phiangle, and \pT\ distributions of the \pizero's reconstructed in the simulation are also shown. 

The \etacoord\ distribution shows the decrease of the calorimeter acceptance at \MM{\etacoord{} = 0} and \MM{\etacoord{} = 1},
because it is likely that one of the decay photons at the calorimeter edges escapes detection.
The asymmetry of the \etacoord\  distribution is due to the fact that the calorimeter half\tsp{0.4}-barrel is positioned 
asymmetrically with respect to the interaction region.
The azimuthal dependence of the calorimeter acceptance was
caused by failing \SMD\ modules (the data used in this paper are from the early years of detector operation, 
in which such failures occured relatively frequently), as well as by dead and hot towers.
The gross features of the data reflecting the calorimeter acceptance are reasonably well reproduced 
by the pure \pizero\ Monte Carlo simulations.

Figure~\ref {fig_data_candidates}(c) shows the \pT\ distribution of the photon pairs, 
separately for the MinBias and for the two HighTower datasets.
The HighTower trigger threshold effects are reasonably well reproduced in the simulation.
It is seen that using the HighTower triggers significantly increased the rate of \pizero\ candidates at high \pT\@.

The \pT-integrated invariant mass distribution in Fig.~\ref {fig_data_candidates}(d) clearly shows
the \pizero\ and \etameson\ peaks, superimposed on a broad background distribution.
This background has a combinatorial and a low-mass component,
discussed in detail in the two following sections.

\subsection {Combinatorial background}

The combinatorial background in the invariant mass distribution originated from pairs of photon clusters 
that were not produced by a decay of a single particle.
To describe the shape of the combinatorial background, we used the event mixing technique,
where photon candidates from two different events were combined.
To avoid the mixing of different event topologies, the data were subdivided into mixing classes
based on the vertex position, \BEMC\ multiplicity, and trigger type (MinBias, \HighTowerOne, and \HighTowerTwo)\@.

\begin {figure} [tb]
\centerline {\hbox {
\includegraphics {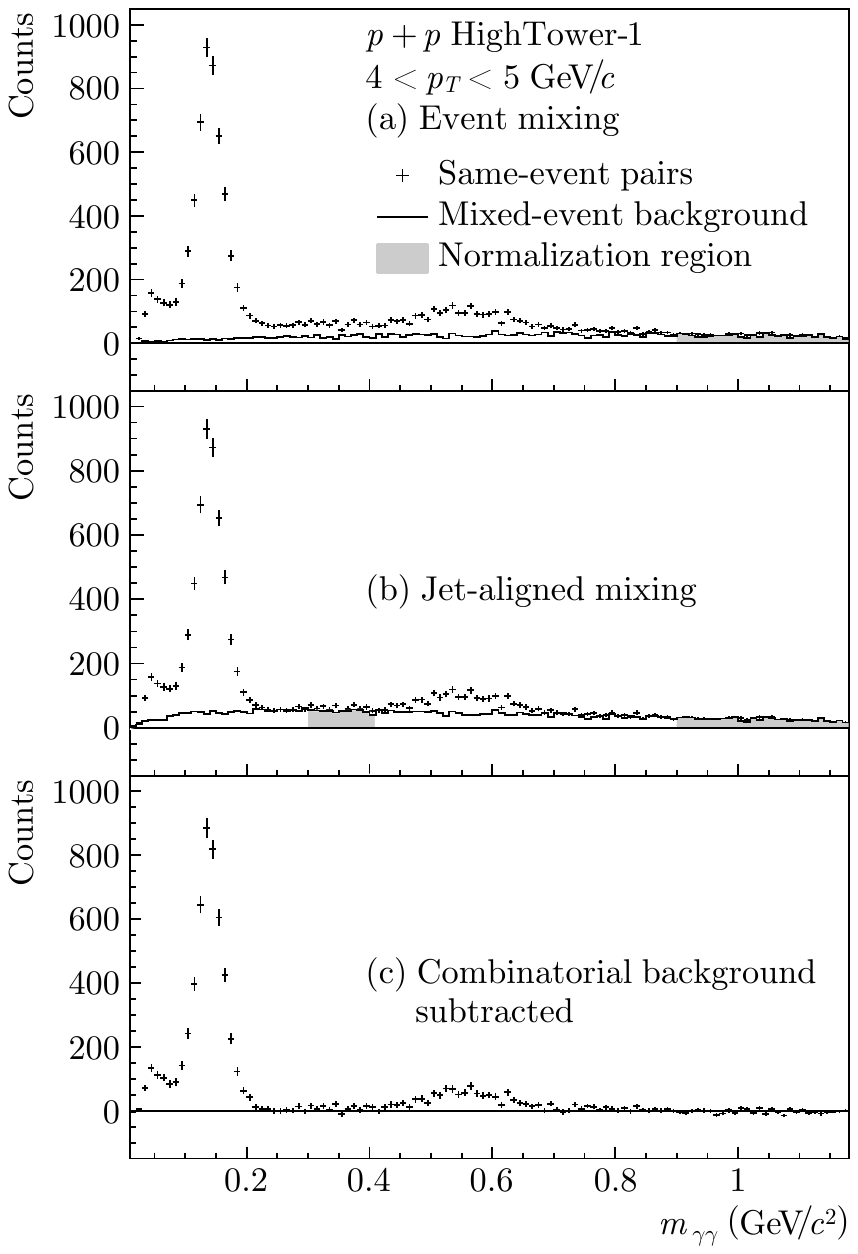}
}}
\caption {The same-event invariant mass distribution (crosses) 
and combinatorial background (histogram) observed in one \pT\ bin of the \protonproton\ data.
(a) Background estimated from random event mixing and
(b) from a linear combination of random and jet-aligned mixing,
and (c) background-subtracted distribution.
The shaded areas indicate the regions where the mixed-event background was normalized to the same-event distributions.
\label {fig_inv_rndmix}
}
\end {figure}

Figure~\ref {fig_inv_rndmix}(a)
shows, as an example, the invariant mass distribution in the 
\MM{4 \LESS{} \pT{} \LESS{} 5\unit{\GeVc}} bin,
obtained from the \protonproton\ \HighTowerOne\ data,
together with the combinatorial background obtained from the event mixing.
The mixed-event background distribution was normalized to the same-event distribution
in the invariant mass region \MM{0.9 \LESS{} \Mgg{} \LESS{} 1.2\unit{\GeVcc}}\@.

There is still some residual background
in the interval \MM{0.2 \LESS{} \Mgg{} \LESS{} 0.4\unit{\GeVcc}}\@.
This background is due to correlation structures (jet structures) in the event, which are not present in the sample of mixed events. 
In order to preserve jet-induced correlations, the jet axes in both events were aligned before mixing~\cite {ref_grebenyuk_thesis}, as described below.

To determine the \etacoord\opdash\phiangle\ position of the most energetic jet in every event,
the cone algorithm was used~\cite {ref_jet_asymmetry}\@.
The mixed-event \pizero\ candidates were constructed by taking two photons from different events,
where one of the events was displaced in \etacoord\ and \phiangle\ by
\MM{\DELTA\etacoord{} = \SUB{\etacoord}{2} - \SUB{\etacoord}{1}}
and \MM{\DELTA\phiangle{} = \SUB{\phiangle}{2} - \SUB{\phiangle}{1}}, respectively.
Here \MM{\SUB{\etacoord}{\RM{1,2}}} and \MM{\SUB{\phiangle}{\RM{1,2}}} are the jet orientations in the two events.

\begin {figure} [tb]
\centerline {\hbox {
\includegraphics {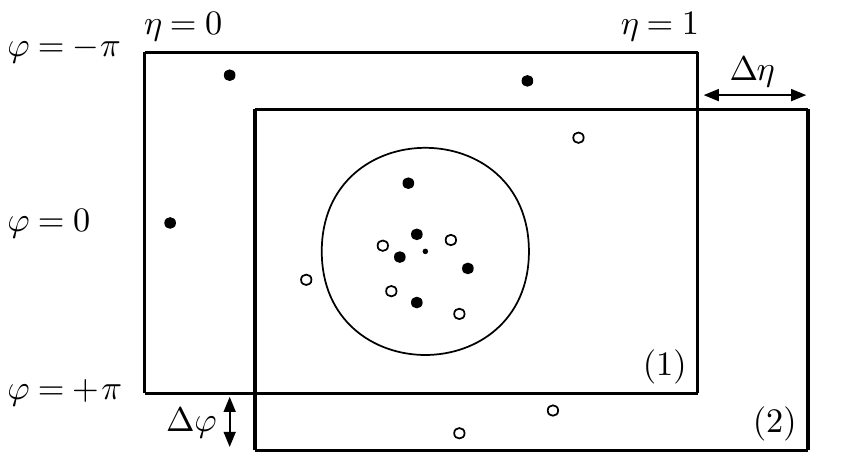}
}}
\caption {\label {fig_jetmix}A schematic view of two superimposed events, where the jet axes are aligned
to preserve the jet-induced correlations in the mixed event.
}
\end {figure}

Figure~\ref {fig_jetmix}
shows a schematic view of two superimposed events where the jet axes are aligned.
In order to minimize acceptance distortions, the events were divided into ten mixing classes in the jet \etacoord\ coordinate.
By mixing only events in the same class, the shift \MM{\DELTA\etacoord} was limited to \MM{0.1}\@.
Because the calorimeter has a cylindrical shape, the shift in \phiangle\ did not induce any significant acceptance distortion.

A side effect of this procedure was that correlations were induced if there was no real jet structure, 
because the jet finding algorithm then simply picked the most energetic track in the event.
To reduce possible bias introduced by such correlations, the combinatorial background was 
taken to be fully random for \MM{\pT{} \LESS{} 1.2\unit{\GeVc}} and fully jet-aligned for \MM{\pT{} \GREATER{} 10\unit{\GeVc}}\@. 
Between these values, the random component decreased linearly with increasing \pT\@. 
We assigned a systematic uncertainty of \MM{10\unitns{\%}} to the random background fraction,
which resulted in a systematic uncertainty of \MM{5\unitns{\%}} of the \pizero\ and \MM{3.5\unitns{\%}}~of~the~\etameson~yields.
\label {syst_mixed_bg}

Figure~\ref {fig_inv_rndmix}(b) shows the same invariant mass spectrum as that shown in panel (a),
with the background estimated by the combined random and jet-aligned event mixing.
The mixed-event background was normalized to the same-event distribution in the ranges
\MM{0.3 \LESS{} \Mgg{} \LESS{} 0.4} and \MM{0.9 \LESS{} \Mgg{} \LESS{} 1.2\unit{\GeVcc}}\@.
\label {syst_comb_bg}
By changing the subtracted background within its normalization uncertainty,
we obtained another component of a systematic error of the \pizero\ and \etameson\ yields,
which was found to increase with increasing \pT\ from \MM{0.5} to \MM{3\unitns{\%}} for the \pizero\ 
and from \MM{10} to \MM{50\unitns{\%}}~for~the~\etameson\@.

Figure~\ref {fig_inv_rndmix}(c) shows the background-subtracted distribution.
It is seen that there is still a residual background component at invariant mass \MM{\Mgg{} \LESS{} 0.1\unit{\GeVcc}}\@.
The origin of this background is described in the next section.

\subsection {Low-mass background}
\label {sec:lowmassbg}

Random fluctuations in the \SMD\ signals occasionally generate a double-peaked hit structure, 
in which case the clustering algorithm incorrectly splits the cluster.
These random fluctuations enhance the yield of pairs with minimal
angular separation and thus contribute to the lowest di-\tsp{-0.2}photon invariant mass region, as can be seen in Fig.~\ref {fig_inv_rndmix}(c)\@.
However, at a given small opening angle, the invariant mass increases with increasing energy of the parent particle,
so that the low-mass background distribution extends to larger values of \Mgg\ with increasing photon \pT\@.

The shape of the low-mass background was obtained from a simulation as follows.
Single photons were generated with flat distributions in \MM{-\PI \LESS{} \phiangle \LESS{} +\PI}, \MM{-0.2 \LESS{} \etacoord{} \LESS{} 1.2}, and \MM{0 \LESS{} \pT{} \LESS{} 25\unit{\GeVc}}\@.
These photons were tracked through a detailed description of the \STAR\ geometry with the \GEANT\ program.
A detailed simulation of the electromagnetic shower development in the calorimeter was used to generate
realistic signals in the towers and in the \SMD\@.
The simulated signals were processed by the same reconstruction chain as the real data. 
Photons with more than one reconstructed cluster were observed, 
and \Mgg\ and \pT\ of such cluster pairs were calculated.
The \Mgg\ histograms were accumulated, with each entry weighted by the \pT\ spectrum of photons in the real data, corrected for the
photon detection efficiency.

\begin {figure} [tb]
\centerline {\hbox {
\includegraphics {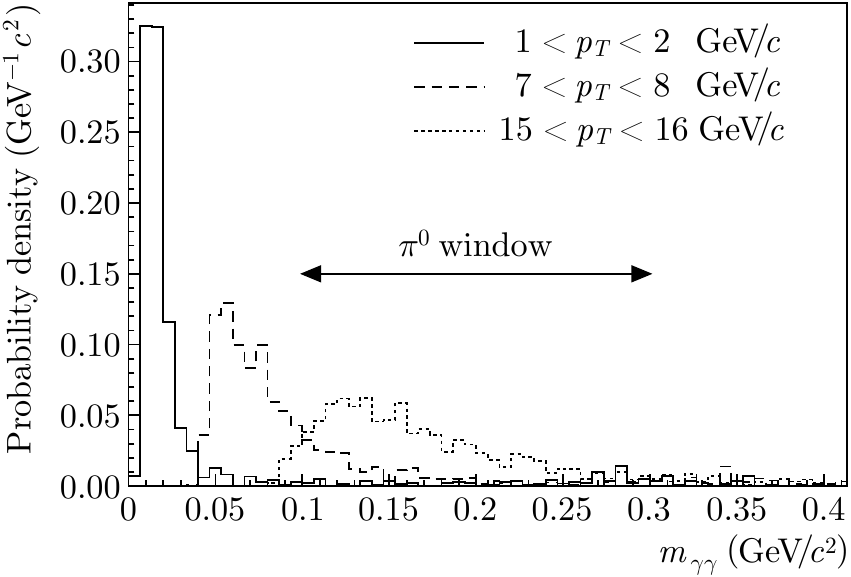}
}}
\caption {The simulated low-mass background distributions from erroneous splitting of single photons,
in three bins of the reconstructed pair \pT\@.
The distributions extend to larger invariant masses with increasing \pT\ and move into the \pizero\ region (shown for \MM{\pT{} = 15\unit{\GeVc}})\@.
\label {fig_inv_gamma}
}
\end {figure}

Figure~\ref {fig_inv_gamma}
shows the low-mass background distributions in three bins of the reconstructed pair \pT\@.
It is seen that the distributions indeed move to larger invariant masses with increasing \pT\ and extend far into the \pizero\ window at high \pT\@.
For this reason, it was not possible to estimate the amount of this background from a phenomenological fit to the data, 
and we had to rely on the Monte Carlo simulation to subtract this background component.

The second significant source of \BEMC\ clusters that passed the \CPV\ cut was the neutral hadrons
produced in the collisions, mostly antineutrons above \MM{2\unit{\GeVc}}\@.
To account for the additional low-mass background from these hadrons,
simulations of antineutrons were performed in the same way as of photons, 
and the reconstructed invariant mass distribution was
added according to the realistic proportion \DIV{\antineutron}{\SUB{\gama}{\RM{incl}}}, where the antineutron yield was 
estimated as described in section~\ref {sec:neutr-hadr-cont} below.

\begin {figure} [tb]
\centerline {\hbox {
\includegraphics {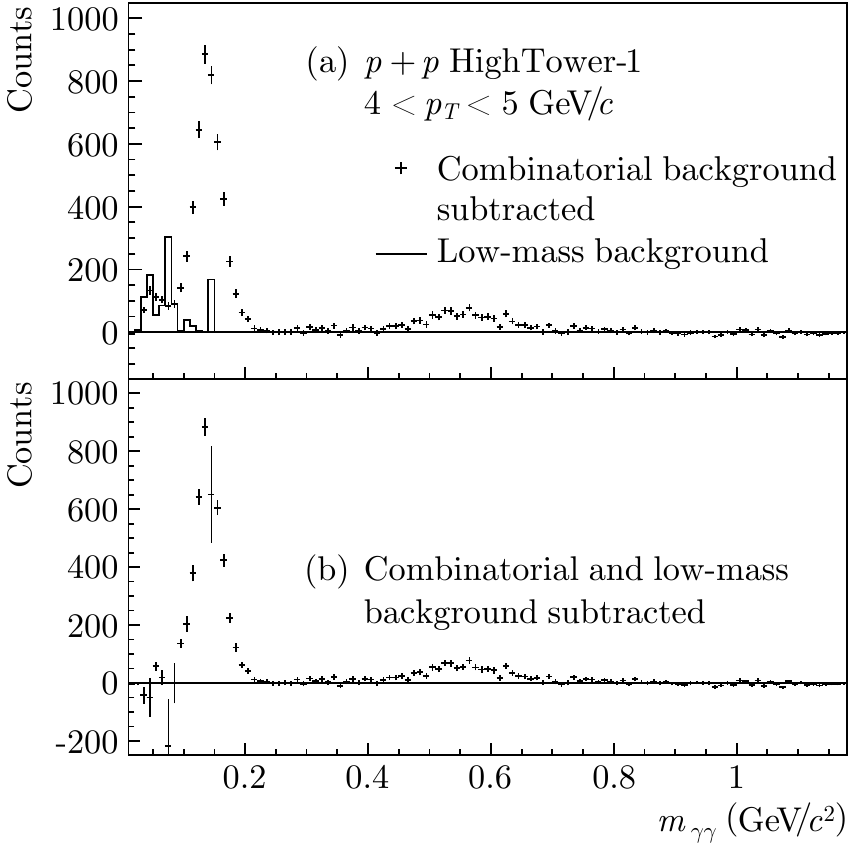}
}}
\caption {The invariant mass distribution observed in one \pT\ bin of the \protonproton\ data (a) before and (b) after the low-mass background subtraction\@.
\label {fig_inv_subtracted}
}
\end {figure}

The low-mass background was normalized by matching the observed \pT\ spectrum of the clusters between simulation and data.
This removal procedure worked well, and Fig.~\ref {fig_inv_subtracted}
shows the invariant mass spectra and the low-mass background component, 
and the final background-subtracted spectrum for the \protonproton\ \HighTowerOne\ data\@.
The normalization uncertainty of the low-mass background contributes to the systematic uncertainty of the \pizero\ cross section,
and reaches \MM{15\unitns{\%}} at the high-\pT\ end of the spectrum.

\subsection {Peak position and width}

\begin {figure} [tb]
\centerline {\hbox {
\includegraphics {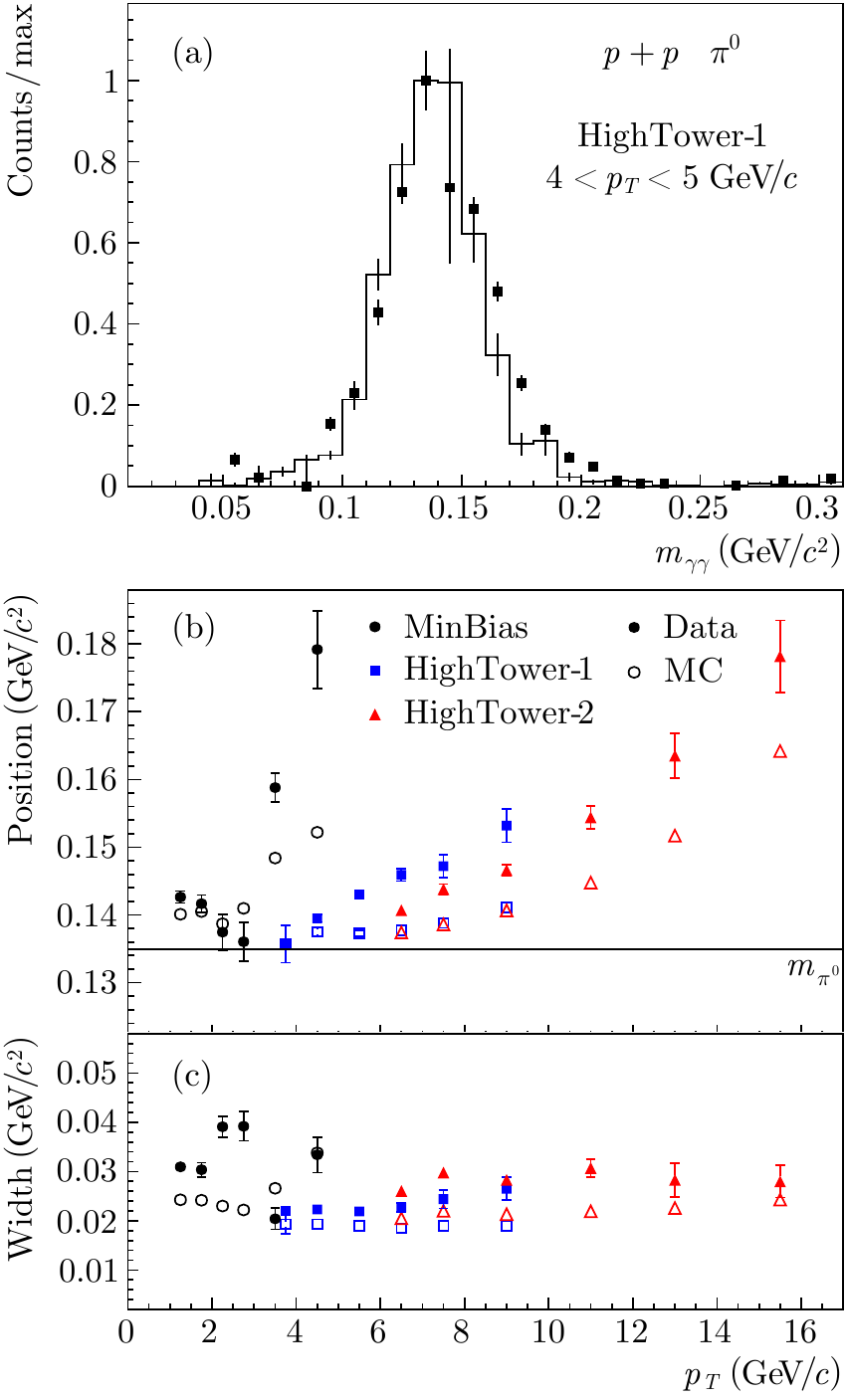}
}}
\caption {\label {fig_compare_m}(color online) (a) Invariant mass spectrum of \pizero's reconstructed in the simulation (histogram) in
comparison to the \protonproton\ \HighTowerOne\ data (symbols) in \MM{4 \LESS{} \pT{} \LESS{} 5\unit{\GeVc}} bin\@.
(b) Peak position and (c) width in the real data (solid symbols) and in the simulation (open symbols).
Horizontal line in panel (b) shows the true \pizero\ mass.
}
\end {figure}

Figure~\ref {fig_compare_m}(a)
shows the background-subtracted \Mgg\ distribution in the region \MM{4 \LESS{} \pT{} \LESS{} 5\unit{\GeVc}} obtained from
the \protonproton\ \HighTowerOne\ data (symbols), together with the corresponding distribution from the detector simulation (histogram)\@.
In order to compare the real and simulated \Mgg\ distributions for all bins in \pT\ and for all datasets,
we estimated the position and width of the peaks using Gaussian fits in the peak region.
Figure~\ref {fig_compare_m}(b) shows the peak positions obtained from the fit to the \protonproton\ data.
It is seen that the peak position shifts towards higher masses with increasing \pT\@.
This shift is a manifestation of the bin migration effect that originates from statistical fluctuations in the calorimeter response.
Due to the steeply falling \pT\ spectrum, the energy resolution causes a net migration towards larger \pT\@.
Since larger values of \pT\ imply larger values of \Mgg, 
the migration effect biases the invariant mass peak towards larger values.

An additional peak shift at the largest values of \pT\ is caused by the \SMD\ strip granularity, 
which imposes a lower limit on the opening angle of the reconstructed photon pairs. The minimal \SMD\ cluster separation
in each dimension that can be resolved by the cluster finder is \MM{1.5} strips, and most clusters contain at least two or three strips. 
Therefore, the pair reconstruction is less efficient for the symmetric decays with the smallest opening angles at 
\MM{\pT{} \GREATERAPPROX{} 10\unit{\GeVc}}\@. This leads to an increased average opening angle and 
\Mgg\ of the reconstructed photon pairs from \pizero\ decays.

The peak position observed in the data is larger than that found in the simulations by \MM{(3.5 \PLMN{} 0.6)\unitns{\%}}, on average.
This difference could be caused by the global energy scale of the \BEMC\ towers being off by a similar amount.
We already accounted for this possibility by assigning a systematic uncertainty of \MM{5\unitns{\%}} 
to the \BEMC\ calibration constants (see section \ref{energy_calibration})\@.

Figure~\ref {fig_compare_m}(c) compares the \pizero\ peak width in the data and in the simulation, 
and it is seen that the peak width in the data is larger than that in the simulation by \MM{(25 \PLMN{} 2)\unitns{\%}}, on average.
This is a sufficiently good agreement for this analysis, because the \pizero\ and \etameson\ yields were counted in the mass windows 
that were adjusted in each \pT\ bin to cover the entire signal peak.

\begin {figure} [tb]
\centerline {\hbox {
\includegraphics {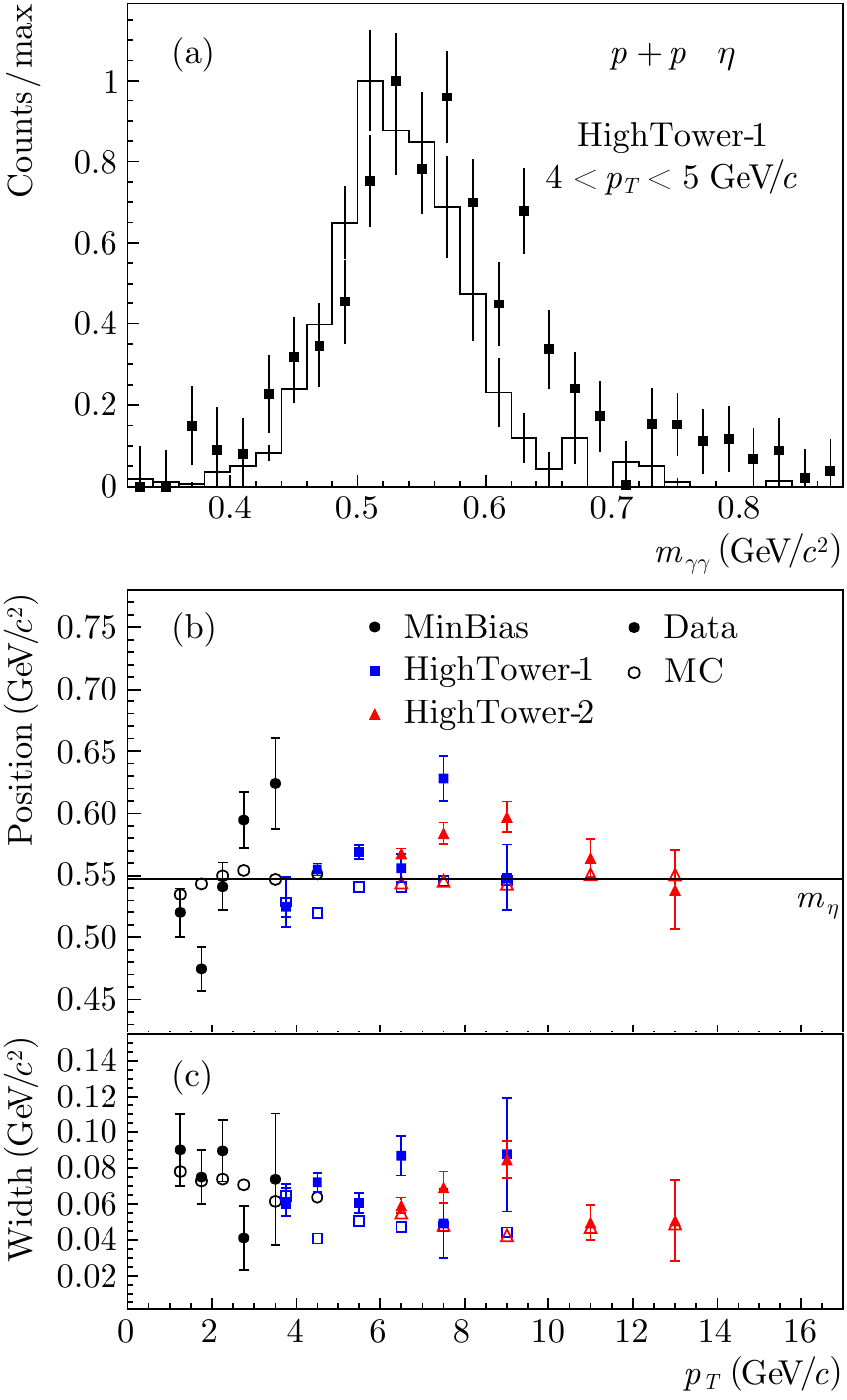}
}}
\caption {\label {fig_compare_m_eta}(color online) (a) Invariant mass spectrum of \etameson\ mesons reconstructed in the simulation (histogram) in
comparison to the \protonproton\ \HighTowerOne\ data (symbols) in \MM{4 \LESS{} \pT{} \LESS{} 5\unit{\GeVc}} bin\@.
(b) Peak position and (c) width in the real data (solid symbols) and in the simulation (open symbols).
Horizontal line in panel (b) shows the true \etameson\ mass.
}
\end {figure}

The peak shape of the \etameson\ meson, as well as its position and width, are 
shown in Fig.~\ref {fig_compare_m_eta},
as a function of the reconstructed \pT\@.
The peak position in the data is larger than in the simulations by \MM{(5.1 \PLMN{} 1.2)\unitns{\%}}, 
and the width by \MM{(34 \PLMN{} 11)\unitns{\%}}, on average.
The difference in the peak position is similar to the \pizero\ case above,
which supports the possibility of both being caused by a small systematic offset in the \BEMC\ calibration.
The observed level of agreement between data and simulations is considered to be sufficient for this analysis.

\subsection {Invariant yield extraction}

The invariant yield of the \pizero\ and \etameson\ mesons per MinBias collision,
as a function of \pT, is given by
\begin {equation}\label {eq:pions_1}
\energy\tsp{0.8} \FRAC{\SUP{\der}{3}\tsp{-0.7}\Number}{\der\SUP{\momentumthree}{3}} =
\FRAC{\SUP{\der}{3}\tsp{-0.7}\Number}{\pT\tsp{0.5}\der\pT\tsp{0.5}\der\rapidity\tsp{0.5}\der\phiangle} =
\FRAC{\SUP{\der}{2}\tsp{-0.7}\Number}{2\PI\pT\tsp{0.5}\der\pT\tsp{0.5}\der\rapidity},
\end {equation}
where in the last equality integration over the full \MM{2\PI} azimuthal coverage of the \STAR\ detector is performed.
Using the experimentally measured quantities, the invariant yield was calculated as
\begin {equation} \label {eq:pions_2}
\energy\tsp{0.8} \FRAC{\SUP{\der}{3}\tsp{-0.7}\Number}{\der\SUP{\momentumthree}{3}} =
\FRAC{1}{2\PI\pT \SUB{\Number}{\RM{trig}} \SUB{\IT{K}}{\RM{trig}}}
\FRAC{\IT{Y}}{\DELTA\pT\DELTA\rapidity}
\FRAC{\SUB{\SUP{\EPS}{\ }}{\RM{vert}}\tsp{0.5}\SUP{\SUB{\IT{c}}{\RM{loss}}}{\PI}}{\SUP{\SUB{\EPS}{\RM{acc}}}{\PI}\tsp{0.5}\SUP{\SUB{\EPS}{\RM{cpv}}}{\PI}}
\FRAC{1}{\SUB{\ensuremath{\mathcal{B}}}{\gama\gama}}
,
\end {equation}
where:
\begin {itemize}
\item {\MM{\IT{Y}}} is the raw yield measured in the bin centered at \pT\ and \rapidity;
\item {\MM{\DELTA\pT}} is the width of the \pT\ bin for which the yield was calculated;
\item {\MM{\DELTA\rapidity}} is the rapidity range of the measurement; 
in this analysis \MM{\DELTA\rapidity{} = 1} for all data points, except for the \etameson\ yields at \MM{\pT{} \LESS{} 3\unit{\GeVc}}, 
where the correction for the difference between rapidity and pseudorapidity reached \MM{7\unitns{\%}};
\item {\MM{\SUB{\Number}{\RM{trig}}}} is the number of triggers recorded;
\item {\MM{\SUB{\IT{K}}{\RM{trig}}}} is the trigger scale factor; 
\MM{\SUB{\IT{K}}{\RM{trig}} \ensuremath{\equiv} 1} for the MinBias events and \GREATER\MM{1} for the HighTower events;
the product \MM{\SUB{\Number}{\RM{trig}}\SUB{\IT{K}}{\RM{trig}}}
is the equivalent number of MinBias events that produced the yield \MM{\IT{Y}};
\item {\MM{\SUB{\EPS}{\RM{vert}}}} is the vertex finding efficiency in MinBias events;
\item {\MM{\SUP{\SUB{\IT{c}}{\RM{loss}}}{\PI}}} is the correction for the missing material in the simulation;
\item {\MM{\SUP{\SUB{\EPS}{\RM{acc}}}{\PI}}} is the \BEMC\ acceptance and efficiency correction factor;
\item {\MM{\SUP{\SUB{\EPS}{\RM{cpv}}}{\PI}}} is a correction for random \TPC\ vetoes;
\item {\MM{\SUB{\ensuremath{\mathcal{B}}}{\gama\gama} = \DIV{\SUB{\ensuremath{\Gamma}}{\gama\gama}}{\tsp{0.2}\ensuremath{\Gamma}}}} is the branching ratio of the di-\tsp{-0.2}photon decay channel
(\MM{0.988} for \pizero\ and \MM{0.392} for \etameson~\cite {ref_pdg})\@.
\end {itemize}

The raw \pizero\ and \etameson\ yields were counted in the \pT-dependent \Mgg\ windows that contained the peaks.
The low-mass border of the \pizero\ peak region was taken to be a linear function of \pT, common for all datasets and triggers. 
This cut was optimized to capture most of the yield and as little of low-mass background as possible.
The high-mass border also linearly increased with \pT\ in order to cover the asymmetric right tail of the peak.
Similarly, the \etameson\ peak region was a \pT-dependent window that captured most of the signal.
For completeness, we give below the parametrization of the \pizero\ and \etameson\ windows:
\begin {equation} \label {eq:mass_windows}
\begin {array} {r@{\ <\ }l@{\ <\ }ll}
75 + 1.7 \tsp{0.5} \pT & \Mgg(\pizero) & 250 + 3.3 \tsp{0.5} \pT & \unitns{\MeVcc}, \\
350 + 3.3 \tsp{0.5} \pT & \Mgg(\etameson) & 750 & \unitns{\MeVcc}, \\
\end {array}
\end {equation}
where \pT\ is measured in \GeVc\@.
The stability of the yields was determined by varying the vertex position cut,
the energy asymmetry cut, and the yield integration windows.
\label {syst_analysis_cuts}
From the observed variations, a point\tsp{0.2}-to\tsp{0.4}-point systematic error of \MM{5\unitns{\%}} was assigned to the \pizero\ and \etameson\ yields.

Within each trigger in the \protonproton\ data, the \pizero\ signal significance decreased 
from \APPROX34 to \APPROX6 standard deviations with increasing \pT, because of the corresponding reduction in statistics\@.
In the \deuterongold\ data, the same trends were observed, 
but the significance was lower than in the \protonproton\ data by a factor of 1.9, on average,
which is mainly caused by the lower integrated nucleon\opdash{}nucleon luminosity in these data.
The significance of the \etameson\ signal was between 18 and 2.5 standard deviations in the \protonproton\ data, 
and between 5.5 and 1.0 standard deviations in the \deuterongold\ data.

\subsection {Acceptance and efficiency correction}
\label {subsect_eff_corr}

To calculate the detector acceptance and reconstruction efficiency correction factor \SUP{\SUB{\EPS}{\RM{acc}}}{\PI},
a Monte Carlo simulation of the detector was used.
The \pizero\ decay photons were tracked through the \STAR\ detector geometry using \GEANT\@.
The simulated signals were passed through the same analysis chain as the real data.

The \pizero's were generated in the pseudorapidity region \MM{-0.3 \LESS{} \etacoord{} \LESS{} +1.3}, which is sufficiently large
to account for edge effects caused by the calorimeter acceptance limits of \MM{0 \LESS{} \etacoord{} \LESS{} 1}\@.
The azimuth was generated flat in \MM{-\PI \LESS{} \phiangle{} \LESS{} +\PI}\@.
The \pT\ distribution was taken to be uniform up to \MM{25\unit{\GeVc}}, which amply covers the measured pion \pT\ range of up to \MM{17\unit{\GeVc}}\@.
The vertex distribution of the generated pions was taken to be Gaussian in \zcoord\ with a spread of \MM{\SIGMA{} = 60\unit{\cm}} and centered at \MM{\zcoord{} = 0}\@.

The generated \pizero's were allowed to decay into two photons, \MM{\pizero\TO\gama\gama}\@.
The \GEANT\ simulation accounted for all interactions of the decay photons with the detector, such as
conversion into \epluseminus\ and showering in the calorimeter or in the material in front of it.

To reproduce a realistic energy resolution of the calorimeter,
an additional smearing had to be applied to the energy deposits calculated by \GEANT\@.
In all simulations, a spread of \MM{5\unitns{\%}} was used to reproduce the \protonproton\ data and \MM{10\unitns{\%}} for the \deuterongold\ data.

To reproduce the \pT\ spectrum of pions in the data, each Monte Carlo event was weighted by a \pT-dependent function.
This weighting technique allowed us to sample the entire \pT\ range with good statistical power, 
while reproducing the bin migration effect caused by the finite detector energy resolution.
An \NLO\ \pQCD\ calculation~\cite {ref_nlo_pqcd} provided the initial weight function,
which was subsequently adjusted in an iterative procedure.

The time dependence of the calorimeter acceptance during data taking was recorded in database tables that were used in the analysis.
In order to reproduce this time dependence in the simulation, the generated events were
assigned timestamps that followed the timeline of the real data taking.
In this way, the geometrical acceptance of the calorimeter (mean fraction of good towers) was reproduced in the Monte Carlo
with a precision of \APPROX\MM{0.5\unitns{\%}}\@.

In the analysis of real data, we used vertices reconstructed from \TPC\ tracks and those derived from \BBC\ time-of-flight measurements.
The former have sub\tsp{0.4}-millimeter resolution, whereas the latter have a precision of only \APPROX\MM{40\unit{\cm}}\@.
To account for the \BBC\ vertex resolution, \MM{35\unitns{\%}} of the generated pions in the \protonproton\ MinBias data had their point 
of origin artificially smeared in the \zcoord\ direction. 
No such smearing was applied to the other simulated data, where \BBC\ vertex was not used.

\begin {figure} [tb]
\centerline {\hbox {
\includegraphics {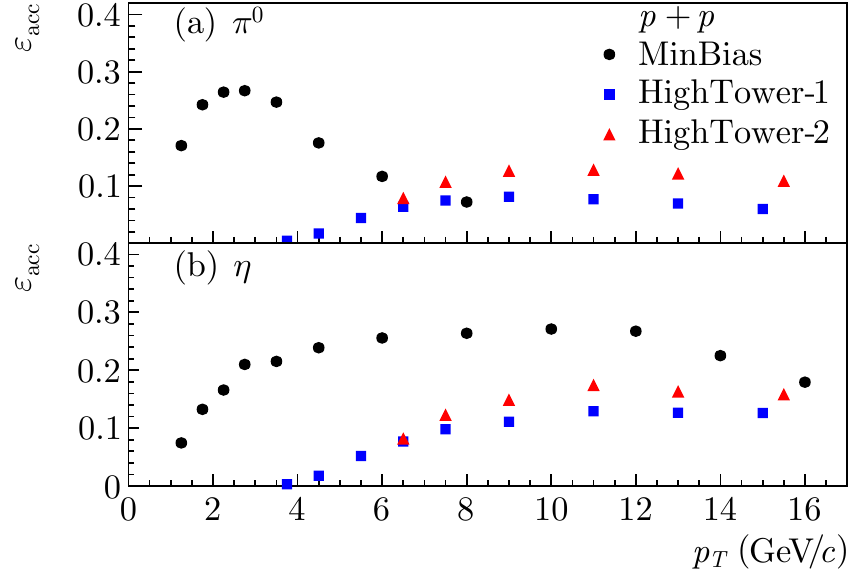}
}}
\caption {\label {fig_eff}(color online) Acceptance and efficiency correction factor \SUB{\EPS}{\RM{acc}} for (a) \pizero\ and (b) \etameson, calculated from the Monte Carlo simulation for the \protonproton\ data.}
\end {figure}

The acceptance and efficiency correction factor was calculated from the simulation
as the ratio of the raw \pizero\ yield reconstructed in a \pT\ bin to the
number of simulated pions with the true \pT\ generated in that bin.
This was done separately for each trigger, using the same \pizero\ reconstruction cuts
as was done in the real data analysis.
In particular, the reconstructed value of pseudorapidity was required to fall in the range \MM{0 \LESS{} \etacoord{} \LESS{} 1} in both
the data and the simulation, while in the latter the generated value of \etacoord\ was also required to fall in this range.
As an example, Fig.~\ref {fig_eff}
shows the \pizero\ and \etameson\ correction factors for the three triggers in \protonproton\ data.

The difference between the MinBias and HighTower correction factors was caused by the \SMD\ requirement in the HighTower data, 
which was absent in the MinBias data.
The absence of the \SMD\ information reduced the \pizero\ reconstruction efficiency in the MinBias data at \MM{\pT{} \GREATER{} 3\unit{\GeVc}},
where the decay photons were separated by less than two towers.
The \etameson\ reconstruction is only affected by this at larger values of \pT\@.

\begin {figure} [tb]
\centerline {\hbox {
\includegraphics {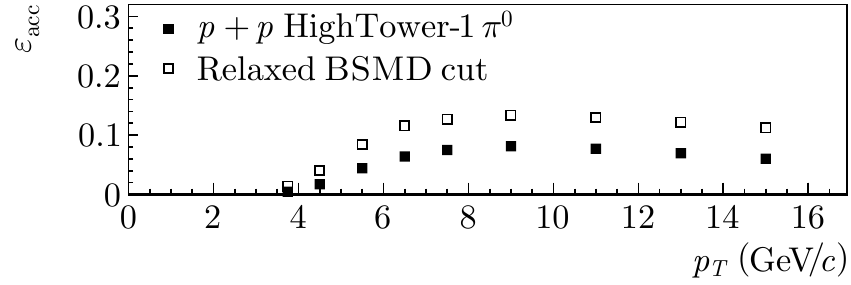}
}}
\caption {Acceptance and efficiency correction factor \SUB{\EPS}{\RM{acc}} for the \protonproton\ \HighTowerOne\ data, with standard set of cuts (solid symbols)
and with \SMD\ quality cut removed (open symbols).
\label {fig_eff_diff}
}
\end {figure}

The effect of the \SMD\ quality requirement of having at least two adjacent strips in a cluster is illustrated in Fig.~\ref {fig_eff_diff},
which shows the correction factor calculated for the \protonproton\ \HighTowerOne\ data
with (solid symbols) and without (open symbols) the \SMD\ quality requirement.
This requirement reduced the number of accepted \pizero\ candidates by \APPROX\MM{45\unitns{\%}}\@.
This explains the difference between the \HighTowerOne\ and \HighTowerTwo\ (no \SMD\ quality cut)
correction factors at high \pT\ seen in Fig.~\ref {fig_eff}\@.

The current simulation framework poorly reproduces the shower shapes in the \SMD\ at the low incident photon energies.
To account for residual bias after applying the \SMD\ quality cut, 
we assigned a systematic uncertainty to the \HighTowerOne\ cross section, which decreases from \MM{15\unitns{\%}} at \MM{\pT{} = 4\unit{\GeVc}} 
to zero at \MM{\pT{} = 7\unit{\GeVc}}\@.

To determine a dependence of the acceptance correction on the track multiplicity \IT{M}, 
and thus on the centrality, 
we analyzed a sample of generated \pizero's embedded into real \deuterongold\ events.
No significant centrality dependence was found. Therefore, the same correction factors were applied to the
different centrality classes in the \deuterongold\ data.
The dependence of the efficiency on the locally higher multiplicity in jets was investigated in a \PYTHIA~\cite{Sjostrand:2001yu} simulation,
and no significant difference in the efficiency was observed relative to a single particle simulation.

\subsection {HighTower trigger normalization}

We have shown in Fig.~\ref {fig_data_candidates} the \pT\ distribution of \pizero\ candidates for the \protonproton\ 
MinBias, \HighTowerOne, and \HighTowerTwo\ data.
To normalize the HighTower spectra to those of the MinBias, \pT-independent scale factors were applied.
These scale factors were estimated as the ratio of observed
MinBias to HighTower event rates,
\begin {equation} \label {eq:HTscale}
\SUB{\IT{K}}{\RM{trig}} = \DIV {\ensuremath{\textstyle\sum} \SUB{\Number}{\RM{\MB}} \SUB{\IT{S}}{\RM{\MB}}} {\ensuremath{\textstyle\sum} \SUB{\Number}{\RM{\HT}} \SUB{\IT{S}}{\RM{\HT}}}.
\end {equation}
Here \SUB{\Number}{\RM{\MB}} and \SUB{\Number}{\RM{\HT}} are the numbers of MinBias and HighTower triggers that passed the selection cuts,
\SUB{\IT{S}}{\RM{\MB}} and \SUB{\IT{S}}{\RM{\HT}} are the online prescale factors adjusted on a run-by-run basis to accomodate the \DAQ\ bandwidth,
and the sums are taken over all runs where both the MinBias and HighTower triggers were active.
We obtained the values \MM{\SUB{\IT{K}}{\RM{trig}} = 4.67\e{3}} and \MM{1.96\e{4}} for the \protonproton\ \HighTowerOne\ and \HighTowerTwo\ triggers, respectively,
and \MM{\SUB{\IT{K}}{\RM{trig}} = 2.87\e{3}} and \MM{2.86\e{4}} for the \deuterongold\ triggers.

To check the scale factors, the HighTower software filter, which simulated the hardware trigger, was applied to the MinBias data.
The scale factors were obtained as the ratio of the total number of MinBias events to the number that passed the filter.
To obtain a more precise \HighTowerOne/\tsp{0.5}\HighTowerTwo\ relative normalization factor, the software filter was applied to the \HighTowerOne\ data.
The results from the two methods agreed within \MM{3\unitns{\%}} for \HighTowerOne\ data and within \MM{5\unitns{\%}} for \HighTowerTwo\ data.
\label {syst_ht_scale}
These numbers were taken as the systematic uncertainties of the HighTower normalization factors.

The difference between vertex finding efficiencies in MinBias and HighTower data was effectively absorbed in the scale factor \SUB{\IT{K}}{\RM{trig}}\@.
Therefore, the vertex finding efficiency correction was applied to the scaled HighTower data, as well as to the MinBias data.

\subsection {Bin centering scale factors}

To assign the yield measured in a \pT\ bin to a single \pT\ value, 
the procedure from Ref.~\cite {ref_where_to_stick_your_data_points} was applied.
The variation of the yield within a bin was approximated by the function \MM{\IT{f}(\pT) = \IT{A} \exp(-\IT{B}\tsp{0.3}\pT)}\@. 
The measured yield in the bin was assigned to the momentum \SUP{\pT}{\tsp{0.7}*} calculated from the equation
\begin {equation}
\IT{f}(\SUP{\pT}{\tsp{0.7}*}) = \FRAC{1}{\DELTA\pT} \SUB{\ensuremath{\int}}{\tsp{-1}\DELTA\pT} \tsp{-5.0} \IT{f}(\pT)\tsp{0.5}\der\pT.
\end {equation}
The procedure was repeated, taking \SUP{\pT}{\tsp{0.7}*} as the abscissa, 
until the \SUP{\pT}{\tsp{0.7}*} values were stable (typically after \MM{3} \mbox{iterations})\@.

To facilitate the comparison of results from the various datasets,
the yields were scaled to the bin centers by the ratio 
\MM{\SUB{\IT{K}}{\RM{bin}} = \DIV {\IT{f}(\SUP{\pT}{\tsp{0.7}*})} {\tsp{-0.5}\IT{f}(\pT\tsp{-0.3})}},
where \pT\ is the center of the bin.
The statistical and systematic errors were also scaled by the same factor.

To estimate a systematic uncertainty introduced by this procedure,
we changed the functional form of \MM{\IT{f}(\pT\tsp{-0.3})} either to a local power law in each bin or to a global power law in the full \pT\ range.
The observed variation in \SUB{\IT{K}}{\RM{bin}} was below \MM{1.5\unitns{\%}} in most \pT\ bins.

\subsection {Fully corrected yields}
\label {corrected_yields}

\begin {figure} [tb]
\centerline {\hbox {
\includegraphics {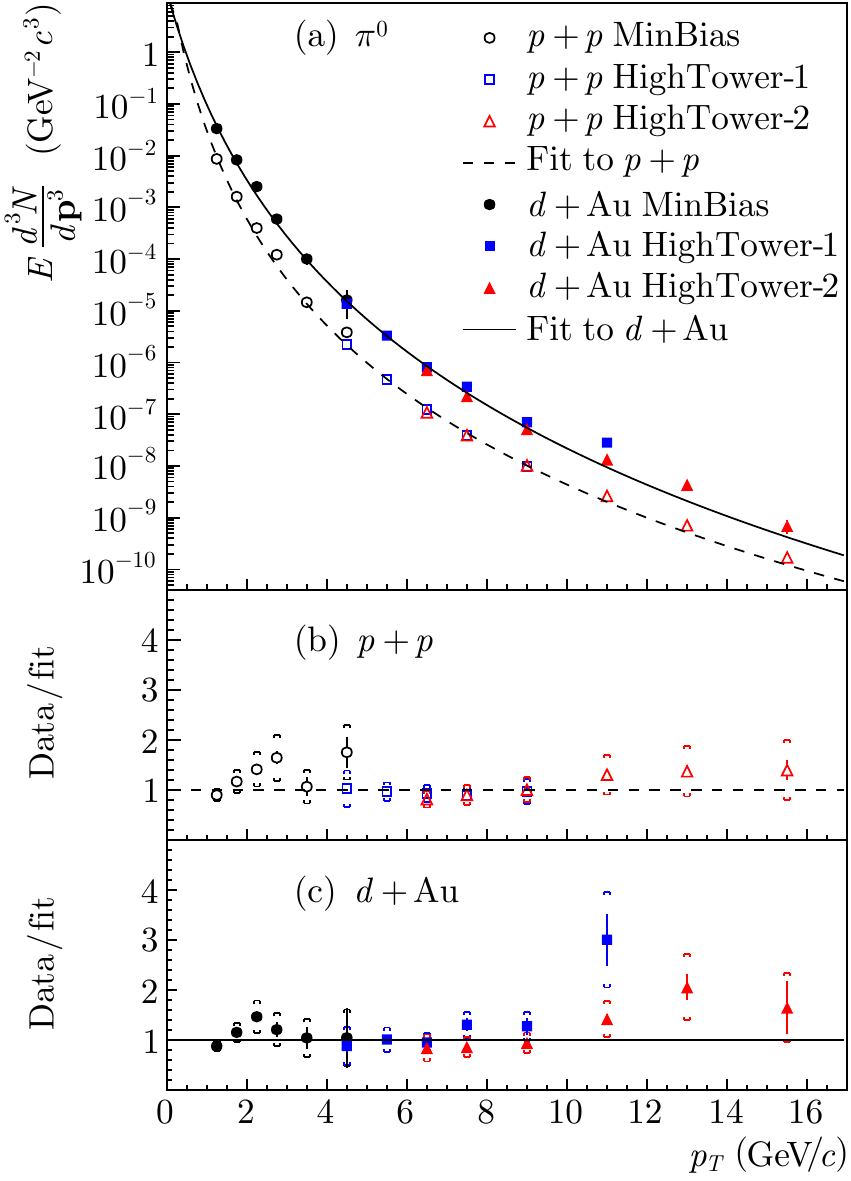}
}}
\caption {(color online) (a) Invariant yield of \pizero\ per MinBias \protonproton\ and \deuterongold\ collision\@.
Curves are the power law fits given in the text.
Invariant yield divided by the fit to the (b) \protonproton\ and (c) \deuterongold\ data.
The error bars are statistical and brackets in the lower panels are the systematic uncertainties.
\label {fig_pi0_invyield}
}
\end {figure}

\begin {figure} [tb]
\centerline {\hbox {
\includegraphics {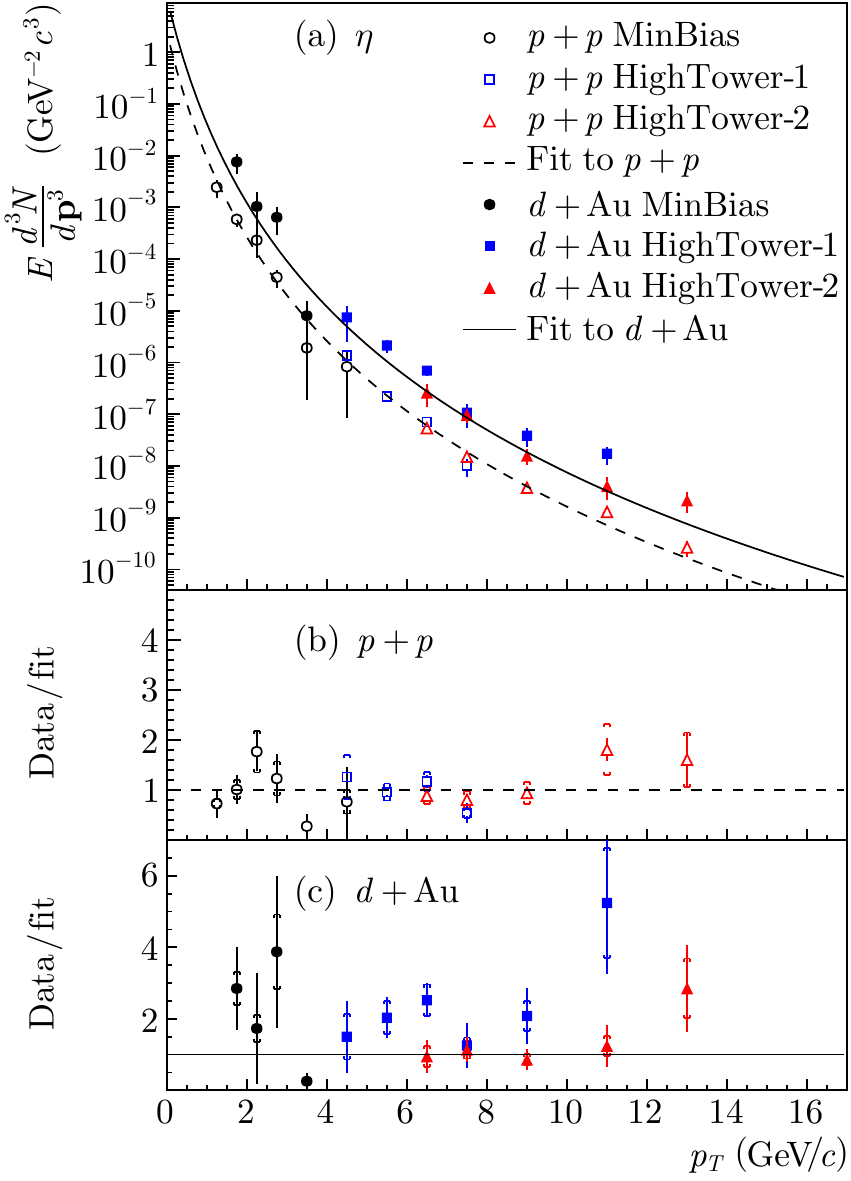}
}}
\caption {(color online) (a) Invariant yield of \etameson\ per MinBias \protonproton\ and \deuterongold\ collision\@.
Curves are the power law fits given in the text.
Invariant yield divided by the fit to the (b) \protonproton\ and (c) \deuterongold\ data.
The error bars are statistical and brackets in the lower panels are the systematic uncertainties.
\label {fig_eta_invyield}
}
\end {figure}

The fully corrected \pizero\ and \etameson\ invariant yields per MinBias event 
[Eq.~(\ref {eq:pions_2})] are shown in the top panels of Figs.~\ref {fig_pi0_invyield} 
and~\ref {fig_eta_invyield},
respectively.
The curves represent power law fits to the data, of the form
\begin {equation} \label {eq:power_law}
\energy\tsp{0.8} \FRAC{\SUP{\der}{3}\tsp{-0.7}\Number}{\der\SUP{\momentumthree}{3}} =
\FRAC {\IT{A}} { \SUP{\ensuremath{\left(} 1 + \DIV{\pT\tsp{-0.1}}{\momentum_{0}}\tsp{-0.1} \ensuremath{\right)}}{\IT{n}}}.
\end {equation}
The ratios between the data and the fits are shown in the lower panels.
The brackets in the lower panels are the systematic uncertainties (see section~\ref{subsect_syst_err} below),
which are partially correlated between different trigger datasets.
The agreement between the data taken with the different triggers is satisfactory,
although a small systematic difference between the \deuterongold\ \HighTowerOne\ and \HighTowerTwo\ yields 
(1.9 standard deviations, on average) was observed.

For the calculation of the final cross sections and their ratios, the data from three triggers were merged. 
The \HighTowerOne\ data were used in the MinBias--\HighTowerOne\ overlap bins,
because the MinBias data in those bins almost entirely represent a subset of \HighTowerOne\ events, 
selected by an online prescale factor \SUB{\IT{S}}{\RM{\MB}}\@.
Similarly, the \HighTowerTwo\ data were used in the \HighTowerOne--\HighTowerTwo\ overlap bins.

\subsection {Summary of systematic uncertainties}
\label {subsect_syst_err}
\begin {table*} [tbp]
\begin {center}
\caption {Systematic error contributions.
The classifications A, B, C, and N are defined in the text.
The error contributions to the \pizero\ cross section, the \etatopi\ ratio, \Rcp, \RdA, 
and the inclusive photon yield are indicated in the respective columns.
}
\label {table_systematic_uncertainty}
\begin {ruledtabular} 
\begin {tabular} {@{\extracolsep{\fill}}l@{\extracolsep{\fill}}c@{\extracolsep{\fill}}c@{\extracolsep{\fill}}c@{\extracolsep{\fill}}c@{\extracolsep{\fill}}c@{\extracolsep{\fill}}c@{\extracolsep{\fill}}c@{\extracolsep{\fill}}}
\noalign{\smallskip}
Source                     & Type & Value at low (high) \pT\ (\unitns{\%}) & \MM{\energy\tsp{0.9}\DIV{\SUP{\der}{3}\tsp{-0.7}\SIGMA\tsp{-0.3}}{\tsp{-0.2}\der\SUP{\momentumthree}{3}}} & \etatopi & \Rcp & \RdA & \SUB{\gama}{\RM{incl}} \\
\noalign{\smallskip}\hline\noalign{\smallskip}
Combinatorial background   & A & 0.5 (3) & + & + & + & + & \\
Mixed-event background     & C & 5 & + & + &  &  &  \\
Low-mass background        & C & 1 (15) & + & + &   &  &  \\
Random vetoes              & N & 2 & + &   & + & + & + \\
HighTower normalization    & B & 3 (5) & + &   & + & + & + \\
Analysis cuts              & A & 5 & + & + & + & + & \\
Conversion correction      & B & 3 for \pizero\ and \etameson, 2 for \SUB{\gama}{\RM{incl}} & + &   &   & + & + \\
Tower energy scale         & B & 15 (35) in \protonproton, 10 (40) in \deuterongold & + &   &   & + & + \\
\SMD\ simulation           & C & 15 (0) at \MM{\pT{} = 4\tsp{1.5}(7)\unit{\GeVc}} & + &   &   &  & \\
\SMD\ energy scale         & B & 3.5 (1) at \MM{\pT{} = 4\tsp{1.5}(11)\unit{\GeVc}} & + &   &   & + &  \\
Bin centering              & C & 1.5 & + &   &   &  & \\
Vertex finding efficiency  & N & 1 in \deuterongold\ MinBias & + &   & + & + & + \\
MinBias cross section      & N & 11.5 in \protonproton, 5.2 in \deuterongold & + &   &   & + & + \\
Glauber model \Ncollmean       & N & 5.3 in \deuterongold\ MinBias, 10.5 in \Rcp &   &   & + & + & \\
\end {tabular}
\end {ruledtabular} 
\end {center}
\end {table*}
The uncertainty of the calorimeter tower calibration was the dominant source of systematic uncertainty in this analysis.
The uncertainty of the uncorrected yield \MM{\IT{Y}\tsp{-0.3}(\tsp{0.4}\pT\tsp{-0.4})} was estimated from
\begin {equation}
\ensuremath{\delta} \IT{Y}\tsp{-0.3}(\tsp{0.4}\pT\tsp{-0.4}) = \ensuremath{\left|} \FRAC{\der\IT{Y}\tsp{-0.3}(\tsp{0.4}\pT\tsp{-0.4})}{\der\pT} \ensuremath{\right|} \ensuremath{\delta}\pT
,
\end {equation}
where \MM{\ensuremath{\delta}\pT} was taken to be \MM{5\unitns{\%}} in the \deuterongold\ and
\protonproton\ data (as derived from the electron calibration, see section~\ref {energy_calibration})\@.
This \pT-dependent systematic uncertainty was, on average, \MM{25\unitns{\%}} in both \protonproton\ and \deuterongold\ data.

Another strongly \pT-dependent term in Eq.~(\ref {eq:pions_2}) is the acceptance and efficiency correction factor \SUB{\EPS}{\RM{acc}}, 
obtained from the Monte Carlo simulation. 
However, this term contributed much less to the cross section uncertainty, because the simulations used 
the inverse of real calibration constants to convert the \GEANT\ energy deposit in each tower to the \ADC\ value.
Therefore, the energies later reconstructed from the \ADC\ values with the same constants were not sensitive to their fluctuations.
This factor was sensitive to the tower calibration only through the HighTower threshold values, 
and contributed a systematic uncertainty of \MM{5\unitns{\%}} in the \protonproton\ data and \MM{8\unitns{\%}} in the \deuterongold\ data, on average, 
correlated with the uncorrected yield uncertainty estimated above.

The uncertainty of the \SMD\ energy scale entered the analysis mainly due to threshold effects in the clustering, where
the loss of a soft photon from the asymmetrically decayed \pizero\ may change the reconstruction efficiency.
This analysis does not depend on the absolute energy calibration of the \SMD, because the main energy measurement was obtained from the towers.
Instead, we estimated the possible disagreement between the \SMD\ scale in the data and in Monte Carlo simulation to be below \MM{20\unitns{\%}}\@.
The resulting variation of \SUB{\EPS}{\RM{acc}} from this source was \MM{3.5\unitns{\%}} in the HighTower data 
at the pion \MM{\pT{} = 4\unit{\GeVc}} and less than \MM{1\unitns{\%}} above \MM{11\unit{\GeVc}}\@.

All systematic error contributions are summarized in Table~\ref {table_systematic_uncertainty},
classified into the following categories:
\label {syst_classification}
\begin {description}
\item [A] point\tsp{0.2}-by-point systematic uncertainty;
\item [B] \pT-correlated systematic uncertainty, but uncorrelated between \protonproton\ and \deuterongold\ datasets;
\item [C] \pT-correlated systematic uncertainty, also correlated between \protonproton\ and \deuterongold\ datasets;
\item [N] normalization uncertainty, uncorrelated between \protonproton\ and \deuterongold\ datasets.
\end {description}

Table~\ref {table_systematic_uncertainty} also lists which measurement is affected by a given source of systematic error.

\section {Direct photon analysis}
\label {sec_direct_photons}

The traditional approach to measuring direct photon production in hadronic collisions uses isolation criteria.
Photons from decays of highly energetic hadrons should be accompanied by other jet fragments.
Therefore, one can reject those by requiring less than a certain amount of background energy in a cone 
around a photon candidate~\cite {ref_recent_critical_study_photons}\@.
However, prompt photon production beyond leading order in \pQCD\ cannot be separated unambiguously from photons from fragmentation processes, 
although the framework for applying isolation cuts in \pQCD\ calculations is established and 
theoretical interpretation of experimental results is possible.
In addition, the use of isolation cuts in the high-multiplicity environment of heavy-ion collisions is not straightforward. 

As the present analysis is intended to provide a baseline measurement for heavy-ion collisions, 
we have chosen to use the method of statistical subtraction to obtain direct photon yields.
For this method, one obtains inclusive photon spectra, which, in addition to the direct contribution, contain
a large background of decay photons, dominantly from \pizero\ decays.
An accurate measurement of \pizero\ and heavier hadrons provides the necessary input 
to subtract the decay background. This method has been successfully used in heavy-ion 
reactions~\cite {ref_wa98_directphotons_PbPb,ref_phenix_direct_photons_AuAu_centrality},
however, it does not provide event-by-event direct photon identification.

The sample of photon candidates served as the main input to the direct photon analysis, 
as in the case of the reconstruction of the \pizero\ spectrum, described in the previous sections.
After subtracting the contamination by charged particles and neutral hadrons,
the raw inclusive photon sample was corrected to account for the limited acceptance and the finite detector resolution. 
In parallel, the total yield of photons from \pizero\!, \etameson, and \omegameson(782) decays was simulated, 
assuming a phenomenological scaling law (\mT\ scaling) for the \etameson\ and \omegameson(782) spectra.

To exploit the fact that the inclusive photon and decay photon yields
have many correlated uncertainties, we studied the direct photon
yield via the double ratio
\begin {equation} \label {eq:rgamma_def}
\Rgamma{} \ensuremath{\equiv} \FRAC 
{\DIV {\SUB{\gama}{\RM{incl}}} {\pizero}} 
{\DIV {\SUB{\gama}{\RM{decay}}} {\pizero}} 
,
\end {equation}
where the numerator equals the point\tsp{0.2}-to\tsp{0.4}-point ratio of the measured spectra
of inclusive photons and \pizero's, as a function of \pT, and the denominator is the simulated
background contribution from decay processes divided by the parametrized
\pizero\ yield. It follows that
\begin {equation}
\Rgamma{} = 1 + \FRAC {\SUB{\gama}{\RM{dir}}} {\SUB{\gama}{\RM{decay}}}
,
\end {equation}
which serves as an indicator of a direct photon signal \SUB{\gama}{\RM{dir}} (\MM{\Rgamma{} \GREATER{} 1})\@.
The absolute direct photon yields can subsequently be determined as
\begin {equation} \label {eq:gamma_direct_def}
\SUB{\gama}{\RM{dir}} = \left( 1 - \SUP{\Rgamma}{\tsp{0.5}-\tsp{-0.4}1} \right) \SUB{\gama}{\RM{incl}}
,
\end {equation}
where the systematic uncertainties, which canceled in the double ratio, have to be included again.

\subsection {Inclusive photons}
\label {sec:correction-factors-photons}
The reconstruction of the inclusive photon spectrum
was in many ways similar to that of the \pizero\@. 
The uncorrected photon spectrum was extracted from the
same data sample using identical event and photon candidate cuts. 
However, in the \pizero\ analysis there were no rigorous constraints on the purity of the photon candidates,
because remaining contributions from charged particles and neutral hadrons were identified
afterwards as the combinatorial background in the mass distributions. 
In contrast, the uncorrected inclusive photon yield \SUB{\IT{Y}}{\RM{incl}} was obtained from an explicit
subtraction of such backgrounds,
\begin {equation} \label {eq:photons:2} 
  \SUB{\IT{Y}}{\RM{incl}} = \left( 1 - \SUB{\IT{C}}{0} \right) \, \left( 1 - \SUB{\IT{C}}{\PLMN}\right) \, \SUB{\IT{Y}}{\RM{cand}} 
,
\end {equation}
where the correction terms \SUB{\IT{C}}{0} and \SUB{\IT{C}}{\PLMN} represent the fractional
contamination by neutral hadrons and charged particles, respectively.
The charged particle contamination \SUB{\IT{C}}{\PLMN} was estimated in section~\ref {subsec:cpv} and found to be smaller than~\MM{5\unitns{\%}}\@.

The invariant yield of inclusive photons was calculated, 
similarly to that of \pizero's in Eq.~(\ref {eq:pions_2}), as
\begin {equation} \label {eq:photons:3} 
\energy\tsp{0.8} \FRAC{\SUP{\der}{3}\tsp{-0.7}\Number}{\der\SUP{\momentumthree}{3}} =
\FRAC{1}{2\PI\pT \SUB{\Number}{\RM{trig}} \SUB{\IT{K}}{\RM{trig}}}
\FRAC{\SUB{\IT{Y}}{\RM{incl}}}{\DELTA\pT\tsp{0.5}\DELTA\rapidity}
\FRAC{\SUB{\SUP{\EPS}{\ }}{\RM{vert}}\tsp{0.5}\SUP{\SUB{\IT{c}}{\RM{loss}}}{\gama}}{\SUP{\SUB{\EPS}{\RM{acc}}}{\gama}\tsp{0.5}\SUP{\SUB{\EPS}{\RM{cpv}}}{\gama}}
.
\end {equation}
Here \SUP{\SUB{\EPS}{\RM{acc}}}{\gama} is the single photon acceptance and efficiency correction factor, discussed in section~\ref {subsec_photon_eff} below.

\subsection {Neutral hadron background}
\label {sec:neutr-hadr-cont}

The term \SUB{\IT{C}}{0} in Eq.~(\ref {eq:photons:2}) was defined as the number of
reconstructed showers generated by neutral hadrons relative to the total
number of showers in the photon candidate sample. The \STAR\ detector
has no means of directly identifying neutrons and antineutrons\@. 
Therefore, this contamination was simulated using the measured (anti)proton spectra as input.

The largest source of neutral contamination was the \antineutron\ annihilation in the calorimeter, 
for example \MM{\antineutronproton{} \TO{} 2\piplus\piminus\pizero}\@.
This initiates a shower that does not necessarily develop in the incident direction of \antineutron\@.
Moreover, the available energy for the reaction products includes twice the rest mass of a nucleon (\APPROX\MM{2\unit{\GeV}})\@. 

\STAR\ has measured the \proton\ and \antiproton\ production in \protonproton\ and \deuterongold\ collisions~\cite {ref_star_idhadrons}\@.
The reported yields, however, were not corrected for the \lambdabaryon\ and \antilambdabaryon\ feed-down, which is expected to have a contribution of
\MM{\SUB{\ensuremath{\delta}}{\lambdabaryon} \APPROX{} 20\unitns{\%}}~\cite{ref_star_idhadrons1}\@.
Therefore, the \antineutron\ yield was estimated as
\begin {equation}
\IT{Y}(\antineutron) = 
\left( 1 - \SUB{\ensuremath{\delta}}{\lambdabaryon} \right) \IT{Y}(\antiproton) + 
\SUB{\ensuremath{\delta}}{\lambdabaryon} \FRAC{\ensuremath{\mathcal{B}}
(\lambdabaryon\TO\neutron\pizero)}{\ensuremath{\mathcal{B}}(\lambdabaryon\TO\proton\piminus)} \IT{Y}(\antiproton)
,
\end {equation}
where the branching ratios are \MM{\ensuremath{\mathcal{B}}(\lambdabaryon\TO\neutron\pizero) = 0.358} 
and \MM{\ensuremath{\mathcal{B}}(\lambdabaryon\TO\proton\piminus) = 0.639}~\cite {ref_pdg}\@.

To study the contamination of the photon candidate spectrum, approximately
\MM{3\e{6}} \antineutron's were generated with an exponentially falling \pT\ spectrum.
This provided sufficient statistics at low \pT, where the \antineutron's constituted a significant source of contamination. 
The \FLUKA\ program~\cite {fasso-2003} was used to describe the particle transport and the interactions in the detector material. 
The parametrizations of the \proton\ and \antiproton\ yields were not only used to assign a weight to the Monte Carlo events, 
but also to determine the absolute contamination of the photon sample. 
The latter was divided by the number of reconstructed photon candidates to calculate the term \SUB{\IT{C}}{0}\@.

\begin {figure} [tb] 
\centerline {\hbox {
\includegraphics {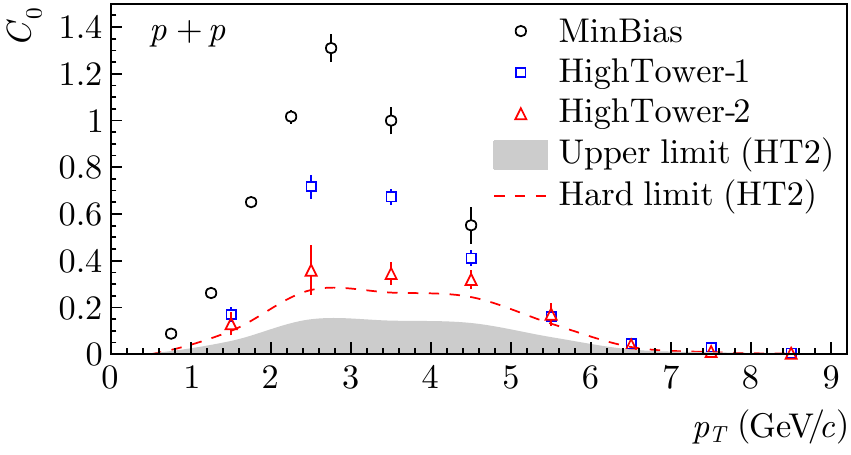}
}}
\caption {\label{fig:Czero}(color online) Relative neutral hadron contribution to the photon candidate yields, \SUB{\IT{C}}{0}\@.
Filled area is the upper limit for the \protonproton\ \HighTowerTwo\ data sample,
dashed curve is the upper limit in the extreme scenario that all photon candidates were a result of showering \antineutron's.
}
\end {figure}

The final contamination factor \SUB{\IT{C}}{0} in the \protonproton\ data is shown in Fig.~\ref{fig:Czero}\@.
In a limited \pT\ range, \SUB{\IT{C}}{0} appears to be larger than unity,
which is not possible unless the associated systematic uncertainties are extremely large.
The violation of this bound, as well as observed discrepancies between the three triggers, 
indicate the large uncertainty of the \antineutron\ simulation.

Two natural limits to the contamination were considered: 
(i) the hard upper limit \MM{\SUB{\IT{C}}{0} \LESSEQ{} 1}, which is not realistic, as it implies that the inclusive photon yield is zero,
and (ii) the limit implied by the assumption that the direct photon signal is zero around the annihilation peak (\APPROX2\unit{\GeVc})\@.
In both cases, a scaling factor for \SUB{\IT{C}}{0} was derived and subsequently applied to the \HighTowerTwo\ data, as shown in Fig.~\ref{fig:Czero}\@.
For the further analysis, we have chosen the second estimate, 
calculated assuming that only background photons were detected in the range \MM{1 \LESS{} \pT{} \LESS{} 4\unit{\GeVc}},
as the upper limit of the neutral hadron contamination \SUB{\IT{C}}{0}\@.
This upper limit was found to be negligible in the range of the present direct photon measurement, \MM{\pT{} \GREATER{} 6\unit{\GeVc}}\@.

The \neutron\ and \SUB{\SUP{\IT{K}}{0}}{\IT{L}} interactions with the \BEMC\ resulted in 
the smaller contamination than that of the \antineutron's, at all values of \pT\@.

\subsection {Photon reconstruction efficiency}
\label {subsec_photon_eff}

We have calculated the photon acceptance and efficiency correction factor \SUP{\SUB{\EPS}{\RM{acc}}}{\gama}
separately for events containing \pizero\ decay photons and for events containing only a single photon.
The latter factor was applied to the fraction of the photon yield from all sources other than the \pizero\ decay.

To determine the acceptance correction \SUP{\SUB{\EPS}{\RM{acc}}}{\gama} for the \pizero\ decay photons,
we used a \GEANT-based Monte Carlo simulation of the \STAR\ detector.
The Monte Carlo events were weighted in such a way that the measured \pizero\ yield was reproduced.
This is important because the photon acceptance depends on the degree of cluster merging. 
This, in turn, depends on the opening angle of the decay photons, and thus on the momentum of the parent \pizero\@.
Furthermore, the simulation included all the possible losses of photon
candidates listed in section~\ref {section_pizero_analysis}, 
except those associated with the invariant mass window and with the cut on the
energy asymmetry \SUB{\IT{Z}}{\gama\gama}\@. 
One important effect affects the showers initiated by daughters of a high-\pT\ \pizero\@. 
Because there was no requirement on the relation between the reconstructed \pT\ and the
Monte Carlo input \pT, the correction implicitly accounted for events in which
one of the two decay photons remained unidentified and the total energy was
assigned to a single cluster. Such merging of photon showers constituted the
main difference between the reconstruction efficiencies of \pizero\ decay products and single photons.

Similarly to the above, the \SUP{\SUB{\EPS}{\RM{acc}}}{\gama} factor for single photons  
was determined using a Monte Carlo sample of \APPROX\MM{1\e{6}} events.
Each event contained a single photon, uniformly distributed in azimuthal angle \MM{-\PI \LESS{} \phiangle{} \LESS{} +\PI},
pseudorapidity \MM{-0.3 \LESS{} \pseudorapidity{} \LESS{} +1.2}, 
and transverse momentum \MM{0 \LESS{} \pT{} \LESS{} 20\unit{\GeVc}}\@.
Events were weighted with a function determined from 
the spectrum of photons from decaying hadrons other than the \pizero, as well  
as from that of the direct photons. However, we will demonstrate in section~\ref {sec:backgr-from-hadr} 
that the shape of the decay photon spectrum and 
the measured \pizero\ spectrum were very similar, at least for the \pT\ range 
of this analysis. Although the direct photon  
spectrum was expected to exhibit a slightly different \pT\ dependence,  
varying the input spectrum correspondingly did not yield quantitatively  
different results. 
 
Finally, we have implemented a correction to the measured photon yields as follows. 
The yield of photons that originated from the decay  
\MM{\pizero{} \TO{} \gama\gama} was determined from the measured \pizero\ spectrum. 
This part of \SUB{\IT{Y}}{\RM{incl}} was corrected with  
the \SUP{\SUB{\EPS}{\RM{acc}}}{\gama} factor calculated for the \pizero\ decay photons. 
The remaining part of \SUB{\IT{Y}}{\RM{incl}} was assumed to consist  
of single photons that were not correlated with the other photon candidates in the event,
and was corrected with the single-photon \SUP{\SUB{\EPS}{\RM{acc}}}{\gama} factor.
This assumption was based on the observation that the reconstruction efficiency for photons from the decay 
\MM{\etameson{} \TO{} \gama\gama}, which is the second largest source of decay photons 
(\APPROX\MM{15\unitns{\%}}), was equal to that of single photons, 
because the opening angle between the two daughter photons is large enough that both are never incident on the same calorimeter tower.
 
\begin {figure} [tb] 
\centerline {\hbox {
\includegraphics {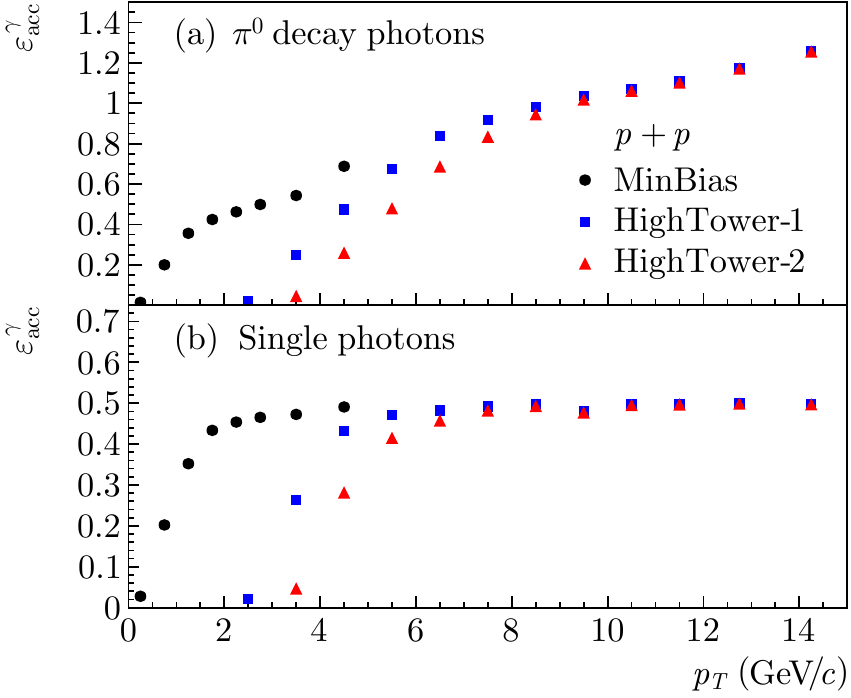}
}}
\caption {\label {fig:reco_eff}(color online) (a) Acceptance and efficiency factor \SUP{\SUB{\EPS}{\RM{acc}}}{\gama} for 
photons exclusively from the decay \MM{\pizero{} \TO{} \gama\gama}\@. 
The observed rise of the efficiency at high \pT\ is caused by the merging of the \pizero\ decay daughters.
(b) The \SUP{\SUB{\EPS}{\RM{acc}}}{\gama} factor for single photons, used to correct the fraction of 
the photon sample that exceeded the simulated \pizero\ decay contribution.}
\end {figure}

Figure~\ref {fig:reco_eff} 
shows the acceptance and reconstruction efficiency 
\SUP{\SUB{\EPS}{\RM{acc}}}{\gama} for the \pizero\ decay photons and for single photons
for the \protonproton\ MinBias, \HighTowerOne, and \HighTowerTwo\ data.
The two HighTower results were found to be very similar in the low-\pT\ region, 
where the angular separation of the decay photons was still large, compared to the size of a \BEMC\ tower.
However, at higher \pT\ the two photons are difficult to separate, particularly in case of the most symmetric decays.
When two photons were merged, the remaining photon candidate was 
erroneously assigned the energy sum of both showers. This led to significantly  
larger reconstruction efficiency, compared to that for single photons.
Eventually, at the largest \pT\ values considered in this analysis, the decay  
photon efficiency even exceeded unity.

\subsection {Fully corrected inclusive yields}
 
The inclusive photon yield \SUB{\IT{Y}}{\RM{incl}} was derived by subtracting 
the charged and neutral backgrounds from the yield of raw photon candidates [Eq.~(\ref {eq:photons:2})]\@. 
The contamination by charged particles was subtracted according to the procedure explained 
in section~\ref {subsec:cpv}, but the neutral hadron correction proved to be difficult.
Although an upper limit for the contamination fraction \SUB{\IT{C}}{0} was derived in section~\ref {sec:neutr-hadr-cont}, 
we did not find any means to reduce the associated 
systematic uncertainty on \SUB{\IT{C}}{0} to a level where a meaningful subtraction could be performed for MinBias data. 
In the HighTower data, the upper limit on the contamination fraction vanishes at higher values of \pT\@. 
Therefore, our final results were obtained in the range  
\MM{6 \LESS{} \pT{} \LESS{} 15\unit{\GeVc}}, and the photon candidates obtained from the MinBias data were discarded.

\begin{figure} [tb] 
\centering 
\includegraphics {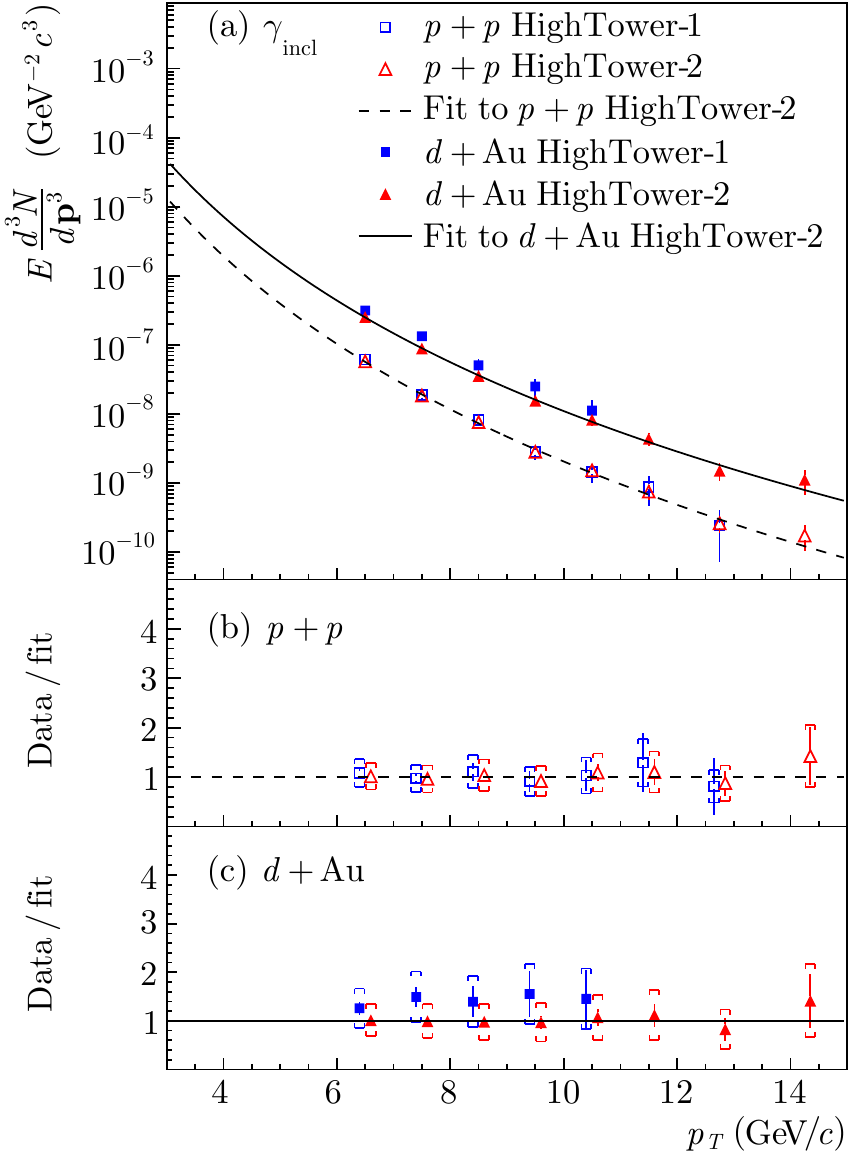}
\caption {(color online) (a) Inclusive photon invariant yield per MinBias \protonproton\ and \deuterongold\ collision\@. 
Curves are the power law fits given in the text.
Invariant yield divided by the fit to the (b) \protonproton\ and (c) \deuterongold\ data.
The neutral hadron contamination was not subtracted (see text) 
and is expected to be significant at \MM{2 \LESSAPPROX{} \pT{} \LESSAPPROX{} 4\unit{\GeVc}}\@.
The error bars are statistical and brackets in the lower panels are the systematic uncertainties.
Data points for the overlapping \pT\ bins in the lower panels are horizontally displaced for clarity.
\label {fig:inclgamma}
}
\end{figure}

Figure~\ref {fig:inclgamma}
shows the corrected inclusive photon spectra in \protonproton\ and \deuterongold\  collisions without the subtraction of the neutral hadron contribution.
The lower panels show the data divided by the corresponding power law fits.
A small systematic difference between the spectra from \deuterongold\ \HighTowerOne\ and \HighTowerTwo\ collisions (1.2 standard deviations, on average) was observed.
However, the measured \HighTowerTwo\ yields is statistically more significant,
because the \HighTowerOne\ events that contained photons in the range \MM{6 \LESS{} \pT{} \LESS{} 10\unit{\GeVc}} were a subset of the \HighTowerTwo\ data. 
We reconstructed 17684 (3738) photon candidates from the \HighTowerTwo\ (\HighTowerOne) \deuterongold\ data in that \pT\ range.
The final direct photon cross sections presented in section~\ref {sec_results} 
were obtained exclusively from the \HighTowerTwo\ data.  
 
\subsection {Background from hadronic decays}
\label {sec:backgr-from-hadr}

The photon yield from hadronic decays was determined with a simulation of the decay processes 
listed in Table~\ref {tab:decaytable}\@. 
\begin {table} [tbp]
\begin {center}
\caption {Dominant hadronic decay contributions to the inclusive photon yield~\protect\cite {ref_pdg}\@.} 
\label {tab:decaytable} 
\begin {ruledtabular} 
\begin {tabular} {@{\extracolsep{\fill}}lD{.}{.}{2.2}} 
\noalign{\smallskip}
Decay & \multicolumn {1} {c} {Branching ratio (\unitns{\%})} \\ 
\noalign{\smallskip}\hline\noalign{\smallskip} 
\MM{\pizero{} \TO{} \gama\gama}           & \quad\quad\quad 98.80 \\ 
\MM{\pizero{} \TO{} \epluseminus\gama}    & \quad\quad\quad 1.20  \\ 
\noalign{\smallskip}\hline\noalign{\smallskip}    
\MM{\etameson{} \TO{} \gama\gama}           & \quad\quad\quad 39.23 \\ 
\MM{\etameson{} \TO{} \piplus\piminus\gama} & \quad\quad\quad 4.78  \\ 
\MM{\etameson{} \TO{} \epluseminus\gama}    & \quad\quad\quad 0.49  \\ 
\noalign{\smallskip}\hline\noalign{\smallskip} 
\MM{\omegameson(782) \TO{} \pizero\gama}         & \quad\quad\quad 8.69  \\ 
\end {tabular}
\end {ruledtabular} 
\end {center}
\end{table} 
The other possible contributions, 
from processes such as \MM{\etamesonprime{} \TO{} \SUP{\rhomeson}{0}\gama}, were found to be
negligible (\LESS\MM{1\unitns{\%}})\@.
A fit of the measured \pizero\ yield in the range \MM{4 \LESS{} \pT{} \LESS{} 15\unit{\GeVc}} to the form
\MM{{\ensuremath{\sim}}\SUP{(1 + \pT)}{-\ensuremath{\alpha}}} served as an input to the simulation. The fit yielded \MM{\ensuremath{\alpha} = 9.1 \PLMN{} 0.1}
and \MM{9.0 \PLMN 0.1} for \deuterongold\ and \protonproton\ collisions, respectively.
The normalization is irrelevant because it cancels in the ratio \DIV{\SUB{\gama}{\RM{decay}}}{\pizero}\@.

\begin{figure} [tb] 
\centering 
\includegraphics {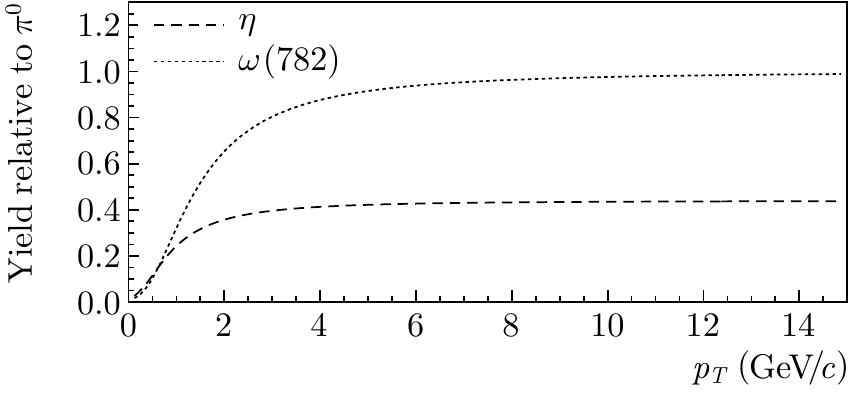} 
\caption {The estimated yield of \etameson\ and \omegameson(782) mesons in 
\deuterongold\ collisions, relative to the measured \pizero\ yield, determined from the \mT\ scaling as described in the text.
\label {fig:mesonspectra}
}
\end{figure}

To estimate the yields of the \etameson\ and \MM{\omegameson(782)}, we used the fact that these scale with the \pizero\ yields
when expressed in terms of the transverse mass 
\MM{\mT{} \ensuremath{\equiv} \SQRT{\SUP{\SUB{\mass}{\phantom{\IT{T}}}}{2} + \SUP{\pT}{2}}} 
instead of \pT~\cite {ref_mT_scaling,schaffner2002,ref_star_resonances_dAu,ref_star_strangepp}\@.
For the \etameson\ spectra, we used the scaling ratios 
\MM{\SUB{\IT{R}}{\etatopisub} = 0.46 \PLMN{} 0.05} for \protonproton\ and
\MM{0.44 \PLMN{} 0.08} for \deuterongold\ data, as followed from our measurement of \etameson\ production.
In case of \MM{\omegameson(782)}, we used \MM{\SUB{\IT{R}}{\omegatopisub} = 1.0 \PLMN{} 0.2}, in agreement with recent measurements at \RHIC~\cite {ref_phenix_omega_pp_dAu}\@.
The estimated \etameson\ and \MM{\omegameson(782)} yields relative to the \pizero\ yield are~shown~in~Fig.~\ref {fig:mesonspectra}\@.

\begin {figure} [tb] 
\centering 
\includegraphics {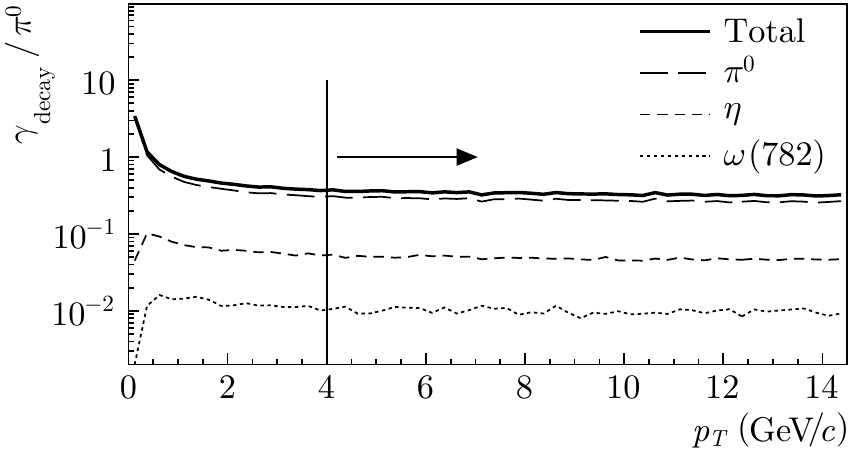} 
\caption {The simulated number of photons per input pion  
\DIV{\SUB{\gama}{\RM{decay}}}{\pizero} from hadronic decays in \deuterongold\ collisions, as a function of \pT\@. 
The included decay processes are listed in Table~\protect\ref {tab:decaytable}\@. The vertical line indicates 
the lower limit of the \pT\ range that was used to fit the \pizero\ spectrum. 
The value of \DIV{\SUB{\gama}{\RM{decay}}}{\pizero} below 
\MM{\pT{} = 4\unit{\GeVc}} is, therefore, less accurate, however, 
it was not used for further analysis (see text)\@.
\label {fig:decay_photons} 
}
\end {figure}

Figure~\ref {fig:decay_photons} 
shows the ratio \DIV{\SUB{\gama}{\RM{decay}}}{\pizero} for \deuterongold\ collisions. 
The curves represent the contributions of the \pizero\!, \etameson, and \omegameson(782), and the total decay photon 
yield, each divided by the parametrization of the measured \pizero\ spectrum. 
The normalization uncertainty cancels upon taking this ratio.
The uncertainty due to the shape of the \pizero\ spectrum
and to the \mT\ scaling factors was estimated by varying the fitted exponents and the scale factors by their errors.

\subsection {Summary of systematic uncertainties}
\label {sec:gamma_systematic}

All systematic error contributions to the double ratio \Rgamma\ [Eq~(\ref {eq:rgamma_def})] 
are summarized in Table~\ref {table_systematic_uncertainty_gamma}\@.
\begin {table} [tbp]
\begin {center}
\caption {Systematic error contributions for the double ratio \Rgamma\@.
The classifications A and B are defined in section~\ref {table_systematic_uncertainty_gamma}\@.
}
\label {table_systematic_uncertainty_gamma}
\begin {ruledtabular} 
\begin {tabular} {@{\extracolsep{\fill}}l@{\extracolsep{\fill}}c@{\extracolsep{\fill}}c@{\extracolsep{\fill}}}
\noalign{\smallskip}
Source                      & Type & Value at low (high) \pT\ (\unitns{\%}) \\
\noalign{\smallskip}\hline\noalign{\smallskip}
\pizero\ yield extraction   & A & 7.1 \\
Beam background             & A & 1 (3) in \deuterongold \\
Tower energy scale          & B & 3 \\
Tower gain spread           & B & 1 \\
\SMD\ energy scale          & B & 12 \\
\SMD\ gain spread           & B & 1 \\
\etatopi                    & B & 2 \\
\pizero\ yield fit          & B & 1.5 \\
\end {tabular}
\end {ruledtabular} 
\end {center}
\end {table}
Expressing the direct photon yield in terms of a double ratio gives a large reduction in the systematic error, 
since the contribution from the \BEMC\ energy scale uncertainty cancels.
Consequently, the largest sources of uncertainty are those associated with the \pizero\ yield extraction and with the \SMD\ energy scale.
The latter leads to the \Rgamma\ variation of \MM{12\unitns{\%}}, independent of \pT\ for \MM{\pT{} \GREATER{} 6\unit{\GeVc}}\@.

The beam background observed in the \deuterongold\ data has a larger effect on single-photon analysis than on the \pizero\ reconstruction, 
since the background-induced showers in the \BEMC\ could not be distinguished from genuine photons originating from the event vertex.
Therefore, we varied the cutoff value for the electromagnetic energy fraction \rBeamBg\ in an event [Eq.~(\ref {eq:rBeamBg_def})] 
in the range \MM{\rBeamBg{} = 0.7}\tsp{-0.5}--\tsp{0.5}\MM{0.9}\@.
This propagated into 1\tsp{-0.2}--\tsp{0.2}3\unitns{\%} point-to-point systematic error of \Rgamma\@. 

\section {Results and Discussion}
\label {sec_results}

\subsection {Cross section for neutral pion production}

\label {cross_section_calculation}
The invariant differential cross section for \pizero\ and \etameson\ production
in inelastic \protonproton\ interactions is given by
\begin {equation} \label {eq:crosssectiondef}
\energy\tsp{0.8} \FRAC{\SUP{\der}{3}\tsp{-0.7}\SUP{\SUB{\SIGMA}{\RM{inel}}}{\protonproton}}{\der\SUP{\momentumthree}{3}} =
\energy\tsp{0.8} \FRAC{\SUP{\der}{3}\tsp{-0.7}\SUP{\SUB{\SIGMA}{\RM{\NSD}}}{\protonproton}}{\der\SUP{\momentumthree}{3}} =
\SUP{\SUB{\SIGMA}{\RM{\NSD}}}{\protonproton}
\tsp{1.0}
\FRAC{\SUP{\der}{2}\tsp{-0.7}\Number}{2\PI\pT\tsp{0.5}\der\pT\tsp{0.5}\der\rapidity}.
\end {equation}
It has been shown that the singly diffractive contribution to the inelastic cross section
is negligible at \MM{\pT{} \GREATER{} 1\unit{\GeVc}}~\cite {ref_star_idhadrons1}\@. 
Therefore, we can assume that the differential inelastic cross section is equal to the
differential \NSD\ cross section in our \pT\ range.
The total \NSD\ cross section in \protonproton\ collisions was found to be
\MM{\SUP{\SUB{\SIGMA}{\RM{\NSD}}}{\protonproton} = 30.0 \PLMN{} 3.5\unit{\mb}},
and the total hadronic cross section in \deuterongold\ collisions was found to be
\MM{\SUP{\SUB{\SIGMA}{\RM{hadr}}}{\deuterongold} = 2.21 \PLMN{} 0.09\unit{\barn}} (see section~\ref {subsec_datasets})\@.

\begin {figure} [tb]
\centerline {\hbox {
\includegraphics {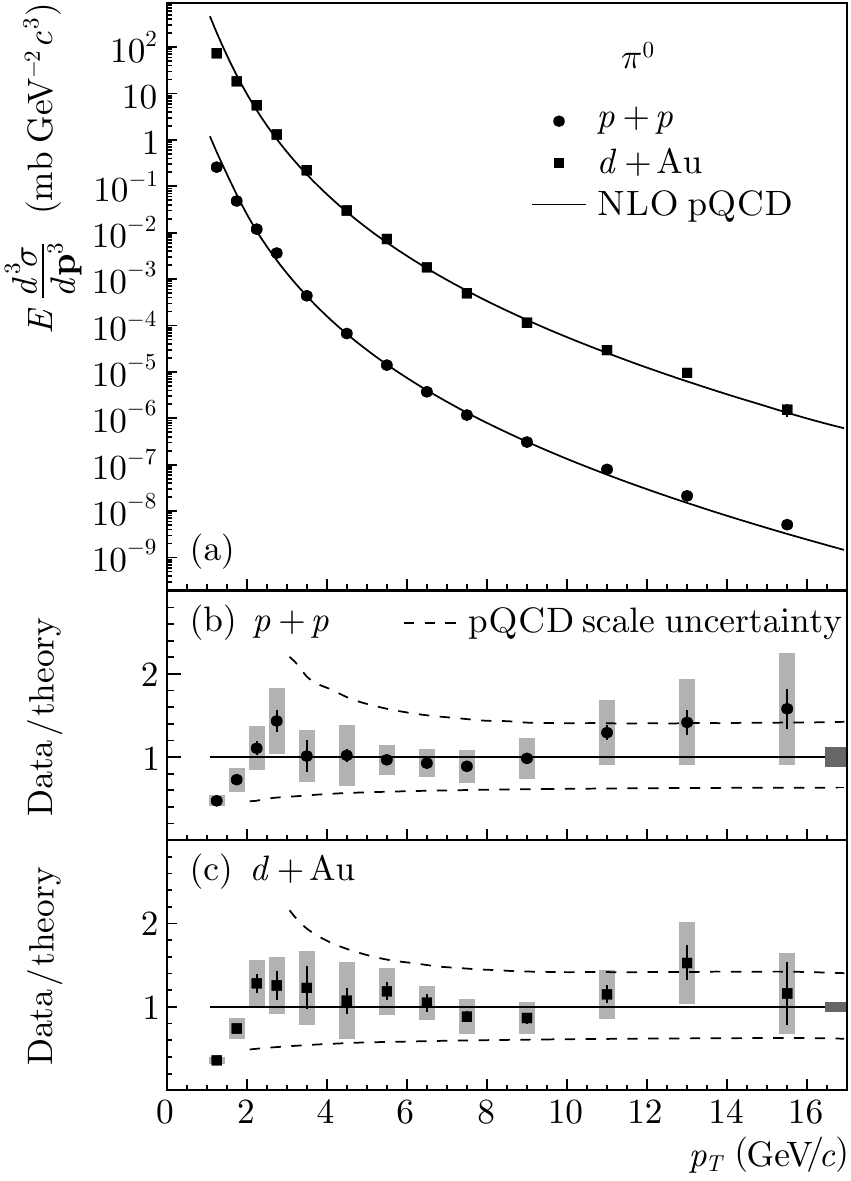}
}}
\caption {(a) Cross section for inclusive \pizero\ production in \protonproton\ and \deuterongold\ collisions at \MM{\sNN{} = 200\unit{\GeV}}, 
divided by the corresponding \NLO\ \pQCD\ calculations~\protect\cite {ref_nlo_pqcd} for (b) \protonproton\ and (c) \deuterongold\ collisions\@.
The error bars are statistical and shaded bands are \pT-correlated systematic uncertainties.
Normalization uncertainties are indicated by shaded bands around unity in the lower panels\@.
\label {fig_crossection_pp_theory}
}
\end {figure}

The measured cross sections for \pizero\ production in the \protonproton\ (presented in Ref.~\cite {ref_star_pi0ALL} and included here for completeness)
and \deuterongold\ collisions are shown in Fig.~\ref {fig_crossection_pp_theory}\@.
The cross sections are compared to the \NLO\ \pQCD\ calculations~\cite {ref_nlo_pqcd}\@.
The \CTEQ6\capsword{M} parton densities~\cite {ref_cteq6} and the \KKP\ fragmentation 
functions~\cite {ref_kkp_ff} were used in the \protonproton\ calculation.
The \deuterongold\ calculation used the nuclear parton distributions for gold~\cite {ref_au_pdf_1,ref_au_pdf_2,ref_au_pdf_3}, in addition.
The factorization scale \ensuremath{\mu} was set equal to \pT\ and was varied by a factor of two to estimate the scale uncertainty,
indicated by the dashed curves in the lower panels of Fig.~\ref {fig_crossection_pp_theory}\@.
These panels show the ratio of the measured cross sections to the corresponding \QCD\ predictions.
The error bars shown in the plot are the statistical and the shaded bands are the systematic uncertainties.
The normalization uncertainties are indicated by shaded bands around unity on the right-hand side of each ratio plot.
The measured \pizero\ cross sections were not corrected for feed-down contributions from 
\MM{\etameson{} \TO{} 3\pizero}, \MM{\etameson{} \TO{} \piplus\piminus\pizero}, and \MM{\SUP{\SUB{\IT{K}}{\IT{S}}}{0} \TO{} \pizero\pizero} decays,
which are expected to be negligible.
It is seen that the measured \pizero\ cross sections in both \protonproton\ and \deuterongold\ collisions
are well described by the \NLO\ \pQCD\ calculations in the fragmentation region \MM{\pT{} \GREATER{} 2\unit{\GeVc}}\@.

\begin {figure} [tb]
\centerline {\hbox {
\includegraphics {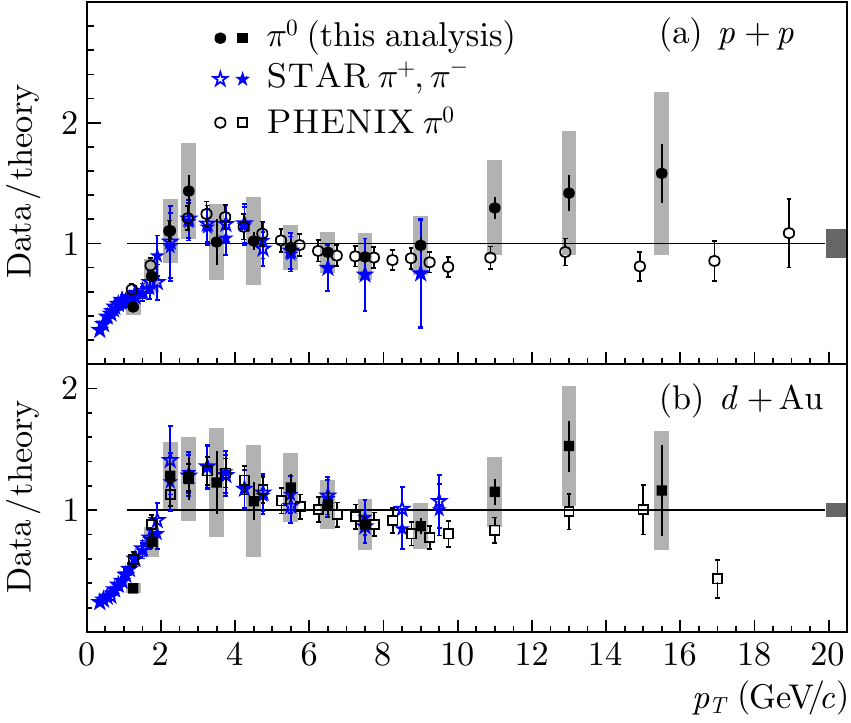}
}}
\caption {(color online) Cross section for inclusive \pizero\ production in (a) \protonproton\ and (b) \deuterongold\  collisions at \MM{\sNN{} = 200\unit{\GeV}}, 
divided by \NLO\ \pQCD\ calculations~\protect\cite{ref_nlo_pqcd} and
compared to the \STAR\ \piplusminus~\protect\cite {ref_star_idhadrons,ref_star_idhadrons1} and \PHENIX\ \pizero~\protect\cite {Adare:2007dg, ref_phenix_pi0_dAu} measurements.
The error bars are statistical and shaded bands are \pT-correlated systematic uncertainties.
Normalization uncertainties are indicated by shaded bands around unity in each panel.
\label {fig_crossection_star_phenix}
}
\end {figure}

In Fig.~\ref {fig_crossection_star_phenix},
we compare the \pizero\ measurements in the \protonproton\ and \deuterongold\ data with the 
previous \piplusminus\ measurements by \STAR~\cite {ref_star_idhadrons,ref_star_idhadrons1} 
and with the \pizero\ measurements by \PHENIX~\cite {Adare:2007dg, ref_phenix_pi0_dAu}\@.
Here, and in all following figures, the cited data are shown with their statistical and systematic uncertainties added in quadrature.
The normalization uncertainties shown by the grey bands in the figure
are largely correlated between the \pizero\ and the \piplusminus\ data points 
and uncorrelated with the \PHENIX\ normalization uncertainties of similar magnitude.
It is seen that the neutral and charged pion spectra from \STAR\ agree very well
in both \protonproton\ and \deuterongold\ data, in spite of different detector subsystems and 
analysis techniques used in these measurements. The present results extend the reach of \STAR\ pion measurements to \MM{\pT{} = 17\unit{\GeVc}}\@.
Comparison to the cross sections measured by \PHENIX\ shows good agreement, within errors, in both collision systems.
However, we note that our data indicate a possible excess over the \PHENIX\ measurements at \MM{\pT{} \GREATER{} 10\unit{\GeVc}} in both cases.

To parametrize the \pT\ dependence, the measured \pizero\ cross section, 
as well as the \etameson\ and \SUB{\gama}{\RM{incl}} cross sections presented in the following sections, were fitted to
the power law function [Eq.~(\ref {eq:power_law})], 
and the resulting parameter values are listed in Table~\ref {table_crossection_fit}\@.
\begin {table} [tbp]
\begin {center}
\caption {The values of the power law fit parameters from Eq.~(\ref {eq:power_law}) for the measured \pizero\!, \etameson, 
and \SUB{\gama}{\RM{incl}} cross sections.}
\label {table_crossection_fit}
\begin {ruledtabular} 
\begin {tabular} {@{\extracolsep{\fill}}l@{\extracolsep{\fill}}c@{\extracolsep{\fill}}c@{\extracolsep{\fill}}c@{\extracolsep{\fill}}c@{\extracolsep{\fill}}}
\noalign{\smallskip}
Data & \IT{A}\ (\unitns{\mb}\SUP{\unit{\GeV}}{-\tsp{-0.3}2}\tsp{-0.5}\SUP{\unit{\cspeed}}{3}) & \SUB{\momentum}{0}\ (\unitns{\GeVc}) & \IT{n} & \ensuremath{\chi^2/\text{ndf}} \\
\noalign{\smallskip}\hline\noalign{\smallskip}
\pizero, \protonproton & (1.69\PLMN{}0.65)\e{3} & 0.723\PLMN{}0.066 & 8.61\PLMN{}0.14 & 65/10 \\
\pizero, \deuterongold & (4.02\PLMN{}1.35)\e{4} & 1.46\PLMN{}0.16 & 9.93\PLMN{}0.32 & 53/10 \\
\etameson, \protonproton & (7.0\PLMN{}5.0)\e{1} & 1.33\PLMN{}0.23 & 9.83\PLMN{}0.44 & 30/9\,\,\, \\
\etameson, \deuterongold & (3.33\PLMN{}0.41)\e{4} & 1.33(fixed) & 9.83(fixed) & 32/10 \\
\SUB{\gama}{\RM{incl}}, \protonproton & (3.1\PLMN{}0.1)\e{0} & 0.941\PLMN{}0.268 & 8.61\PLMN{}0.40 & 2/5 \\
\SUB{\gama}{\RM{incl}}, \deuterongold & (2.4\PLMN{}0.1)\e{1} & 0.697\PLMN{}0.126 & 7.88\PLMN{}0.23 & 2/5 \\
\end {tabular}
\end {ruledtabular} 
\end {center}
\end {table}
Because of the large uncertainties, 
the \SUB{\momentum}{0} and \IT{n} parameters for the \etameson\ cross section in \deuterongold\ data 
had to be fixed at the corresponding \protonproton\ values to achieve a stable fit.
The quoted values of \MM{\DIV{\SUP{\ensuremath{\chi}}{2}}{\RM{ndf}}} indicate that 
these fits provide only a general guidance on the shapes of the spectra and 
do not necessarily describe all features seen in the data.

In addition, the pure power law fit \MM{\ensuremath{\sim}\SUP{\pT}{\IT{-m}}} to the \pizero\ spectra at \MM{\pT{} \GREATER{} 5\unit{\GeVc}} gives
\MM{\IT{m} = 7.5 \PLMN{} 0.1} (\MM{\SUP{\ensuremath{\chi}}{2}/\RM{ndf} = 6/5}) for \protonproton\ and 
\MM{\IT{m} = 7.9 \PLMN{} 0.2} (\MM{\SUP{\ensuremath{\chi}}{2}/\RM{ndf} = 12/5}) for \deuterongold\ collisions.

\subsection {Eta-to-pion ratio}

\begin {figure} [tb]
\centerline {\hbox {
\includegraphics {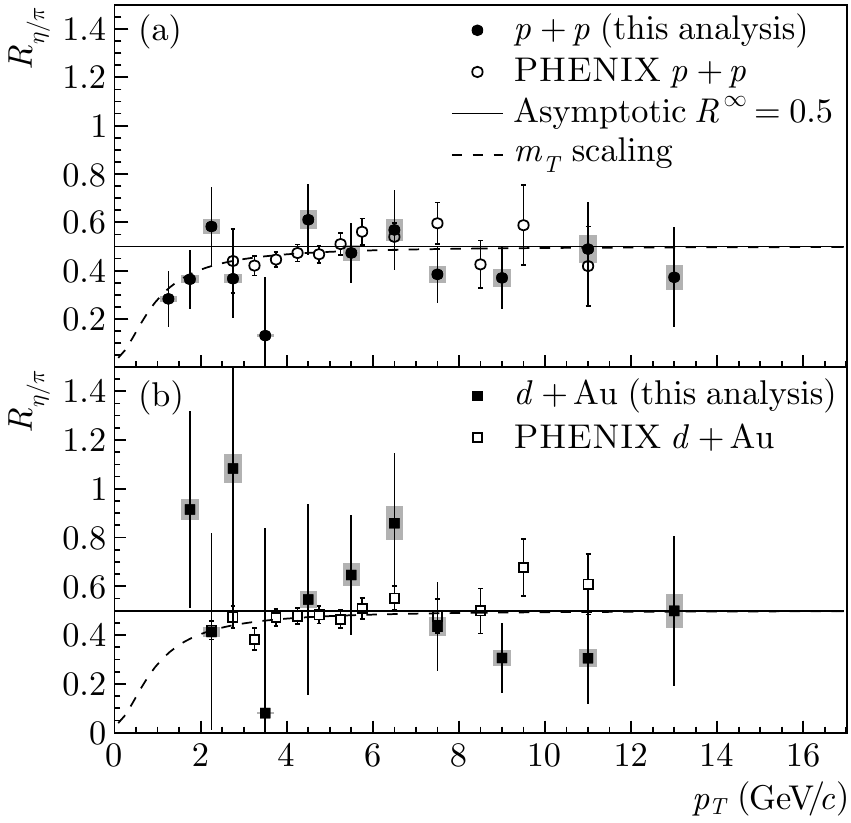}
}}
\caption {The \etatopi\ ratio measured in (a) \protonproton\ and (b) \deuterongold\ collisions at \MM{\sNN{} = 200\unit{\GeV}},
compared to the \PHENIX\ measurements~\protect\cite {ref_phenix_eta} and to the \mT\ scaling predictions\@.
The error bars are statistical and shaded bands are \pT-correlated systematic uncertainties.
\label {fig_etatopi}
}
\end {figure}

The \etameson\ measurement is presented in Fig.~\ref {fig_etatopi} 
as the ratio of \etameson\ to \pizero\ invariant yields (shown in Figs.~\ref {fig_eta_invyield} and~\ref {fig_pi0_invyield}, respectively)\@.
This allows many systematic uncertainties to cancel (see Table~\ref {table_systematic_uncertainty})\@.
The error definitions in the plot are the same as described above for the differential cross sections.
The present measurement agrees very well with previous \PHENIX\ results (open symbols)~\cite {ref_phenix_eta}\@.
The solid lines show the asymptotic ratio 
\MM{\SUP{\IT{R}}{\ensuremath{\infty}} = 0.5}, 
consistent with the world \etatopi\ measurements (see Ref.~\cite {ref_phenix_eta} and references therein)\@.
The fit to our data for \MM{\pT{} \GREATER{} 4\unit{\GeVc}} gives
\MM{\SUB{\IT{R}}{\etatopisub} = 0.46 \PLMN{} 0.05} (\protonproton)
and
\MM{\SUB{\IT{R}}{\etatopisub} = 0.44 \PLMN{} 0.08} (\deuterongold)\@.
The dashed curves in Fig.~\ref {fig_etatopi} show the prediction based on \mT{} scaling~\cite {ref_mT_scaling,schaffner2002,ref_star_resonances_dAu,ref_star_strangepp}\@.
It is seen that the data are consistent with such scaling behavior.

\subsection {Nuclear modification factor}

A convenient way to observe medium-induced modification of particle production is to compare
a nucleus\opdash{}nucleus collision (\collision{A}{B}) with an incoherent superposition of
the corresponding number of individual nucleon\opdash{}nucleon collisions (\nucleonnucleon)\@.
The nuclear modification factor \RAB\ is defined as the ratio of the particle yield in
nucleus\opdash{}nucleus collisions and the yield in nucleon\opdash{}nucleon collisions scaled with the number of binary collisions \Ncoll,
\begin {equation} \label {eq:rabdef}
\RAB{} \ensuremath{\equiv}
\FRAC
    {\DIV{\SUP{\der}{2}\tsp{-0.7}\SUB{\Number}{\IT{AB}}}{\der\pT\tsp{0.5}\der\rapidity}}
    {\TABmean{}\,\DIV{\SUP{\der}{2}\tsp{-0.7}\SUP{\SIGMA}{\protonproton}\tsp{-0.5}}{\der\pT\tsp{0.5}\der\rapidity}}.
\end {equation}
Here \TABmean\ is the nuclear overlap function, which is related to the number
of inelastic \nucleonnucleon\ collisions in one \collision{A}{B}\ collision through
\begin {equation} \label {eq:tabdef}
\TABmean{} \, \SUB{\SUP{\SIGMA}{\nucleon\tsp{-1}\nucleon}}{\RM{inel}} = \Ncollmean.
\end {equation}
In the absence of medium effects, the nuclear modification factor is unity, whereas \MM{\RAB{} \LESS{} 1}
indicates a suppression of particle production in heavy-ion collisions,
compared to an incoherent sum of nucleon\opdash{}nucleon collisions.

We calculated the \RdA\ ratio [Eqs.~(\ref {eq:rabdef}) and~(\ref {eq:tabdef})] as
\begin {equation} \label {eq:rdadef}
\RdA{} =
\FRAC
    {\SUB{\SUP{\SIGMA}{\nucleon\tsp{-1}\nucleon}}{\RM{inel}} \tsp{1} \DIV{\SUP{\der}{2}\tsp{-0.7}\SUB{\Number}{dA}}{\der\pT\tsp{0.5}\der\rapidity}}
    {\Ncollmean\,\DIV{\SUP{\der}{2}\tsp{-0.7}\SUP{\SIGMA}{\protonproton}\ensuremath{\!}}{\der\pT\tsp{0.5}\der\rapidity}}
,
\end {equation}
where the nucleon\opdash{}nucleon inelastic cross section was taken to be \MM{\SUB{\SUP{\SIGMA}{\nucleon\tsp{-1}\nucleon}}{\RM{inel}} = 42\unit{\mb}}
and \MM{\Ncollmean{} = 7.5 \PLMN{} 0.4} was calculated from the Glauber model (see section~\ref {subsec_centralities})\@.

\begin {figure} [tb]
\centerline {\hbox {
\includegraphics {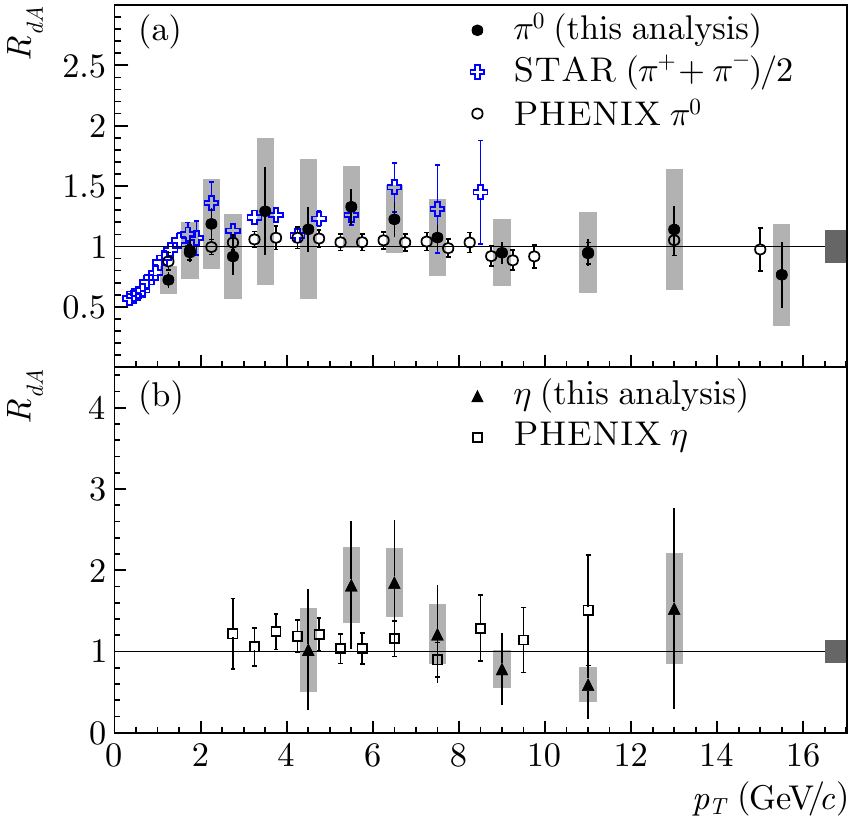}
}}
\caption {(color online) Nuclear modification factor \RdA\ for (a) \pizero\ and (b) \etameson,
compared to the \STAR\ \piplusminus~\protect\cite {ref_star_idhadrons,ref_star_idhadrons1} and 
\PHENIX\ \pizero\ measurements~\protect\cite {ref_phenix_eta,ref_phenix_eta_pi0_dAu}\@.
The error bars are statistical and shaded bands are \pT-correlated systematic uncertainties.
Normalization uncertainties are indicated by shaded bands around unity in each panel.
\label {fig_rda_star}
}
\end {figure}

The nuclear modification factors for \pizero\ and \etameson\ are shown in Fig.~\ref {fig_rda_star}\@.
The definition of the errors is the same as given for the differential cross sections in section~\ref {corrected_yields}\@.
Figure~\ref {fig_rda_star} also shows the \RdA\ for \piplusminus\ measured by \STAR~\cite {ref_star_idhadrons,ref_star_idhadrons1}\@.
A good agreement between neutral and charged pion measurements by \STAR\ is observed.
Our \pizero\ and \etameson\ data also agree reasonably well with the corresponding \PHENIX\ measurements~\cite {ref_phenix_eta,ref_phenix_eta_pi0_dAu}\@.

\begin {figure} [tb]
\centerline {\hbox {
\includegraphics {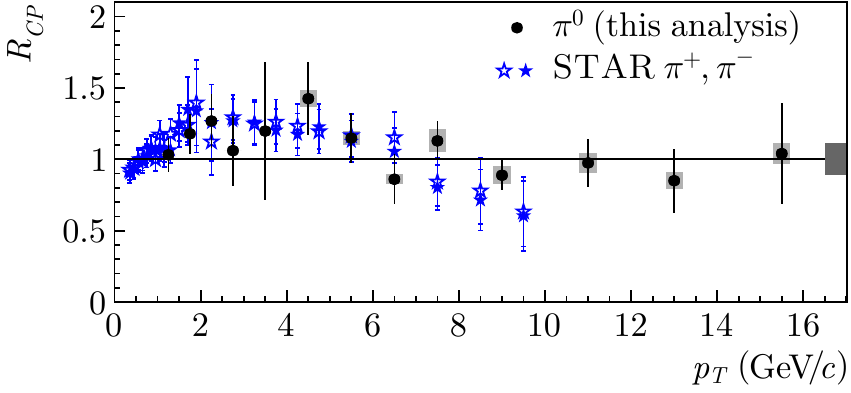}
}}
\caption {(color online) Nuclear modification factor \Rcp\ measured in \deuterongold\ collisions,
compared to \STAR\ \piplusminus\ measurement~\protect\cite {ref_star_idhadrons}\@.
The error bars are statistical and shaded bands are \pT-correlated systematic uncertainties.
Common normalization uncertainty is indicated by a shaded band around unity.
\label {fig_rcp}
}
\end {figure}

In peripheral \deuterongold\ collisions, the number of participant nucleons is small and the creation of a dense medium is not expected.
This suggests that, instead of \protonproton\ interactions, peripheral collisions can be used as a reference.
This was done through the ratio of particle production in \MM{0}--\MM{20}\unitns{\%} central (\IT{C}) 
and \MM{40}--\MM{100}\unitns{\%} peripheral (\IT{P}) events,
\begin {equation} \label {eq:rcpdef}
\Rcp{} = 
\FRAC 
    {\SUB{\Ncollmean}{\IT{P}}} 
    {\SUB{\Ncollmean}{\IT{C}}} 
\FRAC 
    {\DIV{\SUP{\der}{2}\tsp{-0.7}\SUB{\Number}{\IT{C}}}{\der\pT\tsp{0.5}\der\rapidity}} 
    {\DIV{\SUP{\der}{2}\tsp{-0.7}\SUB{\Number}{\IT{P}}}{\der\pT\tsp{0.5}\der\rapidity}}.
\end {equation}
The advantage of this measure is that no \protonproton\ reference data are needed.
The disadvantage is that a stronger model dependence is introduced due to the uncertainty in \Ncollmean\@.
Figure~\ref {fig_rcp} 
shows the \Rcp\ ratio for \pizero, compared to the \STAR\ \piplusminus\ data~\cite {ref_star_idhadrons}\@.
It is seen that the agreement between the neutral and charged pion measurements is very good.
The ratio stays constant at a value consistent with unity beyond \MM{\pT{} = 8\unit{\GeVc}} and,
therefore, does not support a possible decrease of the ratio at high \pT, which was suggested by the \piplusminus\ measurement.

\subsection {Direct photons}

\begin {figure} [tb] 
\includegraphics {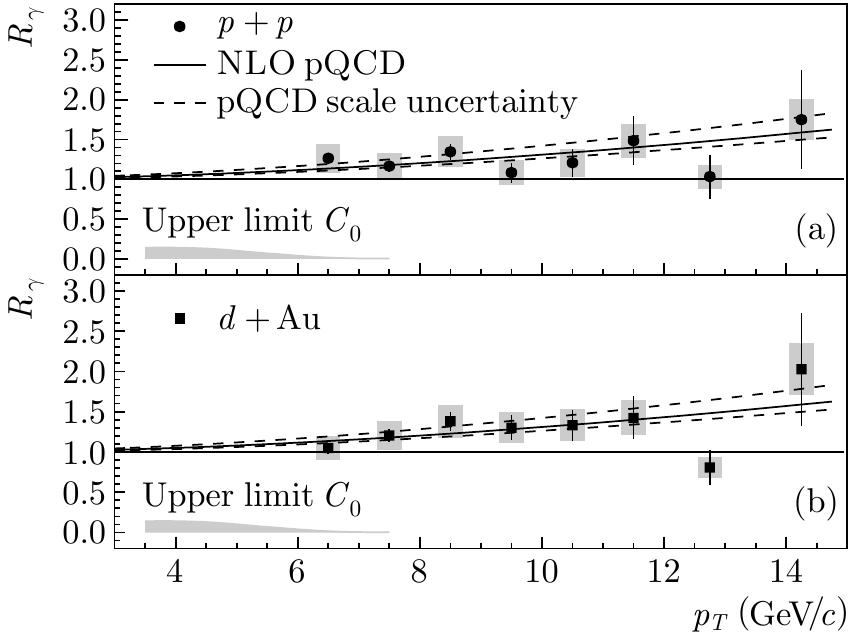}
\caption {The direct photon yield in (a) \protonproton\ and (b) \deuterongold\ collisions at \MM{\sNN{} = 200\unit{\GeV}},
expressed in terms of the double ratio \Rgamma\@. 
The error bars are statistical and the shaded bands are \pT-correlated systematic uncertainties.
The curves correspond to \NLO\ \pQCD\ calculations 
of the differential cross sections for direct photon~\protect\cite {ref_nlo_pqcd_photons} and \pizero~\protect\cite {ref_nlo_pqcd}
production in \protonproton\ collisions for different factorization scales \muscale\ (the upper \pQCD\ curve corresponds to \MM{\muscale{} = \DIV{\pT}{2}}).\@
The upper limit of the fractional neutral hadron contamination \SUB{\IT{C}}{0} is shown as the shaded band at \MM{\Rgamma{} = 0}\@. 
\label {fig:dirphoton_pqcd}
}
\end {figure}

The double ratio \Rgamma\ [Eq.~(\ref {eq:rgamma_def})] measured in \protonproton\ and \deuterongold\ collisions is shown in Fig.~\ref {fig:dirphoton_pqcd}\@.
The shaded band near \MM{\Rgamma{} = 0}
indicates our estimate of the upper limit of the remaining neutral hadron contamination. 
The curves correspond to \NLO\ \pQCD\ calculations~\cite {ref_nlo_pqcd_photons}, which were further evaluated as 
\begin {equation} 
\label {eq:discussion:1} 
\Rgamma\big|_{\RM{theor}} = 1 + 
\FRAC
    {\SUB{\left(\DIV{\SUB{\gama}{\RM{dir}}}{\pizero}\right)}{\RM{\NLO}}} 
    {\SUB{\left(\DIV{\SUB{\gama}{\RM{decay}}}{\pizero}\right)}{\RM{simu}}}
,
\end {equation}
where the numerator is the ratio of the \NLO\ \pQCD\ direct photon and \pizero\ cross sections. 
The denominator is given by the number of decay photons per \pizero, as determined by the simulation  
described in section~\ref {sec:backgr-from-hadr}\@.  
 
The \NLO\ \pQCD\ calculation used the \CTEQ6\capsword{M}~\cite {ref_cteq6} parton  
densities and the \GRV~\cite {ref_grv} parton-to\tsp{0.4}-photon fragmentation functions as an input. 
The scale dependence of this calculation, indicated by the dashed curves in the figure, 
was obtained by changing the scale \muscale\ in the calculation of prompt photon production, 
while keeping the scale corresponding to the \pizero\ cross section fixed at \MM{\muscale{} = \pT}\@. 
In addition, we have varied the factorization scale for both cross sections  
simultaneously. The observed variation was quantitatively similar, although  
in the opposite direction. Since the measured \pizero\ spectrum favors  
the result of the \pQCD\ calculation with \MM{\muscale{} = \pT}, we have used this value  
for all three curves.

Although Fig.~\ref {fig:dirphoton_pqcd} demonstrates that the measured  
values of \Rgamma\ are consistent with the calculated direct photon signal,  
the interpretation in this context has its limitations. First, the   
curves do not follow directly from the theory but depend on our simulation 
of the decay photon yields, as shown by Eq.~(\ref {eq:discussion:1})\@. 
In addition, the \NLO\ \pQCD\ cross section for \pizero\ production is less  
accurately constrained than that for prompt photon production. 
To allow for a more solid comparison to theoretical predictions, as well as to other experimental data, we have  
converted \Rgamma\ to an absolute cross section [Eq.~(\ref {eq:gamma_direct_def})]\@.

The calculation of absolute direct photon yields required that the systematic errors associated with 
inclusive photon yields, which canceled in the ratio \Rgamma, were included again.
We derived the \MM{95\unitns{\%}} confidence limits for the cross section in the \pT\ bins 
where \Rgamma\ did not correspond to a significant direct photon signal,
assuming that the statistical and systematic errors both followed a Gaussian distribution,
and using the fact that \MM{\Rgamma{} \GREATEREQ{} 1} by definition.

\begin {figure} [tb] 
\includegraphics {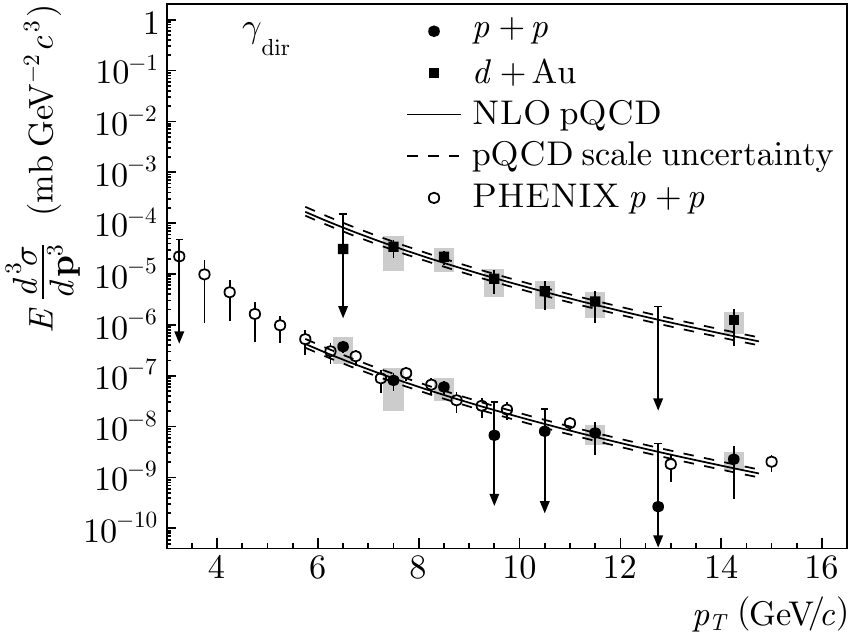}
\caption {The differential cross section for direct photon 
production at midrapidity in \protonproton\ and \deuterongold\ collisions at \MM{\sNN{} = 200\unit{\GeV}},
compared to the \PHENIX\ measurement~\protect\cite{ref_phenix_photons_pp} and 
to the \NLO\ \pQCD\ calculation~\protect\cite{ref_nlo_pqcd_photons},
which was scaled with \TdAmean\ [Eq.~(\protect\ref {eq:tabdef})] in case of \deuterongold\ collisions.
The error bars are statistical the shaded bands are \pT-correlated systematic uncertainties. 
The arrows correspond to the \MM{95\unitns{\%}} confidence limits, as defined in the text. 
\label {fig:xsec_photons}
}
\end {figure}

Figure~\ref {fig:xsec_photons}
shows the invariant cross section for direct photon production in \protonproton\ and \deuterongold\ collisions.
The normalization uncertainties are not explicitly given in the figure. 
The \NLO\ \pQCD\ cross section for direct photon production in \protonproton\ collisions was scaled  
with the nuclear thickness function \TdAmean~[Eq.~(\ref {eq:tabdef})] to account for the number  
of binary collisions in the \deuterongold\ system. 
The precision of the presented measurement is limited by systematic  
uncertainties for \MM{\pT{} \LESSEQ{} 9\unit{\GeVc}} and by statistical uncertainties 
for larger \pT\ values. Nevertheless, our results 
are compatible with the \NLO\ \pQCD\ calculations.
Our data are also in a good agreement with the direct photon cross section in \protonproton\ collisions measured by \PHENIX~\cite {ref_phenix_photons_pp}\@.

\begin{figure}[tb]
\centerline{\hbox{
\includegraphics {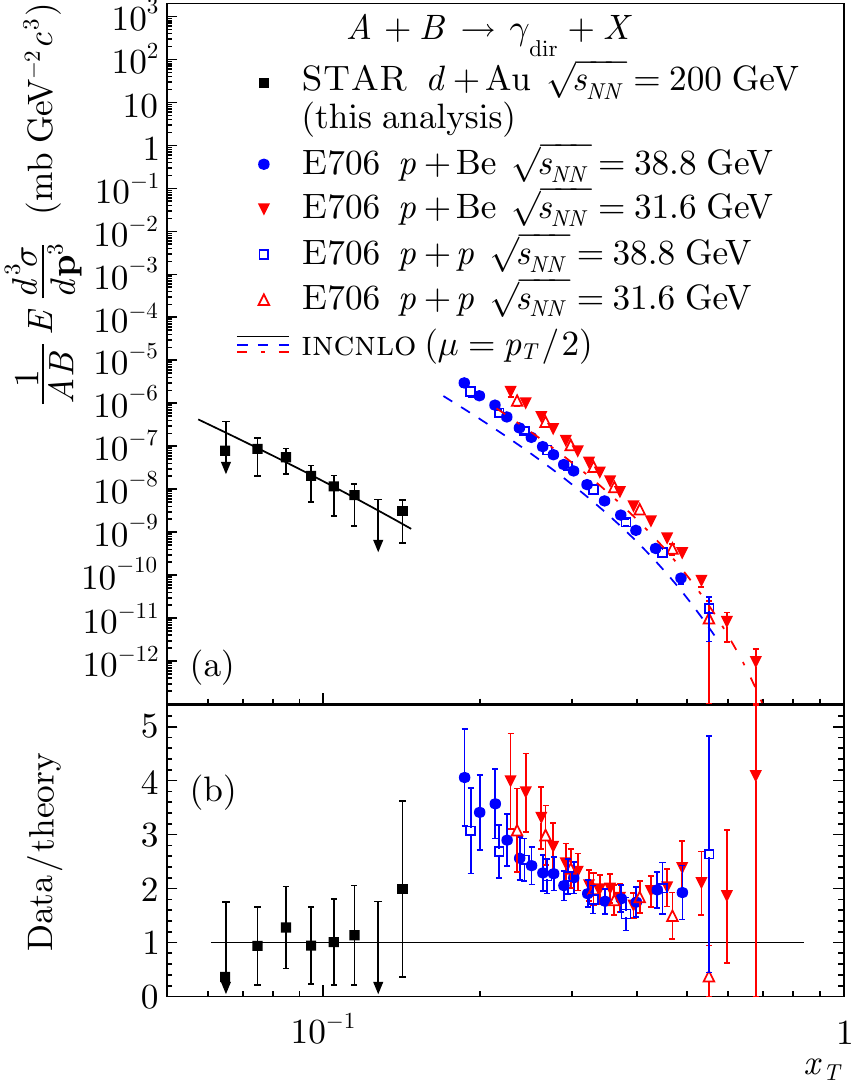}
}}
\caption {(color online) (a) The cross sections per nucleon for direct photon production in \deuterongold\ collisions,
compared to the measurements in \protonproton\ and \protonnucleus\ collisions by E706 experiment~\cite {ref_e706_direct_photons_fixed_target}
at the comparable nucleon\opdash{}nucleon center-of-mass energies.
The theoretical curves were calculated with the \INCNLO\ program~\protect\cite{ref_critical_phenomenological_study_photons} 
with \MM{\ensuremath{\mu} = \DIV{\pT}{2}}\@.
All data have statistical and systematic uncertainties summed in quadrature. 
The vertical arrows indicate our estimate of the \MM{95\unitns{\%}} confidence level.
(b) The ratio of the data and the corresponding calculations.
\label {fig:world_dau}
}
\end{figure}

Earlier measurements of direct photon production in proton\opdash{}nucleus collisions have been 
performed by the E706 experiment~\cite {ref_e706_direct_photons_fixed_target} 
by scattering protons on a fixed beryllium target, with the proton beam energies of \MM{530} and \MM{800\unit{\GeV}}\@.
Those data show a strong discrepancy with \pQCD\ calculations, 
which was attributed to multiple soft gluon radiation 
and phenomenologically described as an additional transverse impulse \kT\ to the incoming partons~\cite{ref_e706_direct_photons_fixed_target}\@.
It has also been argued that this discrepancy might be due to nuclear modifications 
present even in the light berillium nucleus used~\cite{ref_saturated_cronin}, 
although the \protonproton\ data by E706 show similar behavior at the same \sNN\@.
It is, therefore, of interest to compare our \deuterongold\ results to those of E706,
as shown in Fig.~\ref {fig:world_dau}
as a function of \MM{\xT{} \ensuremath{\equiv} \DIV{2\tsp{0.5}\pT\tsp{-0.3}}{\tsp{-0.3}\sqrts}}, 
which is a suitable variable to compare data taken with different beam energies.
Whereas the ratio data/theory from Ref.~\cite {ref_e706_direct_photons_fixed_target} 
shows an increase of up to a factor of 4 towards low \xT,
our results at still lower \xT\ constrain such a potential deviation 
from theory to less than a factor of \MM{2}\@.
It should be noted, however, that the data
have been taken at significantly different \sqrts\@.

\begin{figure}[tb]
\centerline{\hbox{
\includegraphics {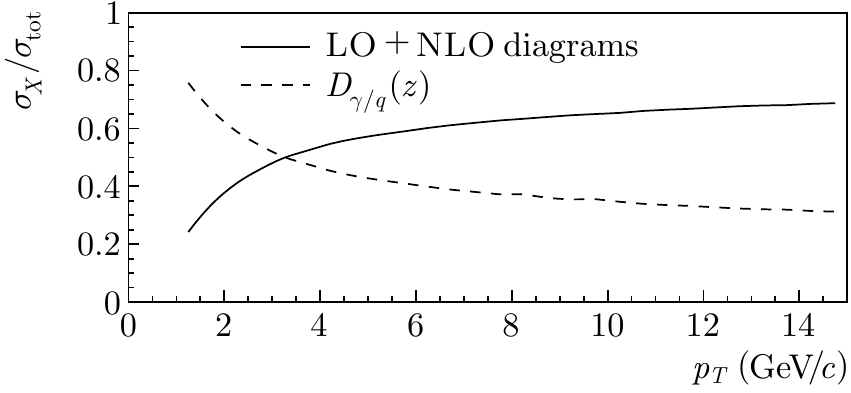}
}}
\caption {The relative contribution of the fragmentation 
(\tsp{-0.5}\SUB{\IT{D}}{\DIV{\gama\tsp{-0.6}}{\tsp{-0.6}\IT{q}}}\tsp{-0.4}) and the \pQCD\ hard scattering processes to
the total direct photon cross section~\protect\cite {ref_nlo_pqcd_photons}\@. 
Both contributions are shown as a function of the \pT\ of the produced photon.
\label {fig:relfrag}
}
\end{figure}

We have included both prompt and fragmentation components 
in the \pQCD\ calculations, since our measurement was based on an inclusive sample of photons. 
The theoretical calculation of these two components is shown in Fig.~\ref {fig:relfrag}\@. 
A first measurement of the contribution from fragmentation photons to the total direct photon cross section 
in \protonproton\ collisions at \RHIC\ was reported in Ref.~\cite {ref_phenix_photons_pp}\@. 

The interest in disentangling photons from the fragmentation process and  
from the initial hard scattering is twofold. First, it has been observed  
that the hot and dense medium produced in central heavy-ion collisions at \RHIC\ causes a suppression of particle yields,
which has been attributed to induced gluon radiation from a parton traversing the medium. 
The same mechanism could lead to a suppression of fragmentation 
photons, although an enhancement of directly produced photons has been proposed as well~\cite {ref_quench_photons}\@. 
Second, a measurement of the identified prompt photons in \protonproton\ collisions is of interest for the \RHIC\ spin program, 
a large part of which is devoted to constraining the gluon spin contribution to the spin of proton.
The isolation criterion selects the quark\opdash{}gluon Compton process and, therefore, enhances the  
sensitivity of the cross section to the gluon content of the proton.  

\subsection {Summary}

The present \pizero\ spectrum complements that of the \piplusminus, which was measured by \STAR\ in 
the transverse momentum range \MM{0.35 \LESS{} \pT{} \LESS{} 10\unit{\GeVc}}, and extends up to \MM{\pT{} = 17\unit{\GeVc}}\@.
There is a good agreement between the neutral and charged pion cross sections in \STAR,
even though very different methods and detector subsystems were used.
The \pizero\ cross section also agrees well with the measurements of \PHENIX,
and with the theoretical \NLO\ \pQCD\ calculations.

This paper presents the first measurements of \etameson\ meson production by \STAR,
which are in agreement with the \PHENIX\ measurements and with the \mT\ scaling assumption.

We present the measurements of the nuclear modification factor \RdA, where the \pizero\ production in \deuterongold\ collisions is
compared to that in \protonproton, and \Rcp, the comparison between central and peripheral \deuterongold\ collisions.
Both results are consistent with unity at high \pT, are in a good agreement with the \piplusminus\ measurements, previously made by \STAR,
and significantly extend the \pT\ range for light meson production measurements.

This paper also reports the first measurement of direct photon production by \STAR\@. 
A direct photon signal consistent with \NLO\ \pQCD\ calculation has been observed at high \pT\ for both systems.
No strong modification of photon production in \deuterongold\ collisions was observed.

The results will provide an important baseline for future \goldgold\ measurements in \STAR\@.

\begin {acknowledgments}
We thank the \RHIC\ Operations Group and \RCF\ at \BNL,  
the \NERSC\ Center at \LBNL\ and the Open Science Grid consortium for providing resources and support.  
This work was supported in part by the Offices of \capsword{NP} and \capsword{HEP} within the U\tsp{-0.6}.\tsp{1}S.\ \capsword{DOE} Office of Science,  
the U\tsp{-0.6}.\tsp{1}S.\ \capsword{NSF}, the Sloan Foundation, the \capsword{DFG} cluster of excellence `Origin and Structure of the Universe',  
\capsword{CNRS/IN2P3}, \capsword{STFC} and \capsword{EPSRC} of the United Kingdom, \capsword{FAPESP} \capsword{CNP}q of Brazil,  
Ministry of Ed.\ and Sci.\ of the Russian Federation,  
\capsword{NNSFC}, \capsword{CAS}, \capsword{M}o\capsword{ST}, and \capsword{M}o\capsword{E} of China,  
\capsword{GA} and \capsword{MSMT} of the Czech Republic,  
\capsword{FOM} and \capsword{NWO} of the Netherlands, \capsword{DAE}, \capsword{DST}, and \capsword{CSIR} of India,  
Polish Ministry of Sci.\ and Higher Ed.,  
Korea Research Foundation, Ministry of Sci., Ed.\ and Sports of the Rep.\ Of Croatia,  
Russian Ministry of Sci.\ and Tech, and RosAtom of Russia. 
\end {acknowledgments}

\end {document}